\documentclass[10pt]{article}
\pdfoutput=1

\usepackage{amsthm,amsfonts,amsmath,amssymb}
\usepackage{pxfonts}
\usepackage{euscript}
\usepackage{hyperref}
\usepackage{xcolor}
\usepackage{tcolorbox}

\usepackage{xypic}
\usepackage[all,2cell,arrow,matrix]{xy} \UseAllTwocells \SilentMatrices
\usepackage{graphicx}
\usepackage{verbatim}
\usepackage{leftidx}

\usepackage{enumerate}
\usepackage{bbold}

\usepackage{sectsty}
\sectionfont{\Large}

\newcommand\void[1]       {}

\theoremstyle{definition}

\newtheorem{thm}{Theorem}[section]

\newtheorem{lem}[thm]{Lemma}

\theoremstyle{definition}
\newtheorem{defn}[thm]{Definition}
\newtheorem{expl}[thm]{Example}
\newtheorem{rem}[thm]{Remark}
\newtheorem{pthm}[thm]{Theorem$^{\mathrm{ph}}$}

\numberwithin{equation}{section}
\numberwithin{thm}{section}

\newcommand\nn             {\nonumber \\}
\newcommand\be            {\begin{equation}}
\newcommand\ee            {\end{equation}}
\newcommand\bea           {\begin{eqnarray}}
\newcommand\eea         {\end{eqnarray}}
\newcommand\bnu          {\begin{enumerate}}
\newcommand\enu          {\end{enumerate}}

\allowdisplaybreaks[1]

\newlength{\fighskip} \fighskip=2pt
\newlength{\figvskip} \figvskip=3pt

\newcommand{\pf}{\begin{proof}}
\newcommand{\epf}{\end{proof}}

\newcommand\Cb            {\mathbb{C}}

\newcommand\Rb            {\mathbb{R}}
\newcommand\Zb            {\mathbb{Z}}
\newcommand\Z             {\mathfrak{Z}}

\newcommand\CA           {\EuScript{A}}
\newcommand\CB           {\EuScript{B}}
\newcommand\CC           {\EuScript{C}}
\newcommand\CD           {\EuScript{D}}
\newcommand\CE          {\EuScript{E}}
\newcommand\CF          {\EuScript{F}}

\newcommand\CI           {\EuScript{I}}
\newcommand\CJ           {\EuScript{J}}
\newcommand\CK         {\EuScript{K}}
\newcommand\CL          {\EuScript{L}}
\newcommand\CM          {\EuScript{M}}
\newcommand\CN         {\EuScript{N}}
\newcommand\CO         {\EuScript{O}}
\newcommand\CP         {\EuScript{P}}
\newcommand\CQ         {\EuScript{Q}}
\newcommand\CR         {\EuScript{R}}
\newcommand\CS         {\EuScript{S}}
\newcommand\CT         {\EuScript{T}}
\newcommand\CU        {\EuScript{U}}

\newcommand\CX         {\EuScript{X}}
\newcommand\CY         {\EuScript{Y}}
\newcommand\CZ         {\EuScript{Z}}

\newcommand\CBs{{\EuScript{B}^\sharp}}

\newcommand\CSs{{\EuScript{S}^\sharp}}

\newcommand\CXs{{\EuScript{X}^\sharp}}

\newcommand{\FZ}{\mathfrak{Z}}

\newcommand{\ising}{\mathbf{Is}}
\newcommand{\toric}{\mathbf{Tor}}
\newcommand{\rep}{\mathrm{Rep}}

 \DeclareMathOperator{\id}{id}

 \DeclareMathOperator{\ev}{ev}

 \DeclareMathOperator{\fun}{Fun}

 \DeclareMathOperator{\Mod}{Mod}
 
 \DeclareMathOperator{\RMod}{RMod}

\newcommand{\rev}{\mathrm{rev}}

\newcommand{\one}{\mathbf1}

\newcommand\bh{\mathbf{H}}

\newcommand\forget  {\mathbf{f}}

\newcommand\op {\mathrm{op}}
\newcommand\bulk {\mathrm{bulk}}
\newcommand\vect {\mathrm{Vec}}

\newcommand\iemfc {{{}^{\mathrm{ind}}\CE\CU\CM\CF}}
\newcommand\nbfc {{{}^{\mathrm{ndg}}\CB\CF^{\mathrm{en-cl}}}}

\newcommand{\ssimeq} { \overset{\mathrm{sp}}{\simeq}}

\begin{document}

\begin{center} \LARGE
A mathematical theory of gapless edges of 2d topological orders. Part II
\end{center}

\vskip 1em
\begin{center}
{\large
Liang Kong$^{a}$,\,
Hao Zheng$^{a,b}$\,
~\footnote{Emails:
{\tt  kongl@sustc.edu.cn, zhengh@sustech.edu.cn}}}
\\[1em]
$^a$ 
Guangdong Provincial Key Laboratory of Quantum Science and Engineering,\\
Shenzhen Institute for Quantum Science and Engineering,\\
Southern University of Science and Technology, Shenzhen 518055, China 
\\[0.7em]
$^b$ Department of Mathematics, Peking University\\
Beijing 100871, China
\end{center}

\vskip 3em

\begin{abstract}
This is the second part of a two-part work on the unified mathematical theory of gapped and gapless edges of 2+1D topological orders. In Part I, we have developed the mathematical theory of chiral gapless edges. In Part II, we study boundary-bulk relation and non-chiral gapless edges. In particular, we explain how the notion of the center of an enriched monoidal category naturally emerges from the boundary-bulk relation. After the study of 0+1D gapless walls, we give the complete boundary-bulk relation for 2+1D topological orders with chiral gapless edges (including gapped edges) and 0d walls between edges. This relation is stated precisely and proved rigorously as a monoidal equivalence, which generalizes the functoriality of the usual Drinfeld center to an enriched setting. We also develop the mathematical theory of non-chiral gapless edges and 0+1D walls, and explain how to gap out certain non-chiral 1+1D gapless edges and 0+1D gapless walls categorically. In the end, we show that all anomaly-free 1+1D boundary-bulk rational CFT's can be recovered from 2d topological orders with chiral gapless edges via a dimensional reduction process. This provides physical meanings to some mysterious connections between mathematical results in fusion categories and those in rational CFT's. 
\end{abstract}

\tableofcontents
\vspace{1cm}

\section{Introduction} \label{sec:introduction}

Throughout this paper, we use $n$d to denote the spatial dimension and $n+$1D to denote the spacetime dimension, and we use \textbf{Theorem}$^{\mathrm{ph}}$ to highlight a physical result and use \textbf{Theorem} to represent a mathematically rigorous result.  

\medskip
This work is a continuation of \cite{kz4}, in which we have developed the mathematical theory of chiral gapless edges of 2d topological orders (without symmetries). Its main result is summarized by the following physical theorem.
\begin{pthm}[\cite{kz4}]
1d chiral gapped/gapless edges of an anomaly-free 2d topological order $(\CC,c)$ is mathematically described and classified by pairs $(V,{}^\CB\CX)$, where
\bnu
\item $V$ is the chiral symmetry, i.e. a unitary rational VOA such that the category $\CB:=\Mod_V$ of $V$-modules is a UMTC. When $V=\Cb$, the edge is gapped and $\CB=\bh$.
\item ${}^\CB\CX$ is a $\CB$-enriched unitary fusion category canonically constructed from the pair $(\CB,\CX)$, where $\CX$ is a UFC and a left fusion $\CB$-module (see Definition\,\ref{def:fusion-bimodule}), and $\CX$ is the underlying category of the enriched category ${}^\CB\CX$. More explicitly, 
\bnu

\item objects $x,y,z \in {}^\CB\CX$ are precisely those in $\CX$, and are topological edge excitations; 

\item the morphism spaces are given by the internal homs, i.e. $\hom_{{}^\CB\CX}(x,y):=[x,y]_\CB$. 

\enu
\enu
Moreover, we have a direct sum decomposation ${}^\CB\CX={}^\CB\CX_1 \oplus \cdots \oplus {}^\CB\CX_n$, where each indecomposable direct summand $\CX_i$ of $\CX$ is the category of boundary conditions of a modular-invariant bulk CFT $A_\bulk^{(i)}$. For $x,y\in\CX_i$, $[x,x]_\CB$ and $[y,y]_\CB$ are boundary CFT's of $A_\bulk^{(i)}$ and $[x,y]_\CB$ is a 0D wall between them. The bulk CFT's $A_\bulk^{(i)}$ and $A_\bulk^{(j)}$ are potentially different for $i\neq j$. The space $[x,y]_\CB$ should also be viewed as the space of instantons bewteen two edge excitations $x$ and $y$. When $V=\Cb$, $\CX=(\Cb,{}^\bh\CX)$ is a gapped edge.
\end{pthm}

One of the consequences of above theorem is that all 1d chiral gapless edges are obtained from {\it topological Wick rotations} \cite[Section\,5.2]{kz4}. It is physically absurd if this result does not generalize to 0d walls between edges. Inspired by this observation, we propose the following correspondence, which plays the role of a guiding principle of this work. 
\begin{quote}
{\bf Gapped-gapless Correspondence}: All gapless edges and 0d walls between edges of 2d topological orders can be obtained from topological Wick rotations plus the information of local quantum (i.e. chiral or non-chiral) symmetries. 
\end{quote}
Actually, some parts of our analysis do not use this principle but leads to results respecting this principle. At the end of the day, all 0d walls respect this principle. This is a low dimensional case of a more general principle for gapless phases in all dimensions proposed in \cite[Section\,7]{kz4}. It provides a powerful tool and a guiding principle for the study of gapless phases in all dimensions (see Section\,\ref{sec:outlooks}).

\medskip
In this work, we develop the mathematical theory of 0+1D walls between two gapless edges, boundary-bulk relation including 0+1D walls and that of non-chiral gapless edges. 
The logic flow and the layout of this work are given below. 

\medskip
In Section\,\ref{sec:bb-relation-1}, we review the boundary-bulk relation for gapped edges. In particular, in Section\,\ref{sec:bfc}, we review some basic mathematical notions, such as a closed module over a multi-fusion category and a closed monoidal modules over a braided fusion category. In Section\,\ref{sec:gapped-0d-wall}, we review the mathematical theory of 0d wall between gapped edges. In Section\,\ref{sec:bbr-gapped}, we review the boundary-bulk relation for gapped edges including 0d walls. In Section\,\ref{sec:fh}, we review the theory of generalized 0d defects and that of factorization homology. The anomaly-free condition discussed there will be used in many places later.  

In Section\,\ref{sec:0d-walls}, we develop the mathematical theory of 0+1D walls between two chiral gapless edges. We start with a careful analysis of observables on the world line of a 0+1D wall in Section\,\ref{sec:0d-defects}. This analysis shows that a natural construction of a 0+1D wall automatically respects the Gapped-gapless Correspondence. In Section\,\ref{sec:0d-1D}, we discuss more general constructions all respecting the Gapped-gapless Correspondence. In particular, we show that a 1+1D chiral symmetry in a neighborhood of the world line and a 0+1D chiral symmetry on the world line are both needed as defining data. Since there is no thermodynamics in 0d, we carefully distinguish the spatial notion of a 0d wall and the spacetime notion of a 0+1D wall (see Definition\,\ref{def:0d-1D-phases}). All 0+1D walls are spatially equivalent to the unique 0d wall. In Section\,\ref{sec:classification-chiral-0d-wall}, using Gapped-gapless Correspondence, we conclude that we have found the mathematical description and the classification of all 0+1D gapless walls between two chiral gapless edges. In Section\,\ref{sec:fusion-anomaly}, we discuss how to fuse two 0+1D walls along a spatial direction and an anomaly associated to it, called spatial fusion anomaly. In Section\,\ref{sec:morita}, we show that the spatial equivalence between two 0+1D walls leads to a mathematical notion of a spatial equivalence between bimodules over enriched multi-fusion categories and the associated spatial Morita theory. As a consequence, two chiral gapless edges are spatially Morita equivalent if and only if they share the same bulk, and the spatial Morita equivalence is precisely defined by a 0+1D gapless wall as a spatially invertible bimodule. 

After the preparation in Section\,\ref{sec:0d-walls}, we are ready to give a complete boundary-bulk relation for chiral gapless edges in Section\,\ref{sec:boundary-bulk}. We warm up to the precise statement by first explaining how the notion of the center of an enriched monoidal category naturally emerges from the physical intuition of the relation between a 2d bulk and a 1d edge in Section\,\ref{sec:bulk} and \ref{sec:drinfeld-center}. In Section\,\ref{sec:Z1}, we add 0+1D walls to the edge and 1+1D gapless walls to the bulk. In Section\,\ref{sec:main-thm}, we give our main mathematical result (see Theorem\,\ref{thm:ff-functor}). It says that assigning the data on the boundary to that in the bulk by taking centers gives a well-defined functor, which is actually a monoidal equivalence. This generalizes our earlier result of the functoriality of Drinfeld center in \cite[Theorem\,3.3.7]{kz1}.

In Section\,\ref{sec:class-gapless-edge}, we develop the mathematical theory of non-chiral gapless edges. The logic flow there is parallel to that of chiral gapless edges. In particular, we provide a classification of non-chiral gapless edges in Section\,\ref{sec:classification-non-chiral-edges}, and discuss its significance in the study of purely edge topological phase transitions in Section\,\ref{sec:edge-tpt}. Different from the chiral cases, two non-chiral gapless edges can have very complicated 0+1D gapless walls. Mathematically, this corresponds to the representation theories of non-chiral symmetries in different categories. We explain this in Section\,\ref{sec:non-chiral-edge-chiral-wall}.

In Section\,\ref{sec:compute}, we show how to use our theory to compute various physical processes. In particular, in Section\,\ref{sec:gappable-edges-morita}, we show how to gap out a non-chiral 0+1D walls; in Section\,\ref{sec:fusion-holes}, we show how to fuse two gapless holes in a 2d topological order; in Section\,\ref{sec:cft}, we show how to recover all 1+1D anomaly-free boundary-bulk CFT's via a dimensional reduction process. At the same time, we clarify some mysterious connections between mathematical results in fusion categories and those in rational CFT's.

In Section\,\ref{sec:outlooks}, we discuss two important lessons we have learned from this work. These lessons are important to the future study of higher dimensional gapped/gapless phases.


\medskip
\noindent {\bf Acknowledgement}: LK and HZ are supported by the Science, Technology and Innovation Commission of Shenzhen Municipality (Grant No. ZDSYS20170303165926217) and by Guangdong Provincial Key Laboratory (Grant No.2019B121203002). LK is also supported by NSFC under Grant No. 11971219. HZ is supported by NSFC under Grant No. 11131008.

\section{Boundary-bulk relation I: gapped edges} \label{sec:bb-relation-1}

In this section, we review some basic mathematical notions and boundary-bulk relation for gapped edges of 2d topological orders. 

\subsection{Basics of braided fusion categories} \label{sec:bfc}

For a unitary multi-fusion category (UMFC) $\CC$, we denote its tensor product by $\otimes$, its tensor unit by $\one_\CC$ and the identity morphisms by $1_x:x\to x$ for $x\in\CC$. A UMFC is called indecomposable if it is a not direct sum of two non-zero UMFC's. A unitary fusion category (UFC) is a UMFC with a simple tensor unit. We use $\CC^\rev$ to denote the same category as $\CC$ but equipped with the tensor product $\otimes^\rev$ defined by $a\otimes^\rev b:=b\otimes a$; and use $\CC^\op$ to denotes the opposite category. The simplest UFC is the category $\bh$ of finite dimensional Hilbert spaces. Deligne tensor product is denoted by $\boxtimes$.

\begin{defn} \label{def:closed-module}
For UMFC's $\CB$ and $\CC$, a left $\CB$-module is a finite unitary category $\CM$ equipped with a unitary monoidal functor 
\be \label{eq:module-phi-M}
\phi_\CM: \CB \to \fun(\CM, \CM),
\ee
where $\fun(\CM,\CM)$ denotes the category of unitary functors from $\CM$ to $\CM$; a right $\CC$-module is a left $\CC^\rev$-module; a $\CB$-$\CC$-bimodule is a left $\CB\boxtimes\CC^\rev$-module. A (left, right, bi-)module is called {\it closed} if $\phi_\CM$ is also an equivalence. 
\end{defn}

\begin{rem}
For a left $\CB$-module module, for $b\in\CB$, we often denote the endo-functor $\phi_\CM(b): \CM \to \CM$ by $b\odot -$, where $\odot: \CB \times \CM \to \CM$ is a well-defined $\CB$-action on $\CM$. Two $\CB$-modules $\CM$ and $\CN$ are equivalent if there exists an equivalence between $\CM$ and $\CN$ intertwining the $\CB$-actions. 
\end{rem}

For UMFC's $\CA,\CB,\CC$, an $\CA$-$\CB$-bimodule $\CM$ and a $\CB$-$\CC$-bimodule $\CN$, the relative tensor product $\CM\boxtimes_\CB\CN$ is a well-defined $\CA$-$\CC$-bimodule. We have a well-defined symmetric mononoidal category:
\begin{itemize}
\item ${}^{\mathrm{ind}}\CU\CM\CF$: objects are indecomposable UMFC's; morphisms are the equivalence classes of bimodules; the composition maps are defined by relative tensor products; the symmetric tensor product is the Deligne tensor product. 
\end{itemize}

For a UMTC $\CC$, we use $\overline{\CC}$ to denote the same unitary fusion category but with the braidings defined by the anti-braidings of $\CC$.  

\begin{defn}[\cite{kz1}] \label{def:fusion-bimodule}
For UMTC's $\CB$ and $\CC$, a (multi-)fusion right $\CC$-module $\CM$ is a UFC (or UMFC) equipped with a unitary braided monoidal functor $\phi_\CM: \CC \to \FZ(\CM)$; a (multi-)fusion left $\CB$-module is a (multi-)fusion right $\overline{\CB}$-module; a (multi-)fusion $\CB$-$\CC$-bimodule is a (multi-)fusion right $\overline{\CB}\boxtimes \CC$-module. Such a (left, right, bi-)module is called {\it closed} if $\phi_\CM$ is also an equivalence. 
\end{defn}

Given a multi-fusion right $\CC$-module $\CM$, by composing $\phi_\CM$ with the forgetful functor $\forget: \FZ(\CM) \to \CM$, we obtain the following commutative diagram: 
$$
\xymatrix{ \CC \ar[r]^{\phi_\CM}  \ar[dr]_{f_\CM=\forget \circ \phi_\CM} & \FZ(\CM) \ar[d]^{\forget} \\
& \CM
}
$$
A functor $\CC \to \CM$ factoring through the forgetful functor $\forget$ is called a central functor. The action functor $\odot: \CC \times \CM \to \CM$ defined by $(a,m) \mapsto f_\CM(a)\otimes m$ is a monoidal functor. 
\begin{defn} \label{def:monoidal-module-map}
For a UMFC $\CC$ and two right multi-fusion $\CC$-modules $\CM,\CN$, a {\em monoidal $\CC$-module functor $F:\CM\to\CN$} is a unitary monoidal functor equipped with an isomorphism of monoidal functors $F\circ f_\CM\simeq f_\CN: \CC\to\CN$ rendering the following diagram
\be  \label{diag:m-m-map}
\raisebox{2em}{\xymatrix{
  F(f_\CM(a)\otimes x) \ar[r]^\sim \ar[d]_\sim & F(x\otimes f_\CM(a)) \ar[d]^\sim \\
  f_\CN(a)\otimes F(x) \ar[r]^\sim & F(x)\otimes f_\CN(a) \\
}}
\ee
commutative for $a\in\CC$ and $x\in\CM$. $\CM$ and $\CN$ are said to be equivalent if $F$ is also an equivalence.
\end{defn}

For UMTC's $\CA,\CB,\CC$, a multi-fusion $\CA$-$\CB$-bimodule $\CM$ and a multi-fusion $\CB$-$\CC$-module $\CN$, the relative tensor product $\CM\boxtimes_\CB\CN$ is a well-defined multi-fusion $\CA$-$\CC$-bimodule. By \cite[Theorem 3.3.6.]{kz1}. we have a well-defined symmetric monoidal category: 
\begin{itemize}
\item $\CU\CM\CT^{\mathrm{cl}}$: objects are UMTC's; morphisms are the equivalence classes of closed multi-fusion bimodules; the composition maps are defined by relative tensor products; the symmetric tensor product is the Deligne tensor product.
\end{itemize}

\begin{defn} \label{def:en-UMFC}
A {\em $\CB$-enriched unitary (multi-)fusion category} is an enriched monoidal category ${}^\CB\CX$ obtained by the canonical construction from a pair $(\CB,\CX)$, where $\CB$ is a UMTC and $\CX$ is a (multi-)fusion left $\CB$-module. 
\end{defn}

\subsection{Gapped edges and 0d walls} \label{sec:gapped-0d-wall}

An anomaly-free 2d topological order (without symmetry) can be described by a pair $(\CC,c)$ (see \cite[Appendix\,E]{kitaev} for a review), where $\CC$ is a UMTC and $c$ is the chiral central charge. The pair $(\bh,0)$ describes the trivial 2d topological oder.  

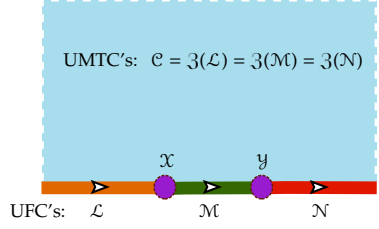
\begin{figure}
$$
\raisebox{-0pt}{
  \begin{picture}(95,80)
   \put(-20,10){\scalebox{1.2}{\includegraphics{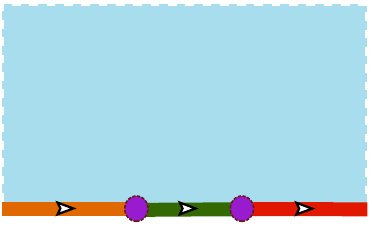}}}
   \put(-20,10){
     \setlength{\unitlength}{.75pt}\put(-18,-19){
     \put(3,12) {\scriptsize UFC's:}
     \put(43, 12)       {\scriptsize $\CL$}
     \put(155, 12)     {\scriptsize $ \CN $}
     \put(98, 12)     {\scriptsize $ \CM $}
     \put(78,37) {\scriptsize $\CX$}
     \put(127,37) {\scriptsize $\CY$}
     \put(30, 88)     {\scriptsize UMTC's:\,\, $ \CC=\Z(\CL)=\Z(\CM)=\Z(\CN)$}
          }\setlength{\unitlength}{1pt}}
  \end{picture}}
$$
\caption{This picture depicts a 2d topological order $(\CC,0)$ with three different gapped edges given by UFC's $\CL,\CM,\CN$ separated by two 0d walls $\CX$ and $\CY$. The 2d bulk is oriented as the usual $\Rb^2$ with the normal direction pointing out of the paper in readers' direction. The arrows indicate the induced orientation on the edge.}
\label{fig:bulk=center-1}
\end{figure}

\begin{pthm} \label{pthm:gapped-edge}
As illustrated in Figure.\,\ref{fig:bulk=center-1}, we have the following results. 
\bnu
\item A gapped edge of a 2d topological order $(\CC,0)$ is described mathematically by a closed right fusion $\CC$-module $\CL$. 

\item Different gapped edges $\CL,\CM,\CN$ (as UFC's) share the same bulk (as their Drinfeld centers) if and only if they are Morita equivalent \cite{eno2008}. 

\enu
A 2d topological order $(\CC,0)$ admitting gapped edges is called {\it a non-chiral 2d topological order}. In these cases, the central functor $f_\CL$ describes how excitations in the bulk are fused into those on the edge, thus will be called the bulk-to-boundary map.  
\end{pthm}

\begin{rem} \label{rem:multi-fusion}
Unstable 1d topological orders naturally occur in dimensional reduction processes. They can be described by an indecomposable UMFC \cite{kong-wen-zheng-1,ai}. 
\end{rem}

\begin{rem} \label{rem:gapped-walls}
By the folding trick, Theorem$^{\mathrm{ph}}$\,\ref{pthm:gapped-edge} implies that a gapped 1d wall between two 2d topological orders $(\CA,c)$ and $(\CB,c)$ (see the second picture in (\ref{fig:label-disk})) is described by a closed fusion $\CA$-$\CB$-bimodule or a closed multi-fusion $\CA$-$\CB$-bimodule if we allow unstable gapped walls. 
\end{rem}

A 1d gapped edge $\CL$ of a 2d topological order should itself be viewed as an anomalous 1d topological order, described mathematically by a UFC $\CL$. Its anomaly is completely captured by its bulk, which is described by the Drinfeld center $\FZ(\CL)$. It is anomaly-free if $\FZ(\CL)\simeq \bh$. 

\begin{pthm}[\cite{kong-wen-zheng-1,ai}] \label{pthm:0d-wall-gapped-edge}
A 0d wall between two gapped edges $\CL$ and $\CM$ (i.e. UFC's) of the same 2d topological order $(\CC,c)$ as depicted in Figure\,\ref{fig:bulk=center-1} is mathematically described by the unique closed left $\CL\boxtimes_\CC\CM^\rev$-bimodule $\CX$. 
\void{ satisfying the following
\begin{itemize}
\item {\it anomaly-free condition}: the following functor 
\begin{align} \label{eq:LMX}
\CL \boxtimes_\CC \CM^\rev &\xrightarrow{\simeq} \fun(\CX,\CX) \\
x\boxtimes_\CC y &\mapsto x\odot - \odot y \nonumber
\end{align}
is a well-defined unitary monoidal equivalence. As a consequence, $\CX$ is an invertible $\CL$-$\CM$-bimodule and is uniquely determined by the unitary monoidal equivalence (\ref{eq:LMX}) as the unique (up to equivalences) indecomposable left $\fun(\CX,\CX)$-module.
\end{itemize}
}
\end{pthm}

\begin{figure}[bt]
$$
\raisebox{-50pt}{
  \begin{picture}(95,100)
   \put(0,0){\scalebox{1}{\includegraphics{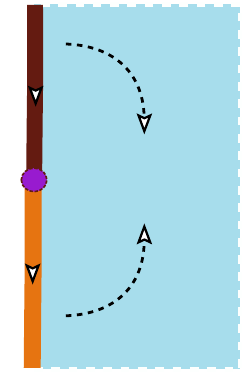}}}
   \put(0,0){
     \setlength{\unitlength}{.75pt}\put(-18,-19){
     \put(15, 150)       {\scriptsize $ \CL $}
     \put(15, 30)     {\scriptsize $ \CM $}
     \put(0, 93)   {\scriptsize $(\CX, x)$}
     \put(80, 90)     {\scriptsize $ \CC $}
          }\setlength{\unitlength}{1pt}}
  \end{picture}}
\quad\quad\quad\quad\quad\quad
\raisebox{-60pt}{
  \begin{picture}(105,100)
   \put(0,35){\scalebox{1}{\includegraphics{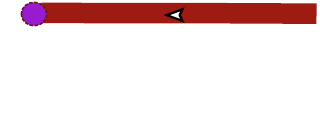}}}
   \put(0,35){
     \setlength{\unitlength}{.75pt}\put(-18,-19){
     \put(0, 60)       {\scriptsize $ (\CX,x) $}
     \put(40, 70) {\scriptsize  $\CL \boxtimes_\CC \CM^\rev = \fun(\CX,\CX)$}
     }\setlength{\unitlength}{1pt}}
  \end{picture}}
$$
$$
(a) \quad\quad\quad\quad \quad \quad\quad\quad\quad \quad
\quad\quad\quad\quad  (b)
$$
\caption{These two picture depicts a dimensional reduction process from $(a)$ to $(b)$.
}
\label{fig:dim-reduction}
\end{figure}
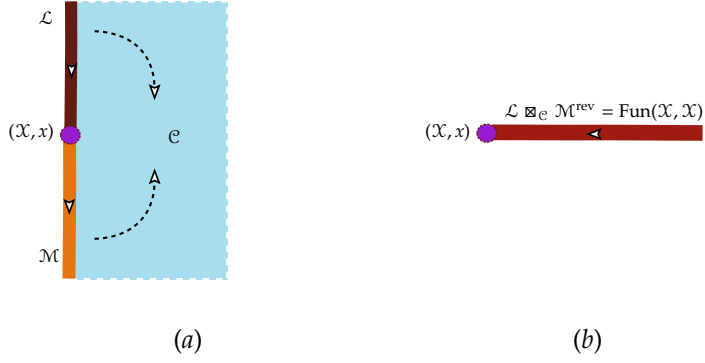

\begin{rem}
Physically, the $\CL$-$\CM$-bimodule structure on $\CX$ is provided by the fusion of topological excitations in $\CL$ and $\CM$ to $\CX$ from two sides. The closedness condition is an anomaly-free condition, which says that the 1d topological order $\CL \boxtimes_\CC \CM^\rev$, obtained from the dimensional reduction process depicted in Figure\,\ref{fig:dim-reduction} \cite{fsv,anyon}, should be nothing but the unique 1d bulk of $\CX$ given by $\fun(\CX,\CX)$ (see \cite{kong-wen-zheng-1,ai} for more details). This condition determines $\CX$ uniquely (up to equivalences). 
\end{rem}

If we consider the entire 0+1D world line of the 0d wall, it makes no sense to specify a wall excitation $x\in\CX$ because it can be changed to other excitations on the world line. But if we want to specify a particular spatial slide of the 0+1D wall, we can further specify a distinguished wall excitation $x\in\CX$. This leads to a new description of the 0d wall as a pair $(\CX,x)$, which is useful in the calculation of global observables or factorization homology on space manifolds (see Section\,\ref{sec:fh}).

\begin{rem}
Mathematically, such a pair $(\CX,x)$ can be viewed as an $E_0$-algebra in the 2-category of categories in the sense of Lurie \cite{lurie}. In the same 2-category, a monoidal category is an $E_1$-algebra; a braided monoidal category is an $E_2$-algebra; a symmetric monoidal category is an $E_3$-algebra or $E_\infty$-algebra. $\fun(\CX,\CX)$ is the $E_0$-center of $(\CX,x)$; the Drinfeld center is an $E_1$-center; the M\"{u}ger center is an $E_2$-center. 
\end{rem}

\begin{expl}
A topological excitation $u\in\CC$ in the 2d bulk $(\CC,c)$ (resp. on a gapped edge $\CL$) can be viewed as an anomalous 0d topological order, which can be mathematically described by $(\CC, u)$ (resp. $(\CL,u)$) in a spatial slide, where $\CC$ (resp. $\CL$) is viewed as a finite unitary category by forgetting its monoidal structures \cite{kong-wen-zheng-1,ai}. In these cases, the closed (or anomaly-free) condition holds automatically, i.e.
\begin{align}
\CC\boxtimes_{\CC\boxtimes\overline{\CC}} \CC^\rev &\xrightarrow{\simeq} \fun(\CC,\CC), \quad\quad\quad\quad
\CL\boxtimes_{\FZ(\CL)} \CL^\rev \xrightarrow{\simeq} \fun(\CL,\CL)  \nn
a\boxtimes_{\CC\boxtimes\overline{\CC}} b &\mapsto a\otimes - \otimes b, 
\quad\quad\quad\quad\quad\,\,\,\,
l\boxtimes_{\FZ(\CL)} m \mapsto l\otimes - \otimes m, 
\end{align}
both of which are special cases of a general formula (\ref{eq:KZ}). We will discuss the anomaly-free condition for general 0d defect junctions in Section\,\ref{sec:fh}. 
\end{expl}

\subsection{Boundary-bulk relation for gapped edges} \label{sec:bbr-gapped}

It turns out that the boundary-bulk relation discussed in the previous subsections is only the first layer of a heirarchic structure. 

\begin{figure}[tb]
 \begin{picture}(150, 100)
   \put(100,10){\scalebox{2}{\includegraphics{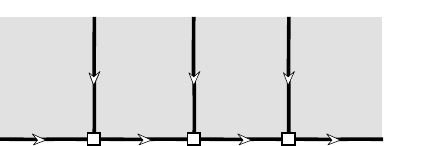}}}
   \put(60,-55){
     \setlength{\unitlength}{.75pt}\put(-18,-19){
     \put(95, 98)       {\scriptsize $\CL$}
     \put(175, 98)       {\scriptsize $\CM$}
     \put(250, 98)       {\scriptsize $\CN$}
     \put(325, 98)       {\scriptsize $\CO$}
     \put(140,98)      {\scriptsize $\CX$}
     \put(218,98)      {\scriptsize $\CY$}
     \put(290,98)      {\scriptsize $\CZ$}
     \put(85, 160)    {\scriptsize $\Z(\CL)$}
     \put(165, 160)    {\scriptsize $\Z(\CM)$}
     \put(240, 160)    {\scriptsize $\Z(\CN)$}
     \put(310, 160)    {\scriptsize $\Z(\CO)$}
     \put(130,212)     {\scriptsize $\Z^{(1)}(\CX)$}
     \put(210,212)     {\scriptsize $\Z^{(1)}(\CY)$}
     \put(285,212)     {\scriptsize $\Z^{(1)}(\CZ)$}
     }\setlength{\unitlength}{1pt}}
  \end{picture}
\caption{The picture illustrates the complete boundary-bulk relation, which is the physical meaning of Theorem\,\ref{thm:KZ} . The arrows indicate the orientation of the edges/walls and the order of the fusion product of topological excitations on the edges/walls. 
}
\label{fig:bbr-gapped}
\end{figure}
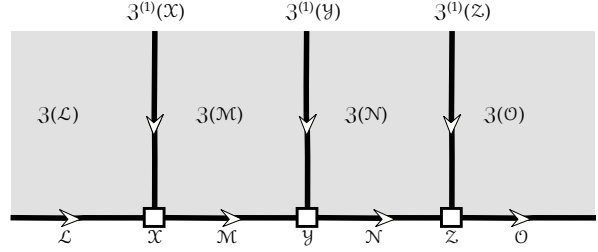

\medskip
A most general situation for the boundary-bulk relation is depicted in Figure\,\ref{fig:bbr-gapped}. The 0d gapped defect labeled by $\CX$ is a junction of three 1d gapped defects labeled by $\CL,\CM,\Z^{(1)}(\CX)$. In this case, $\CX$ is an $\CL$-$\CM$-bimodule but not invertible in general. The 1d gapped wall labeled by $\Z^{(1)}(\CX)$ is a closed multi-fusion $\Z(\CL)$-$\Z(\CM)$-bimodule (recall Remark\,\ref{rem:gapped-walls}). By the unique bulk principle proposed in \cite{kong-wen-zheng-1}, the gapped 1d wall $\Z^{(1)}(\CX)$, which should be viewed as a 1d ``relative bulk'' of $\CX$, is uniquely determined by $\CX, \CL,\CM$ as follows: 
$$
\Z^{(1)}(\CX)=\fun_{\CL|\CM}(\CX,\CX),
$$
where $\fun_{\CL|\CM}(\CX,\CX)$ is the category of unitary $\CL$-$\CM$-bimodule functors. Moreover, we should have a canonical monoidal equivalence:  
$$
\CL\boxtimes_{\Z(\CL)} \Z^{(1)}(\CX) \boxtimes_{\Z(\CM)} \CM^\rev \simeq \fun(\CX,\CX)
$$
which is a consequence of the formula (\ref{eq:KZ}).

\begin{rem} \label{rem:left-right-convention}
Note that our convention of the left and right action in the definition of a fusion bimodule is that if the orientation of the wall is the same (resp. the opposite) as the induced orientation with respect to a bulk phase, then this bulk phase acts on the wall from right (resp. left). 
We will use this convention throughout this work. 
\end{rem}

Now we consider the fusion of two gapped walls, say $\Z^{(1)}(\CX)$ and $\Z^{(1)}(\CY)$. This fusion gives a new gapped wall $\Z^{(1)}(\CX)\boxtimes_{\Z(\CB)} \Z^{(1)}(\CY)$ between $\Z(\CA)$ and $\Z(\CC)$. On the other hand, it should also be viewed as the 1d ``relative bulk'' of a new 0d wall between $\CA$ and $\CC$ obtained by fusing $\CX$ and $\CY$, i.e. $\CX\boxtimes_\CM \CY$. Hence, we should expect a monoidal equivalence: 
\be \label{eq:KZ}
\fun_{\CL|\CM}(\CX,\CX)\boxtimes_{\Z(\CM)} \fun_{\CM|\CN}(\CY,\CY) \simeq \fun_{\CL|\CN}(\CX\boxtimes_\CM \CY, \CX\boxtimes_\CM \CY). 
\ee
This monoidal equivalence was rigorously proved in \cite[Theorem\,3.1.7.]{kz1}. It simply says that the assignment $\CL \mapsto \Z(\CL)$ and $\CX \mapsto \Z^{(1)}(\CX)$ is functorial. This functoriality, stated more precisely in Theorem\,\ref{thm:KZ}, provides a complete mathematical description of the boundary-bulk relation for 2d topological orders with gapped edges. 
\begin{thm} \label{thm:KZ}
The functor $\FZ: {}^{\mathrm{ind}}\CU\CM\CF \to \CU\CM\CT^{\mathrm{cl}}$ defined by 
$$
\CL \mapsto \Z(\CL) \quad\quad \mbox{and} \quad\quad
\CX \mapsto \Z^{(1)}(\CX):=\fun_{\CL|\CM}(\CX,\CX)
$$ 
is a well-defined fully faithful symmetric monoidal functor. 
\end{thm}

\begin{rem}
The non-unitary version of above theorem was proved in \cite[Theorem\,3.3.7]{kz1}. Its proof can be generalized to unitary cases. The physical meaning of Theorem\,\ref{thm:KZ} (as depicted in Figure\,\ref{fig:bbr-gapped}) can be all realized by Levin-Wen type of lattice models \cite{kk}. 
\end{rem}

One of the main goals of this work is to generalize above result to gapless edges (see Theorem\,\ref{thm:ff-functor}).

\subsection{Factorization homology on space manifolds} \label{sec:fh}

The integration of the local observables on space manifolds is achieved by the mathematical theory of factorization homology \cite{ai} (see \cite{af} for a recent mathematical review). In this subsection, we review some basic results that will be useful later.

\begin{defn} \label{def:coefficient}
A coefficient system $A$ of an oriented disk-stratified 2-manifold $\Sigma$ (see \cite{aft}) is an assignment of each $i$-cell to an $i$-dimensional topological order for $i=0,1,2$. A coefficient system $A$ is called {\it anomaly-free} if the following conditions 
are satisfied:
\be \label{fig:label-disk}
\raisebox{-3em}{\begin{picture}(60, 60)
   \put(10,10){\scalebox{1.2}{\includegraphics{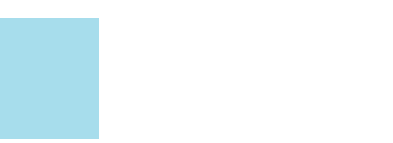}}}
   \put(-30,-54){
     \setlength{\unitlength}{.75pt}\put(-18,-19){
     \put(90, 130)    {\scriptsize $\CA$}
   }\setlength{\unitlength}{1pt}}
  \end{picture}}
\quad\quad
\raisebox{-3em}{\begin{picture}(60, 60)
   \put(10,10){\scalebox{1.2}{\includegraphics{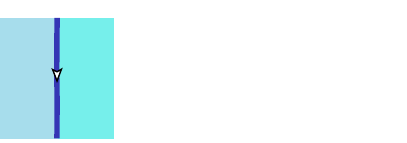}}}
   \put(-30,-54){
     \setlength{\unitlength}{.75pt}\put(-18,-19){
     \put(77, 130)    {\scriptsize $\CA$}
     \put(109, 130)    {\scriptsize $\CB$}
     \put(93,166)     {\scriptsize $\CM$}
   }\setlength{\unitlength}{1pt}}
  \end{picture}}
\quad\quad
\raisebox{-3em}{  \begin{picture}(70,60)
   \put(10,6){\scalebox{1.2}{\includegraphics{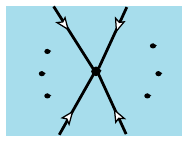}}}
   \put(10,6){
     \setlength{\unitlength}{.75pt}\put(-8,-9){
     \put(57,41)   {\scriptsize $(\CX,x)$}
     \put(62, 77)     {\scriptsize $ \CM_1 $}
     \put(25, 77)     {\scriptsize $ \CM_n $}
     \put(26,5)      {\scriptsize $ \CM_{i+1}$}
     \put(64,5)      {\scriptsize$\CM_i$}
     \put(45,65)      {\scriptsize $\CA_0$}
     \put(45,17)      {\scriptsize $\CA_i$}
          }\setlength{\unitlength}{1pt}}
 \end{picture}}         
\ee
\bnu
\item each 2-cell is assigned to a UMTC $\CA$ (or an anomaly-free 2d topological order $(\CA,c)$, where $c$ is fixed for each connecting component of $\Sigma$, thus can be ignored);
\item each oriented 1-cell between two adjacent 2-cells (as illustrated in the second picture in (\ref{fig:label-disk})) is assigned to a closed multi-fusion $\CA$-$\CB$-bimodule $\CM$; 
\item each 0-cell as the one depicted in the third picture in (\ref{fig:label-disk}) is assigned to a pair $(\CX,x)$, where $\CX$ is a closed $\CP$-module for $\CP:=\CM_1\boxtimes_{\overline{\CA_0} \boxtimes \CA_1} (\CM_2 \boxtimes_{\CA_2} \cdots \boxtimes_{\CA_{n-1}} \CM_n)$ (recall Theorem$^{\mathrm{ph}}$\,\ref{pthm:0d-wall-gapped-edge}) and $x$ is an object in $\CX$.
\enu
\end{defn}

Such a coefficient system describes a physical configuration of 0d,1d,2d topological orders on $\Sigma$. Anomaly-free condition means that the corresponding physical configuration can be realized by a 2-dimensional local Hamiltonian lattice model on $\Sigma$. The ``closed'' condition determines $\CX$ uniquely. In other words, an anomaly-free 0d defect is determined by the physics of its neighborhood uniquely (up to the choices of the distinguished object $x\in\CX$).

\begin{rem}
Note that if we flip the orientation of a 1-cell and replaced its assignment $\CM$ by $\CM^\rev$ at the same time, then the physics configuration remains the same. Therefore, it must defines an equivalent coefficient system.  
\end{rem}

The factorization homology of a coefficient system $A$ on an oriented disk-stratified 2-manifold $\Sigma$ is well-defined and is denoted by $\int_\Sigma A$. 

\begin{thm}[\cite{ai}] \label{thm:akz} 
If $\Sigma$ is compact and $A$ is anomaly-free, then $\int_\Sigma A = (\bh, u_\Sigma)$, where $u_\Sigma$ is a distinguished object in $\bh$.  
\end{thm}

The physical meaning of $u_\Sigma$ is nothing but the space of ground states of the associated physical configuration on $\Sigma$. 
This integral is well-defined on any submanifold of $\Sigma$ as well. In particular, the integral over any open 2-disk like region $D$ in $\Sigma$ gives a pair $(\CX,x)$, i.e. $\int_D A = (\CX,x)$. The following result will be useful later. 

\begin{thm}[\cite{ai}] \label{thm:ai}
By shrinking an open 2-disk like region $D$ in $\Sigma$ to a 0-cell and assigning $\int_D A$ to this 0-cell, we obtain a coefficient system $A'$ on a new oriented disk-stratified 2-manifold $\Sigma'$. The coefficient system $A'$ on $\Sigma'$ is again anomaly-free. 
\end{thm}

\begin{expl} \label{expl:ai}
Consider the following open 2-disk like region $D$ on $\Sigma$ with an anomaly-free coefficient system: 
$$
\raisebox{-60pt}{ \begin{picture}(90, 70)
   \put(-20,0){\scalebox{1.2}{\includegraphics{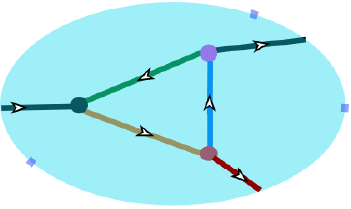}}}
   \put(21,26){
     \setlength{\unitlength}{.75pt}\put(-18,-19){
     \put(-18, 41)        {\scriptsize $(\CP,p) $}
     \put(-30,35)       {\scriptsize $\CI$ }
     \put(40, -1)       {\scriptsize $(\CR,r)$}
     \put(40, 63)      {\scriptsize $(\CQ,q)$}
     \put(110, 28)     {\scriptsize  $\CB$ }
     \put(64, 28)     {\scriptsize $ \CN $}
     \put(31, 30)     {\scriptsize $ \CC $}
     \put(20, 50)     {\scriptsize $ \CM $}
     \put(22, 10)     {\scriptsize $\CL$ }
     \put(75,3)     {\scriptsize $\CK$ }
     \put(0,-5)      {\scriptsize $\CA$ }
     \put(86,52)   {\scriptsize $\CJ$ }
     \put(0,62)   {\scriptsize $\CD$ }
     }\setlength{\unitlength}{1pt}}
  \end{picture}}
$$
If $\int_D A=(\CX,x)$, then $\CX$ is uniquely determined by $\CS:=\CI \boxtimes_{\overline{\CA}\boxtimes \CD} (\CJ^\rev \boxtimes_\CB \CK^\rev)$ as the unique closed left $\CS$-module. In other words, $\CX$ is independent of other data: $(\CP,p),\CM,(\CQ,q),\CN,(\CR,r),\CL, \CC$, but the distinguished object $x\in\CX$ depends on them. The physical meaning is that if we view from far away,  this open 2-disk like region can simply be viewed a 0d defect junction, defined by the pair $(\CX,x)$, connecting three 1d defects labeled by $\CI,\CJ,\CK$.   
\end{expl}

\section{0d walls between chiral gapless edges} \label{sec:0d-walls}

In this section, we develop the theory of 0d walls between two chiral gapless edges. 

\subsection{Observables on the world line of a 0d wall} \label{sec:0d-defects}

Consider a 0d gapless wall between two chiral gapless edges as depicted in Figure\,\ref{fig:0d-defects}. The 2d bulk topological order is $(\CC,c)$. Two chiral gapless edges are $(V_\CA,{}^{\CA}\CX)$ and $(V_\CB,{}^{\CB}\CY)$, where both chiral symmetries $V_\CA$ and $V_\CB$ have the same central charge $c$. If $c=0$, then it is necessary that $V_\CA=V_\CB=\Cb$. Throughout this work, we use vertical planes in Figures to represent 1+1D world sheets of gapless edges/walls.


If $m$ is a topological excitations living on the 0d wall, by fusing topological excitations in the bulk and edges with $m$, we obtain different topological excitations. All such wall excitations can be labeled by the objects in a category $\CM$, which is called the category of topological excitations. Similar to the analysis of the observables on the 1+1D world sheet of a chiral gapless edge, using the same dimensional reduction trick as depicted in \cite[Figure\,5]{kz4}, by the ``No-Go Theorem'' \cite[Section\,3.3]{kz4}, chiral fields on the 0+1D world line supported on $m$ form 1D boundary CFT's and 0D walls between them. Topological excitations can also be viewed as the boundary conditions of these boundary CFT's. Therefore, we can label these boundary CFT's as $A_m$ for a given topological excitation $m$ on the wall and a 0D wall by $M_{m,m'}$ for two topological excitations $m,m'$. The space $M_{m,m'}$ consists of boundary condition changing operators and we have $M_{m,m}=A_m$. These chiral fields can have OPE $M_{m',m''} \otimes_\Cb M_{m,m'} \to M_{m,m''}$. 

\void{
\begin{figure} 
$$
 \raisebox{-50pt}{
  \begin{picture}(150,140)
   \put(-50,7){\scalebox{0.7}{\includegraphics{pic-0d-defect.eps}}}
   \put(-30,0){
     \setlength{\unitlength}{.75pt}\put(0,0){
     \put(56,155)  {\footnotesize$(\CA,c)$}
     \put(64,47)   {\footnotesize$\CX$}
     \put(123,155) {\footnotesize $\CP$}
     \put(162,155)  {\footnotesize$(\CB,c)$}
     \put(170,47)   {\footnotesize$\CY$}
     \put(15,30)  {\footnotesize $x\in \CX$}
     \put(219,30) {\footnotesize $y\in \CY$}
     
     \put(0,183)  {\footnotesize $(V_\CA,{}^{\CA}\CX)$}
     \put(210, 183) {\footnotesize $(V_\CB,{}^{\CB}\CY)$}
     \put(105,183)   {\footnotesize $(V,{}^\CP\CM)$}    
    
     \put(285,68) {\footnotesize $(\CC,c)$}
     \put(105,30)  {\footnotesize $m\in \CM$}
     
     \put(82,84)  {\footnotesize $M_{m,m'}$}
     \put(77,129)  {\footnotesize $M_{m',m''}$}
    
    \put(-18,107)  {\footnotesize $[x,x']_\CA$}
    \put(234,107) {\footnotesize $[y,y']_\CB$}
   
    \put(284,163)  {\footnotesize $t$}
    \put(310,135) {\footnotesize $x$}

     }\setlength{\unitlength}{1pt}}
  \end{picture}}
$$
\caption{This picture depicts the back side (so it is shaded) of the 1+1D world sheet of two potentially different chiral gapless edges $(V_\CA,{}^{\CA}\CX)$ and $(V_\CB,{}^{\CB}\CY)$ connected by a gapless 0d wall. The vertical black line is the 0+1D world sheet of the wall. 
Note that the complex coordinate $z=t+ix$ is given and determines the orientation of the world sheet. 
}
\label{fig:0d-defects}
\end{figure}
}

\begin{figure} 
$$
 \raisebox{-50pt}{
  \begin{picture}(150,140)
   \put(-70,7){\scalebox{0.7}{\includegraphics{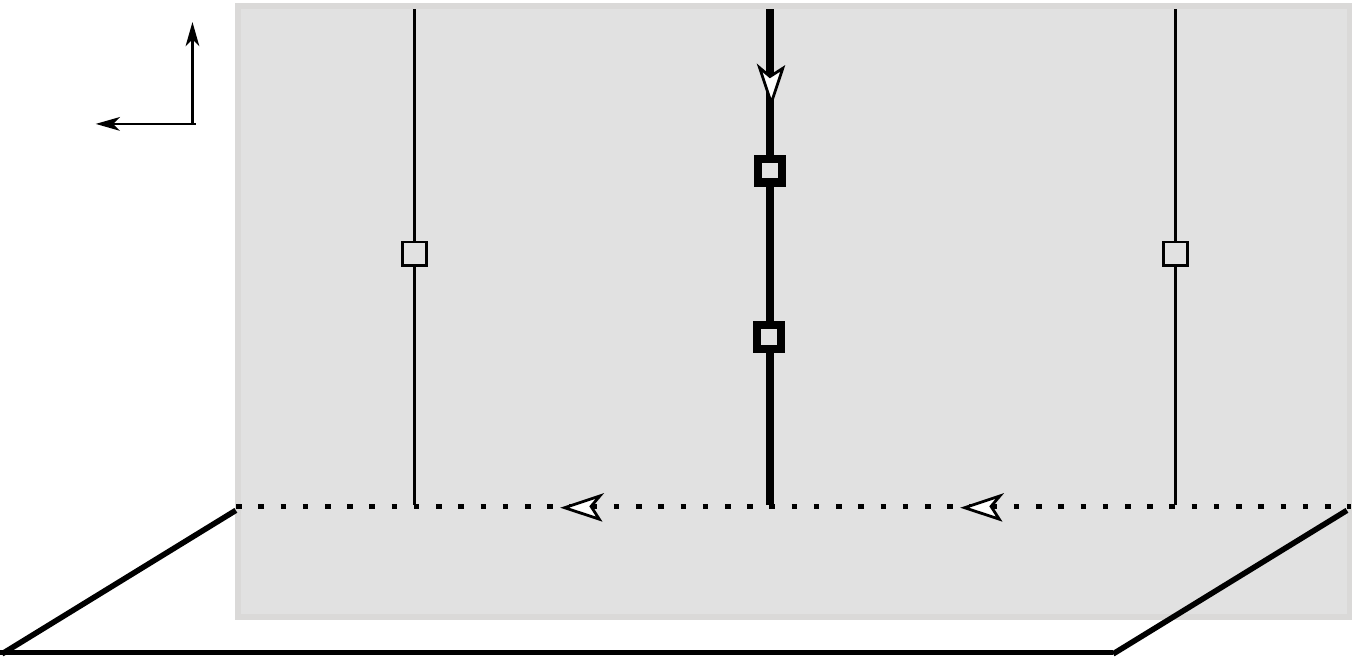}}}
   \put(-2,7){
     \setlength{\unitlength}{.75pt}\put(0,0){
     \put(56,155)  {\footnotesize$(\CA,c)$}
     \put(64,47)   {\footnotesize$\CX$}
     \put(123,155) {\footnotesize $\CP$}
     \put(162,155)  {\footnotesize$(\CB,c)$}
     \put(170,47)   {\footnotesize$\CY$}
     
     \put(15,30)  {\footnotesize $x\in \CX$}
     \put(219,30) {\footnotesize $y\in \CY$}
     
     \put(0,183)  {\footnotesize $(V_\CA,{}^{\CA}\CX)$}
     \put(210, 183) {\footnotesize $(V_\CB,{}^{\CB}\CY)$}
     \put(105,183)   {\footnotesize $(V,{}^\CP\CM)$}    
    
     \put(-60,10) {\footnotesize $(\CC,c)$}
     \put(105,30)  {\footnotesize $m\in \CM$}
     
     \put(82,84)  {\footnotesize $M_{m,m'}$}
     \put(77,129)  {\footnotesize $M_{m',m''}$}
    
    \put(-18,107)  {\footnotesize $[x,x']_\CA$}
    \put(234,107) {\footnotesize $[y,y']_\CB$}
   
    \put(-35,163)  {\footnotesize $t$}
    \put(-65,135) {\footnotesize $x$}

     }\setlength{\unitlength}{1pt}}
  \end{picture}}
$$
\caption{This picture depicts the 1+1D world sheet of two chiral gapless edges $(V_\CA,{}^{\CA}\CX)$ and $(V_\CB,{}^{\CB}\CY)$ connected by a 0+1D gapless wall (i.e. the vertical black line). The complex coordinate $z=t+ix$ is given and determines the orientation of the world sheet. 
}
\label{fig:0d-defects}
\end{figure}

By fusing chiral fields in $U_\CA=[\one_\CX,\one_\CX]_\CA$ and $U_\CB=[\one_\CY,\one_\CY]_\CB$ into the world line, we obtain two natural maps $\iota_L: V_\CA \to A_m$ and $\iota_R: V_\CB \to A_m$, respectively. These maps are clearly preserving OPE. Hence, they are homomorphisms of OSVOA's \cite{osvoa}. Let $\omega_\CA, \omega_m, \omega_\CB$ be the Viraroso elements in $V_\CA, A_m, V_\CB$, respectively. The minimal requirement for a consistent boundary CFT is to satisfy the following condition: 
\begin{itemize}
\item {\bf Conformal invariant boundary condition}: $\langle \omega_\CA\rangle \xrightarrow{\iota_L} \langle \omega_m \rangle \xleftarrow{\iota_R} \langle \omega_\CB\rangle$ are isomorphisms. 
\end{itemize} 
More generally, we require: 
\begin{itemize}
\item {\bf $V$-invariant boundary condition}: There is a VOA $V$ embeded in $V_\CA, A_m, V_\CB$ rendering the following diagrams commutative: 
\be \label{eq:VAB}
\raisebox{2em}{\xymatrix{
& V \ar@{^{(}->}[ld] \ar@{^{(}->}[d] \ar@{^{(}->}[rd] & \\
V_\CA \ar[r]^{\iota_L} & A_m & V_\CB \ar[l]_{\iota_R}  
}} 
\quad\quad\quad \forall m\in\CM. 
\ee
\end{itemize}
This VOA $V$ is called the 1+1D chiral symmetry of the wall (defined in the neighborhood of the world line), and is assumed to be a unitary rational VOA such that $\Mod_V$ is a UMTC. It is clear that $M_{m,m'}\in \Mod_V$. The path independent embedding $V \hookrightarrow A_m$ becomes a canonical morphism $\id_m: \one_{\Mod_V} \to A_m$ called the identity morphism. The OPE of defect fields defines a composition morphism in $\Mod_V$:
\be \label{eq:OPE-to-composition}
M_{m',m''}\otimes_V M_{m,m'} \xrightarrow{\circledcirc} M_{m,m''}, 
\ee
which is associative and unital as illustrated in the following commutative diagrams:  
\be
\raisebox{2em}{\xymatrix{
M_{m'',m'''} \otimes_V M_{m',m''} \otimes_V M_{m,m'} \ar[r]^-{1\circledcirc} \ar[d]_{\circledcirc 1} & 
M_{m'',m'''} \otimes_V M_{m,m''} \ar[d]^{\circledcirc} \\
M_{m',m''} \otimes_V M_{m,m'} \ar[r]^-{\circledcirc} & M_{m,m'''}
}}
\ee
\be
\raisebox{2em}{\xymatrix@C=0em{
& M_{m',m'} \otimes_V M_{m,m'} \ar[rd]^{\circledcirc} & \\
M_{m,m'} \ar[ur]^{\id_{m'} 1} \ar[rr]^{1} & & M_{m,m'}
}},
\quad\quad
\raisebox{2em}{\xymatrix@C=0em{
& M_{m,m'} \otimes_V M_{m,m} \ar[rd]^{\circledcirc} & \\
M_{m,m'} \ar[ur]^{1\id_{m}} \ar[rr]^{1} & & M_{m,m'}
}}
\ee

Therefore, the chiral fields on the 0+1D world line of this wall form a category enriched in $\Mod_V$. Its underlying category is precisely the category $\CM$ of topological excitations on the 0d wall. The background category $\Mod_V$, however, does not have a direct physical meaning because it is not the correct choice of the background category as we will show next.

\medskip
It is easy to see that all $M_{m,m'}$ are $V_\CA$-$V_\CB$-bimodules, i.e. $M_{m,m'}\in (\Mod_V)_{V_\CA|V_\CB}$, where $(\Mod_V)_{V_\CA|V_\CB}$ is the category of $V_\CA$-$V_\CB$-bimodules in $\Mod_V$. 
Note that a left $V_\CA$-module $X$ is automatically a right $V_\CA$-module with the right action defined by 
$$
X \otimes_V V_\CA \xrightarrow{c_{X, V_\CA}} V_\CA\otimes_V X \to X.
$$
Similarly, a right $V_\CB$-module $Y$ is automatically a left $V_\CB$-module with the left action defined by 
$$
V_\CB \otimes_V Y \xrightarrow{c_{V_\CB,Y}} Y\otimes_V V_\CB. 
$$
Therefore, a $V_\CA$-$V_\CB$-bimodules $X$ in $\Mod_V$ is canonically a $(V_\CA\otimes_V V_\CB)$-$(V_\CA\otimes_V V_\CB)$-bimodule in $\Mod_V$. In the definition of this bimodule structure, whenever we exchange the order of $V_\CA, X, V_\CB$ via braidings,  the object $V_\CA$ always stay on the top and $V_\CB$ always stay at the bottom according to our braiding convention \cite[Remark\,3.11]{kz4}\footnote{A warning is that \cite[Figure\,7]{kz4} explaining the braiding convention in \cite[Remark\,3.11]{kz4} was drawn in the opposite perspective of Figure\,\ref{fig:0d-defects} in this work.}. Therefore, the category $(\Mod_V)_{V_\CA|V_\CB}$ has a fusion product defined by the relative tensor product 
$$
X \otimes_{V_\CA\otimes_V V_\CB} Y, \quad\quad \forall X,Y\in (\Mod_V)_{V_\CA|V_\CB}.
$$ 
The algebra $V_\CA\otimes_V V_\CB$ in $\Mod_V$ is not commutative unless $V_\CB$ is in the centralizer of $V_\CA$, thus should be viewed as an open-string VOA (OSVOA) extension of $V$. Moreover, since both $V_\CA$ and $V_\CB$ are $\dagger$-SSSFA in $\Mod_V$, $V_\CA\otimes_V V_\CB$ is a (not necessarily simple) symmetric special $\dagger$-Frobenius algebra. As a consequence, the category $(\Mod_V)_{V_\CA|V_\CB}$ is an indecomposable UMFC.

\medskip
First, notice that we have 
\be  \label{eq:one-Mmm}
\hom_{\Mod_V}(\one_{\Mod_V}, M_{m,m'}) \simeq \hom_{(\Mod_V)_{V_\CA|V_\CB}}(V_\CA \otimes_V V_\CB, M_{m,m'}). 
\ee
Therefore, the identity morphism $\id_m: \one_{\Mod_V} \to M_{m,m}$ defines a canonical $V_\CA$-$V_\CB$-bimodule map $\id_m: V_\CA \otimes_V V_\CB \to M_{m,m}$ for $m\in \CM$. Secondly, from Figure\,\ref{fig:0d-defects}, it is easy to see that the composition morphism $\circledcirc$ defined in Eq.\,(\ref{eq:OPE-to-composition}) should intertwine both the $V_\CA$-action and the $V_\CB$-action. Therefore, it is a morphism in $(\Mod_V)_{V_\CA|V_\CB}$. In other words, the chiral fields on the 0+1D world line on the wall form a category enriched in $(\Mod_V)_{V_\CA|V_\CB}$. 

Note that these two different choices of background categories: $\Mod_V$ and $(\Mod_V)_{V_\CA|V_\CB}$ are gauge choices. They describe exactly the same physics because objects in $(\Mod_V)_{V_\CA|V_\CB}$ can be viewed automatically as objects in $\Mod_V$ via the forgetful functor $\forget: (\Mod_V)_{V_\CA|V_\CB} \to \Mod_V$. However, the new background category $(\Mod_V)_{V_\CA|V_\CB}$ has a direct physical meaning. More precisely, it describes a fictional gapped wall between two fictional bulk phases $(\CA,c)$ and $(\CB,c)$. This can be seen from a physical construction. Consider a 2d topological order $(\Mod_V, c)$. By condensing two condensable algebras $V_\CA$ and $V_\CB$ in $\Mod_V$, we obtain two new UMTC's 
$$
(\Mod_V)_{V_\CA}^0=\Mod_{V_\CA} \quad\quad \mbox{and} \quad\quad
(\Mod_V)_{V_\CB}^0=\Mod_{V_\CB}, \quad \mbox{respectively}. 
$$
Two gapped walls, defined by UFC's $(\Mod_V)_{V_\CA|V}$ and $(\Mod_V)_{V|V_\CB}$, are also produced during these two condensation processes as shown in the first of the following pictures. 
\be \label{eq:construction-P}
\raisebox{-35pt}{
 \begin{picture}(150, 75)
   \put(20,10){\scalebox{1.5}{\includegraphics{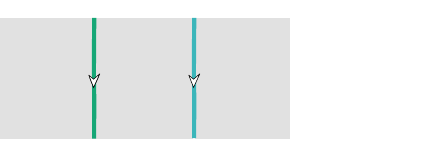}}}
   \put(20,-55){
     \setlength{\unitlength}{.75pt}\put(-18,-19){
     \put(55, 183)       {\scriptsize $(\Mod_V)_{V_\CA|V}$}
     \put(115,96)       {\scriptsize $(\Mod_V)_{V|V_\CB}$}
     \put(90, 140)    {\scriptsize $\Mod_V$}
     \put(155, 140)    {\scriptsize $\Mod_{V_\CB}$}
     \put(20, 140)    {\scriptsize $\Mod_{V_\CA}$}
     }\setlength{\unitlength}{1pt}}
  \end{picture}}
\quad\quad\xrightarrow{fusing} \quad\quad
\raisebox{-35pt}{ 
\begin{picture}(150, 75)
   \put(20,10){\scalebox{1.5}{\includegraphics{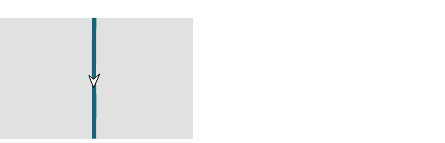}}}
   \put(20,-55){
     \setlength{\unitlength}{.75pt}\put(-18,-19){
     \put(40, 183)       {\scriptsize $\CP:=(\Mod_V)_{V_\CA|V_\CB}$}
     \put(95, 140)    {\scriptsize $\Mod_{V_\CB}$}
     \put(25, 140)    {\scriptsize $\Mod_{V_\CA}$}
     }\setlength{\unitlength}{1pt}}
  \end{picture}}  
\ee
Then we fuse these two walls. We obtain a new wall 
$$
(\Mod_V)_{V_\CA|V} \boxtimes_{\Mod_V} (\Mod_V)_{V|V_\CB} \xrightarrow{\simeq} (\Mod_V)_{V_\CA|V_\CB},
$$ 
where the functor is defined by $x\boxtimes_{\Mod_V} y \mapsto x\otimes_V y$, and is an equivalence (see for example in \cite[Theorem\ 2.2.3]{kz1}). Moreover, this equivalence is clearly a monoidal equivalence. Therefore, $\CP:=(\Mod_V)_{V_\CA|V_\CB}$ has a physical meaning as a gapped wall between two 2d topological orders $(\CA,c)$ and $(\CB,c)$, thus a better choice of the background category. Mathematically, $\CP$ is a closed multi-fusion $\Mod_{V_\CA}$-$\Mod_{V_\CB}$-bimodule \cite[Theorem\ 3.20]{dmno}. This fact provides an evidence of the Gapped-gapless Correspondence for 0d walls. 

The key to the understanding of the enriched category describing the 0d wall is to work out the relation between $\CP$ and $\CM$. The analysis is entirely similar to that in \cite[Section\,6.1]{kz4}. We will not repeat it here. Instead, we will take the advantage of what we have already shown. Since all gapless edges can be obtained from topological wick rotations, it is only reasonable if all 0d gapless walls can also be obtained from topological wick rotations. This is precisely the Gapped-gapless Correspondence stated in Section\,\ref{sec:introduction}. As a consequence, the category $\CM$ of wall excitations is uniquely determined. More precisely, by Definition\,\ref{def:coefficient}, the underlying category $\CM$ is uniquely determined by $\CX,\CA,\CP,\CB,\CY,\CC$ via the following canonical monoidal equivalence: 
\be \label{eq:anomaly-free-condition-of-M}
(\CX^\rev\boxtimes_\CA \CP\boxtimes_\CB \CY)\boxtimes_{\FZ(\CC)} \CC \simeq \fun(\CM,\CM). 
\ee
Therefore, the 0d wall depicted in Figure\,\ref{fig:0d-defects} can be characterized by a pair $(V,{}^\CP\CM)$.

\subsection{General cases: 0d phases vs. 0+1D phases} \label{sec:0d-1D}

For a fixed 1+1D chiral symmetry $V$, is $\CP:=(\Mod_V)_{V_\CA|V_\CB}$ the only choice for the background category? Note that we have shown that $M_{m,m'} \in \CP$, and all the identity morphisms and composition morphisms are morphisms in $\CP$. Therefore, the only other possibilities are subcategories of $\CP$, or equivalently, categories that map into $\CP$ faithfully. By Gapped-gapless Correspondence, these categories must be UMFC's that are Morita equivalent to $\CP$. Such a category is precisely given by $\CP_{X|X}$ for a (not necessarily simple) symmetric special $\dagger$-Frobenius algebra $X$ in $\CP$. It is equipped with a forgetful functor $\forget: \CP_{X|X} \to \CP$, which is faithful. When $X=\one_\CP=V_\CA\otimes_V V_\CB$, $\CP_{X|X}=\CP$. In general, $X$ is an OSVOA extension of $V_\CA\otimes_V V_\CB$. Note that $\CP_{X|X}$ realize all closed multi-fusion $\CA$-$\CB$-bimodule up to equivalences. Once we fix the background category to be $\CP_{X|X}$, the category of 0d wall excitations is determined uniquely by Gapped-gapless Correspondence and the anomaly-free condition in Definition\,\ref{def:coefficient}. 

\begin{figure} 
$$
 \raisebox{-50pt}{
  \begin{picture}(150,140)
   \put(-70,7){\scalebox{0.7}{\includegraphics{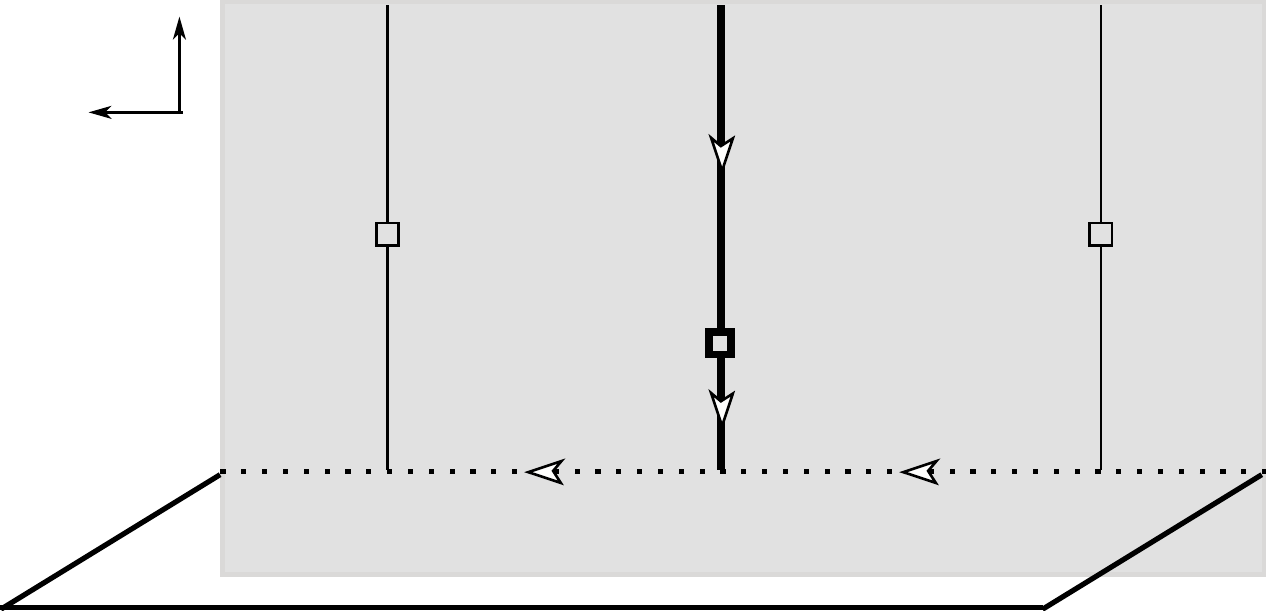}}}
   \put(-10,-3){
     \setlength{\unitlength}{.75pt}\put(0,0){
     \put(56,155)  {\footnotesize$(\CA,c)$}
     \put(64,38)   {\footnotesize$\CX$}
     \put(123,135) {\footnotesize $\CQ$}
     \put(123,60) {\footnotesize $\CP$}
     \put(162,155)  {\footnotesize$(\CB,c)$}
     \put(170,38)   {\footnotesize$\CY$}
     
     
     \put(0,183)  {\footnotesize $(V_\CA,{}^{\CA}\CX)$}
     \put(210, 183) {\footnotesize $(V_\CB,{}^{\CB}\CY)$}
     \put(100,183)   {\footnotesize $(V,{}^\CQ(\CK\boxtimes_\CP\CM))$}    
    
     \put(-60,18) {\footnotesize $(\CC,c)$}
     \put(110,39)  {\footnotesize $\CM$}
     
     \put(99,82)  {\footnotesize $\CK$}
    
   
    \put(-38,168)  {\footnotesize $t$}
    \put(-58,139) {\footnotesize $x$}

     }\setlength{\unitlength}{1pt}}
  \end{picture}}
$$
\caption{This picture depicts the 0+1D world sheet of a 0d wall $(V,{}^\CQ(\CK\boxtimes_\CP\CM))$. $\CK$ is depicted as a fictional 0D defect placed very close to another fictional 0D defect $\CM$ such that $\CK\boxtimes_\CP\CM$ should be viewed as a single fictional 0D defect that defines the category of excitations of $(V,{}^\CQ(\CK\boxtimes_\CP\CM))$.}
\label{fig:0d-defects-equivalence}
\end{figure}

\medskip
Do different background categories $\CP_{X|X}$ and $\CP$ produce different 0d walls? For $Q:=\CP_{X|X}$ and an invertible $\CQ$-$\CP$-bimodule $\CK:=\CP_{X|\one_\CP}$, we obtain a 0d wall defined by $(V,{}^\CQ(\CK\boxtimes_\CP\CM))$ as illustrated in Figure\,\ref{fig:0d-defects-equivalence}, in which $\CK$ is depicted very close to $\CM$ such that $\CK\boxtimes_\CP\CM$ should be viewed as a single fictional defect junction in the figure and the category of 0d wall excitations. We want to compare $(V,{}^\CQ(\CK\boxtimes_\CP\CM))$ with $(V,{}^\CP\CM)$. 
\bnu

\item We compare the boundary CFT's on the wall $(V,{}^\CQ(\CK\boxtimes_\CP\CM))$ with those on $(V,{}^\CP\CM)$. Note that ${}^\CP\CM={}^\CP\CM_1 \oplus \cdots \oplus {}^\CP\CM_n$, where $\CM_i$ are the indecomposable components of $\CM$ as a left $\CP$-module. On the one hand, for $0\neq m\in\CM_i$, we have $\CM_i \simeq \CP_{[m,m]_\CP}$ and $[m,m]_\CP$ is a boundary CFT on the 0+1D world sheet of $(V,{}^\CP\CM)$. On the other hand, $\CK\boxtimes_\CP\CM_i\simeq \CP_{X|[m,m]_\CP}$ is an indecomposable $\CQ$-module. For $x\in \CK\boxtimes_\CP\CM_i$, $[x,x]_\CQ$ is a boundary CFT on the 0+1D world sheet of $(V,{}^\CQ(\CK\boxtimes_\CP\CM))$. Regarding $[x,x]_\CQ$ as an algebra in $\CP$ via the forgetful functor $\forget: \CP_{X|X} \to\CP$, one can easily show that $[x,x]_\CQ$ and $[m,m]_\CP$ are Morita equivalent. Therefore, the set of boundary CFT's on $(V,{}^\CQ(\CK\boxtimes_\CP\CM))$ is a subset of those on $(V,{}^\CP\CM)$. 

\item We compare the categories of wall excitations. They are obviously different. What causes this difference? Recall that the previous notion of the chiral symmetry $V$ is a 1+1D notion. It is a VOA that is transparent in a neighborhood of the world line except at 0D defects. Mathematically, it just means that a VOA is a conformal analogue of an $E_2$-algebra (or a 2-disk algebra) \cite{lurie,aft,af}. On the 0+1D world line of the wall, we can impose a new 0+1D chiral symmetry, which is only transparent on the world line except at 0D defects, and is potentially different from the chiral symmetry $V$. This 0+1D chiral symmetry should be given by an observable algebra only defined on an open 1-disk. Mathematically, it is a conformal analogue of an $E_1$-algebra. In our case, it is nothing but an OSVOA, or more precisely, a symmetric special $\dagger$-Frobenius algebra $A$ in $\CP$. For a fixed 0+1D chiral symmetry $A$, it is clear that a 0D defect living on the world line must be an $A$-$A$-bimodule. For example, for $(V,{}^\CP\CM)$, this 0+1D chiral symmetry is just $V_\CA\otimes_VV_\CB$; for $(V,{}^\CQ(\CK\boxtimes_\CP\CM))$, it is $X$. Their difference in the category of wall excitations is due to the fact that larger 0+1D chiral symmetry allows fewer wall excitations and fewer morphisms between wall excitations. Moreover, Figure\,\ref{fig:0d-defects-equivalence} shows that one can change the 0+1D chiral symmetry by introducing a 0D wall (e.g. $\CK$) on the world line. 



\item Although $(V,{}^\CQ(\CK\boxtimes_\CP\CM))$ and $(V,{}^\CP\CM)$ differ in their 0+1D chiral symmetry and wall excitations, this 
difference is superficial from the usual condensed matter physics point of view because there is no thermodynamics limit in 0d. It means that changing the 0+1D chiral symmetry, or equivalently, introducing 0D fictional defects (e.g. $\CK$) onto the 0+1D world line, does not trigger a real space phase transition. From this point of view, a 0d wall or an anomalous 0d phase should automatically include all possible 0+1D chiral symmetries and 0D defects (e.g. $\CK$) on the world line, and $(V,{}^\CQ(\CK\boxtimes_\CP\CM))$ with $(V,{}^\CP\CM)$ should be viewed as two gauge equivalent descriptions of the same 0d wall.

\item Consider gapped 0d walls between two 1d gapped edges. In Theorem$^{\mathrm{ph}}$\,\ref{pthm:0d-wall-gapped-edge}, the background category is fixed to $\bh$, as a consequence, the category of 0d wall excitations is unique. As we will show in Section\,\ref{sec:gappable-edges-morita}, in certain length scale, not only it makes sense to talk about ``a gappable gapless 0d wall'', but also it has a precise mathematical description. Gapping it out does not trigger any 0d phase transition because there is no thermodynamics limit in 0d. Since there are precise mathematical descriptions before and after the gapping-out process, it is useful to introduce a notion before the gapping and a notion of ``gauge equivalence'' between the description of a 0d gapped wall and that of a gappable 0d gapless wall. 

\enu

\begin{rem}
1+1D and 0+1D chiral symmetries are both local quantum symmetries. We believe that their relation presented here catches some general features of gapless phases of all dimensions (see Section\,\ref{sec:outlooks}). 
\end{rem}

Similar phenomena also occur if we vary the chiral symmetry $V$. In general, there are more than one VOA $V$ rendering Diagram (\ref{eq:VAB}) commutative. It means that we can impose different 1+1D chiral symmetries on the 0d wall. We denote them by $V_i$, $i=1, 2, \cdots$. We obtain different pairs $(V_i, {}^{\CP_i}\CM_i)$, where $\CP_i=(\Mod_{V_i})_{V_\CA|V_\CB}$ and $\CM_i$ is uniquely determined. By the construction of $\CP$ (recall (\ref{eq:construction-P})), it is clear that $\CP_i$ is again a closed multi-fusion $\CA$-$\CB$-bimodules. Hence, $\CP_i$ and $\CP_j$ are Morita equivalent. By introducing a 0D wall $\CK_{ji}$ on the world line between $\CP_j$ and $\CP_i$, we break/change the chiral symmetry from $V_i$ to $V_j$. Due to the lack of thermodynamics limit in 0d, this breaking/changing of 1+1D chiral symmetries does not trigger a real space phase transition. 

\medskip
Note that the usual notion of a phase in condensed matter physics is a spatial notion. From this perspective, all possible $(V_i, {}^{\CP_i}\CM_i)$ and 0D walls among them should be included in the complete definition of the spatial notion of a 0d phase. On the other hand, $(V_i, {}^{\CP_i}\CM_i)$ and $(V_j, {}^{\CP_j}\CM_j)$ for $i\neq j$ define two different sets of boundary CFT's preserving different 0+1D chiral symmetries, and can be obtained from two different topological Wick rotations. It becomes convenient, or physically important, to introduce and carefully distinguish two concepts: a 0d phase (a spatial notion) and a 0+1D phase (a spacetime notion). 
\begin{defn} \label{def:0d-1D-phases}
There are two different notions associated to a gapless 0d wall or a potentially anomalous 0d gapless phase.
\bnu

\item By a (potentially anomalous) ``0+1D phase'', we mean a 0d defect in a physical system with a fixed 1+1D chiral (resp. non-chiral) symmetry $V$ defined in a 1+1D neighborhood of the world line and a fixed 0+1D chiral (resp. non-chiral) symmetry $X$ defined on the world line. 

\item By a (potentially anomalous) ``0d phase'', we mean a 0d defect in a physical system such that all possible 1+1D (resp. 0+1D) chiral (or non-chiral) symmetries are realized in a neighborhood of (resp. on) the world line. More precisely, two 0+1D phases are called {\it spatially equivalent} if they can be transformed from one to the other by introducing a 0D defect (e.g. $\CK$ in Figure\,\ref{fig:0d-defects-equivalence}) on the world line. Then a 0d phase (or wall) is just a spatial equivalence class of 0+1D phases (or walls). 

\enu

\end{defn}

In the context of this subsection, we can denote the 0+1D walls constructed in this subsection by $(V,V,{}^\CP\CM)$, $(V,X,{}^\CQ(\CK\boxtimes_\CP\CM))$ and $(V_i,V_i,{}^{\CP_i}\CM_i)$, respectively. 

\begin{rem}
We will introduce the notion of a non-chiral symmetry in Section\,\ref{sec:classification-non-chiral-edges}. Above definition also applies to the study of 0d walls between two non-chiral gapless edges (see Section\,\ref{sec:non-chiral-edge-chiral-wall}). 
\end{rem}

\begin{rem}
The spatial equivalence leads us to the mathematical notion of a spatial equivalence between two bimodules over enriched multi-fusion categories in Definition\,\ref{def:spatial-functor}. In that context, ${}^\CP\CM$ and ${}^\CQ(\CK\boxtimes_\CP\CM)$ are spatially equivalent ${}^\CA\CX$-${}^\CB\CY$-bimodules. 
\end{rem}

\begin{rem}
The subtle difference between the spatial and spacetime notions in Definition\,\ref{def:0d-1D-phases} is unique in 0d, and disappear in higher dimensions. For example, introducing a 1-codimensional wall on the 1+1D world sheet of a gapped or gapless edge triggers a real space phase transition. 
\end{rem}

\begin{rem} \label{rem:gappable-excluded}
Actually, the spatial equivalence class of a 0+1D wall is much more than the 0+1D walls constructed in this subsection. For example, consider a 1+1D gapless phase defined by a RCFT defined on a cylinder $S^1 \times \Rb^1$, where $S^1$ is the space manifold and $\Rb^1$ is the time. Assume that the size of $S^1$ is small. Physically, we know that if we shrink $S^1$ to a point, the spectrum of the RCFT become gapped in this limit. Mathematically, by integrating the RCFT on the cylinder (via factorization homology), we obtain a mathematical description of a 0+1D wall, which is still gapless because this integration (or factorization homology) does not know the size of $S^1$. But this wall is gappable. Its gappability can be characterized by spatial equivalences as we will show in Section\,\ref{sec:gappable-edges-morita}.  We denote it by $(\Cb,Y,\CSs)$, where $\Cb$ is the 1+1D local quantum symmetry and $Y$ is the 0+1D non-chiral symmetry. By attaching this gappable 0+1D wall to any one of 0+1D chiral gapless walls constructed in this subsection, say $(V,V,{}^\CP\CM)$, we get a new 0+1D gapless wall $(V,V,{}^\CP\CM) \boxtimes (\Cb,Y,\CSs)$. This type of 0+1D gapless walls is beyond previous constructions. We would like to ignore such gappable 0+1D walls for our classification of 0+1D walls. Note that $Y$ is infinite dimensional and does not live in $\bh$. By requiring the 0+1D chiral symmetry to be a symmetric special $\dagger$-Frobenius algebra in $(\Mod_V)_{V_\CA|V_\CB}$, we ensure that the 0+1D wall does not contain any gappable factors or parts. We will explain in details how to gap out a 0+1D gappable gapless wall in Section\,\ref{sec:gappable-edges-morita}. 
\end{rem}

\subsection{Classification of 0+1D walls and examples} \label{sec:classification-chiral-0d-wall}

As a consequence of Definition\,\ref{def:0d-1D-phases} and Gapped-gapless Correspondence, we obtain the following classification of 0+1D gapless walls without any gappable parts (see Remark\,\ref{rem:gappable-excluded}) stated as a physical theorem.
\begin{pthm} \label{pthm:wall-chiral-edges}
All 0+1D gapless walls (without any gappable parts) between two chiral gapless edges $(V_\CA,{}^\CA\CX)$ and $(V_\CB,{}^\CB\CY)$ of the same 2d topological order $(\CC,c)$ are mathematically described and classified by triples $(V,X,{}^\CP\CM)$: 
\bnu
\item $V$ is the 1+1D chiral symmetry (defined in the neighborhood of the world line of the wall), i.e. a unitary rational VOA of central charge $c$, and $X$ is the 0+1D chiral symmetry (defined on the world line), i.e. a symmetric special $\dagger$-Frobenius algebra $X$ in $(\Mod_V)_{V_\CA|V_\CB}$. They are equipped with algebras homomorphisms between algebras in $\Mod_V$ rendering the following diagram commutative:
\be \label{diag:relation-V-X}
\raisebox{4em}{\xymatrix@R=2em{
& V \ar@{^{(}->}[ld] \ar@{^{(}->}[d] \ar@{^{(}->}[rd] & \\
V_\CA \ar[r]^{\iota_L} \ar@{^{(}->}[dr] & X & V_\CB \ar[l]_{\iota_R} \ar@{^{(}->}[dl] \\
& V_\CA \otimes_V V_\CB \, ,\ar[u]_{\iota_X} & 
}} 
\ee
where $\iota_X$ is an algebraic homomorphism between two algebras in $(\Mod_V)_{V_\CA|V_\CB}$. 

\item ${}^\CP\CM$ is an enriched category defined by the canonical construction from the pair $(\CP,\CM)$, where
\bnu
\item the background category $\CP$ is a closed multi-fusion $\CA$-$\CB$-bimodule defined by 
\be \label{eq:def-P}
\CP=((\Mod_V)_{V_\CA|V_\CB})_{X|X};
\ee 
\item the underlying category $\CM$ is the category of topological excitations in the 0d wall, and is mathematically defined by a finite unitary category equipped with a unitary monoidal equivalence: 
\be \label{eq:anomaly-free-condition-of-M}
(\CX^\rev\boxtimes_\CA \CP\boxtimes_\CB \CY)\boxtimes_{\FZ(\CC)} \CC \xrightarrow{\simeq} \fun(\CM,\CM),
\ee
where $(\CX^\rev\boxtimes_\CA \CP\boxtimes_\CB \CY)$ is a closed multi-fusion right $\FZ(\CC)$-module. Note that $\CM$ is uniquely determined by the monoidal equivalence in (\ref{eq:anomaly-free-condition-of-M}) and has a canonical left $\CP$-module structure defined by
$$
\CP \to (\CX^\rev\boxtimes_\CA \CP\boxtimes_\CB \CY)\boxtimes_{\FZ(\CC)} \CC \simeq \fun(\CM,\CM). 
$$
The space of chiral fields living on the world line between two wall excitations $m,m'\in\CM$ is given by $M_{m,m'}=[m,m']_\CP$ for $m,m'\in\CM$. 
\enu 
\enu
Moreover, all these 0+1D walls are spatially equivalent and define the same 0d wall. When $V_\CA=V_\CB=\Cb$, we must have $V=\Cb$, and this 0d wall is gapped. For many purposes, it is convenient to abbreviate the triple to ${}^\CP\CM$ for simplicity (see Remark\,\ref{rem:add-information-to-P},\ref{rem:abbreviation},\ref{rem:include-gappable}). 
\end{pthm}

\begin{rem}
If we want to emphasize or study a particular spatial slide of the 0+1D wall, we can specify a wall excitation $m\in\CM$ in the spatial slide, thus obtain a quadruple $(V,X,{}^\CP\CM,m)$. 
\end{rem}

\begin{rem} \label{rem:add-information-to-P}
If we naively apply topological wick rotation, the background category $\CP$ does not have any direct physical meaning. It is necessary to set $\CP=((\Mod_V)_{V_\CA|V_\CB})_{X|X}$ instead of only requiring an equivalence. Strictly speaking, by requiring ``='', we add the information of $(V,V_\CA,V_\CB,X)$ to $\CP$. For this reason, we will sometimes abbreviate the triple $(V,X,{}^\CP\CM)$ to ${}^\CP\CM$ for simplicity. 
\end{rem}

\begin{rem} \label{rem:include-gappable}
If we ignore $V$ and $X$ in $(V,X,{}^\CP\CM)$, it turns out that the pure categorical description ${}^\CP\CM$ automatically covers all spatially equivalent 0+1D walls including those gappable factors or parts discussed in Remark\,\ref{rem:gappable-excluded} (see Remark\,\ref{exam:fun-CC-H}). 
\end{rem}

\begin{expl} 
We discuss a few special cases and examples of Theorem$^{\mathrm{ph}}$\,\ref{pthm:wall-chiral-edges}.   
\bnu

\item When $V_\CA=V_\CB=\Cb$, we have $V=\Cb$, and $X$ can be a finite direct sum of matrix algebras and $\CP=\Mod_X(\bh)$.
If, in addition, $X=\Cb$, then we recover the gapped cases in Theorem$^{\mathrm{ph}}$\,\ref{pthm:0d-wall-gapped-edge}. If $X\neq \Cb$, then it is already beyond the usual description of a 0d wall in Theorem$^{\mathrm{ph}}$\,\ref{pthm:0d-wall-gapped-edge}.   


\item When $V_\CA=V_\CB=V$, if $X=V$, then $(V,X,{}^\CP\CM)$ gives the trivial 0+1D wall in the gapless edge $(V,{}^\CA\CX)$; if $X=a\otimes a^\ast \nsimeq V$ for $a\in\Mod_V$, then $(V,X,{}^\CP\CM)$ gives a non-trivial 0+1D wall. For example, when $V_\CA=V_\CB=V$ is the minimal model unitary rational VOA $V_\ising$ of central charge $c=\frac{1}{2}$, the UMTC $\Mod_V$ has three simple objects $1,\psi,\sigma$ with fusion rule $\sigma \otimes \sigma = 1\oplus \psi$. When $X=\sigma\otimes \sigma^\ast$, $(V,X,{}^\CP\CM)$ gives a non-trivial 0+1D wall. 

\item If $V_\CA \hookrightarrow V_\CB$ and $X=V=V_\CB$, then we have $\CP=\CA_X$ (i.e. the category of right $X$-modules in $\CA$).  

\item If $V_\CA \neq \Cb = V_\CB$, then there is no 0d wall, i.e. no wall exists between a non-trivial chiral gapless edges and a gapped edge. 

\enu

\end{expl}

\begin{rem}
0d wall between gapless edges were also studied in \cite{ccbcn}. It will be interesting to explore how examples there fit into the mathematical theory developed here. 
\end{rem}


\subsection{Spatial fusion anomalies}  \label{sec:fusion-anomaly}
\begin{figure} 
$$
 \raisebox{-50pt}{
  \begin{picture}(180,100)
   \put(-10,0){\scalebox{0.5}{\includegraphics{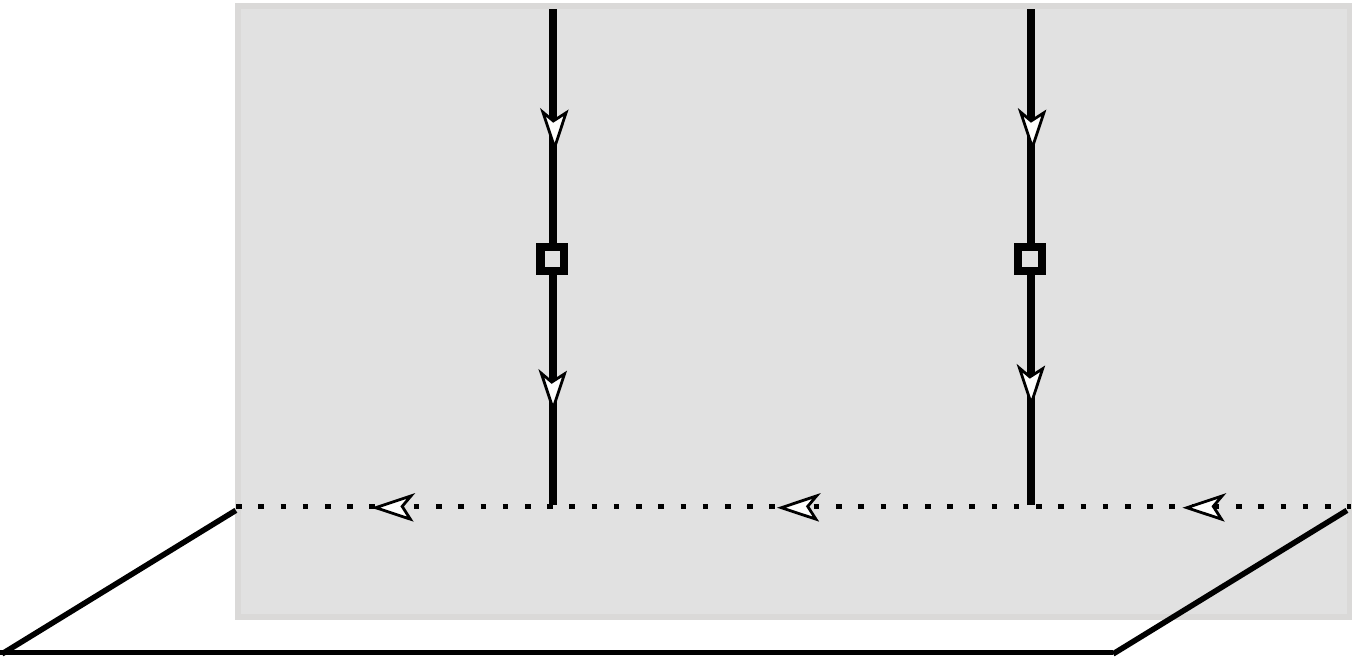}}}
   \put(-10,0){
     \setlength{\unitlength}{.75pt}\put(0,0){
     \put(22,6) {\footnotesize $(\CC,c)$}
     \put(73,37)  {\footnotesize $\CX_1$}
     \put(15,100)  {\footnotesize $(V_1,{}^{\CB_1}\CX_1)$}
     \put(147,37)  {\footnotesize $\CX_2$}
     \put(130,100){\footnotesize $(V_2,{}^{\CB_2}\CX_2)$}
     \put(228,37)  {\footnotesize $\CX_3$}
     \put(240,100) {\footnotesize $(V_3,{}^{\CB_3}\CX_3)$}
     \put(95,20) {\footnotesize $m\in\CM$}
     \put(95,100)  {\footnotesize $\CP$}
     \put(203,100)  {\footnotesize $\CQ$}
     \put(190,20) {\footnotesize $n\in \CN$}
     \put(82,130)  {\footnotesize $(V_{12},X,{}^{\CP}\CM)$}
     \put(175,130)  {\footnotesize $(V_{23},Y,{}^{\CQ}\CN)$}
     \put(63,75)  {\footnotesize $[m,m']_{\CP}$}
     \put(205,75) {\footnotesize $[n,n']_{\CQ}$}
     }\setlength{\unitlength}{1pt}}
  \end{picture}}
  \quad\quad \longrightarrow \quad\quad
 \raisebox{-50pt}{
  \begin{picture}(150,100)
   \put(-10,0){\scalebox{0.5}{\includegraphics{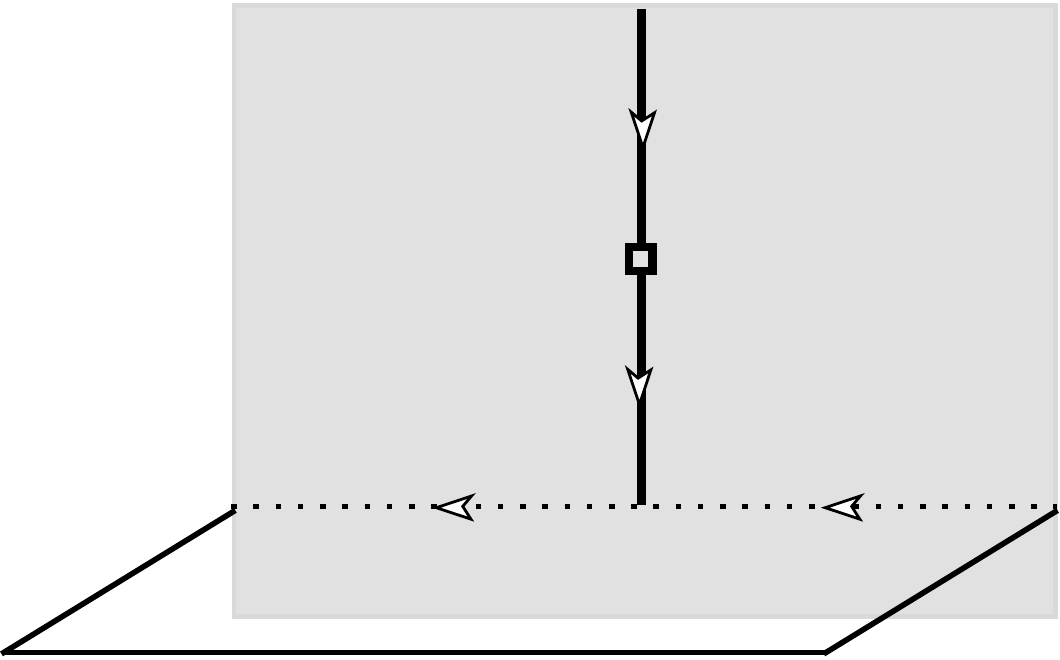}}}
   \put(-10,0){
     \setlength{\unitlength}{.75pt}\put(0,0){
      \put(22,6) {\footnotesize $(\CC,c)$}
     \put(60,37)  {\footnotesize $\CX_1$}
     \put(25,103)  {\footnotesize $(V_1,{}^{\CB_1}\CX_1)$}
     \put(178,37)  {\footnotesize $\CX_3$}
     \put(175,103) {\footnotesize $(V_3,{}^{\CB_3}\CX_3)$}
     \put(105,20) {\footnotesize $m\boxtimes_{\CX_2} n$}
     \put(80,130)  {\footnotesize $(V_{123}, X\boxtimes_{\CB_2}Y, {}^{\CP\boxtimes_{\CB_2}\CQ}(\CM\boxtimes_{\CX_2}\CN))$}
     \put(23,75)  {\footnotesize $[m\boxtimes_{\CX_2} n,m'\boxtimes_{\CX_2} n']$}
     }\setlength{\unitlength}{1pt}}  \end{picture}}
$$
\caption{This picture illustrates the fusion of two 0d gapless walls $(V_{12},X,{}^\CP\CM)$ and $(V_{23},Y,{}^\CQ\CN)$. This fusion is defined by (\ref{eq:fusion-0d-walls}).
}
\label{fig:0d-walls}
\end{figure}

In Figure\,\ref{fig:0d-walls}, we depict three chiral gapless edges $(V_i,{}^{\CB_i}\CX_i)$, $i=1,2,3$ of a 2d topological order $(\CC,c)$. They are connected by two 0d gapless walls $(V_{12},X,{}^\CP\CM)$ and $(V_{23},Y,{}^\CQ\CN)$. We would like to study the spatial fusion of these two gapless walls. 

\begin{rem}\label{rem:left-right-convention} 
According to the orientation of the edge, the spatial fusion is from right to left. It is against our usual convention of the order of tensor product unless we look at Figure\,\ref{fig:0d-walls} from the back. Using the following two canonical (monoidal) equivalences
$$
\CP\boxtimes_{\CB_2}\CQ \simeq \CQ\boxtimes_{\overline{\CB_2}}\CP; \quad\quad
\CM\boxtimes_{\CX_2^\rev} \CN \simeq \CN\boxtimes_{\CX_2}\CM, 
$$
we can write the spatial fusion product from left to right (see (\ref{eq:fusion-0d-walls})). Note that this convention is the opposite to the one we used in the gapped cases. 
\end{rem}

\medskip
We first restrict ourselves to the special cases $V=V_{12}=V_{23}$ and the spatial fusion preserves the 1+1D chiral symmetry $V$. Similar to the discussion in \cite[Section\,6.3]{kz4}, the naive fusion of observables on two world lines:
$$
[m,m']_\CP \boxtimes_{\CB_2} [n,n']_\CQ \in \CP\boxtimes_{\CB_2} \CQ
$$
does not give the correct fusion in general. What happens is again a quantum quenching process. This fusion of two 0d walls causes the fusion between the topological excitations $m\in \CM$ and those in $n\in \CN$ according to the following fusion functor:
\begin{align}
\CM \boxtimes \CN &\xrightarrow{\boxtimes_{\CX_2^\rev}} \CM \boxtimes_{\CX_2^\rev} \CN \nn
m\boxtimes n &\mapsto  m\boxtimes_{\CX_2^\rev} n. \nonumber
\end{align}
Therefore, the underlying category, or the category of topological excitations, of the resulting 0+1D wall is given by $\CM \boxtimes_{\CX_2^\rev} \CN$. By the boundary-bulk relation (see Theorem\,\ref{thm:KZ}), we obtain that the background category of the resulting 0+1D wall has to be the UMFC $\CP\boxtimes_{\CB_2} \CQ$. Fusing topological excitations $m\in \CM$ and $n\in\CN$ causes a change of the microscopic physics so that it is pushed away from an RG fixed point, but then must flow to a new RG fixed point. More precisely, when two excitations are getting close, certain non-local operators acting on two excitations are becoming local. As a consequence, local observables on the 0+1D world line supported on $m\boxtimes_{\CX_2^\rev} n$ are more than the naive fusion of those on the world lines supported separately on $m$ and $n$. According to the {\bf Principle of Universality at RG fixed points}, introduced in \cite[Section\,6.3]{kz4}, observables on the 0+1D world line supported on $m\boxtimes_{\CX_2^\rev} n$ at the new RG fixed point must be the universal one, i.e. the internal hom $[m\boxtimes_{\CX_2^\rev} n, m\boxtimes_{\CX_2^\rev} n] \in \CP\boxtimes_{\CB_2} \CQ$. More generally, at the new RG fixed point, the space of boundary-condition changing operators between two boundary conditions $m\boxtimes_{\CX_2^\rev} n$ and $m'\boxtimes_{\CX_2^\rev} n'$ is given by the internal hom:
$$
[m\boxtimes_{\CX_2^\rev} n, m'\boxtimes_{\CX_2^\rev} n'] \in \CP\boxtimes_{\CB_2} \CQ. 
$$
Therefore, in this case, we obtain the following fusion formula:  
\be \label{eq:fusion-0d-walls}
( V, X, {}^\CP\CM)
 \boxtimes_{(V_2, {}^{\CB_2}\CX_2)} ( V, Y, {}^\CQ\CN )  := ( V, X\boxtimes_{\CB_2} Y, {}^{(\CP\boxtimes_{\CB_2} \CQ)}(\CM \boxtimes_{\CX_2^\rev} \CN)),  
\ee
where $X\boxtimes_{\CB_2} Y$ is naturally a symmetric separable $\dagger$-Frobenius algebra in $(\Mod_V)_{V_1|V_3}$, and we have a natural unitary monoidal equivalence $\CP\boxtimes_{\CB_2}\CQ \simeq ((\Mod_V)_{V_1|V_3})_{X\boxtimes_{\CB_2} Y|X\boxtimes_{\CB_2} Y}$.

\begin{expl} \label{expl:fusion-two-opposite-walls}
We consider a special case of Figure\,\ref{fig:0d-walls}: $(V_3,{}^{\CB_3}\CX_3)=(V_1,{}^{\CB_1}\CX_1)$, $V=V_{12}=V_{23}$ and $(V_{23},Y,{}^{\CQ}\CN)=(V_{12},X^*,{}^{\CP^\rev}\CM^\op)$, where $X^\ast$ is the dual symmetry separable $\dagger$-Frobenius algebra of $X$ in $(\Mod_{V})_{V_2|V_1}$ and can be viewed as the tensor unit of $\CP^\rev$ via the canonical monoidal equivalence: 
\begin{align*}
\CP^\rev = ((\Mod_{V})_{V_1|V_2})_{X|X})^\rev &\xrightarrow{\simeq} (\Mod_{V})_{V_2|V_1})_{X^\ast|X^\ast}^\op
\end{align*}
defined by $x \mapsto x^\ast$. Actually, as Frobenius algebras in $\Mod_{V}$, $X^\ast$ and $X$ are isomorphic thus defines
the same 0+1D chiral symmetry. In this case, the spatial fusion of the two walls gives
\be \label{eq:fusion-0d-wall-inverse}
(V,X,{}^\CP\CM) \boxtimes_{(V_2,{}^{\CB_2}\CX_2)} (V_{12},X^\ast,{}^{\CP^\rev}\CM^\op) \simeq (V, X\boxtimes_{\CB_2}X^\ast, {}^{\CP\boxtimes_{\CB_2} \CP^\rev}(\CM\boxtimes_{\CX_2^\rev}\CM^\op)),
\ee
where ${}^{\CP\boxtimes_{\CB_2} \CP^\rev}(\CM\boxtimes_{\CX_2^\rev}\CM^\op)$
is a ${}^{\CB_1}\CX_1$-${}^{\CB_1}\CX_1$-bimodule (see Definition\,\ref{def:module-enriched-monoidal-cat}). Since the 0d wall between $(V_1,{}^{\CB_1}\CX)$ and $(V_1,{}^{\CB_1}\CX)$ is unique, the wall after the spatial fusion should be spatially equivalent to the trivial 0d wall $(V_1,V_1,{}^{\CB_1}\CX_1)$. We will explain the fact in Example\,\ref{expl:edge-are-morita}. 
\end{expl}

In some interesting cases, $[m,m']_\CP \boxtimes_{\CB_2} [n,n']_\CQ \simeq [m\boxtimes_{\CX_2^\rev} n,m'\boxtimes_{\CX_2^\rev} n']$. For example, when $\CB_1=\CX_1=\CB_2=\CX_2=\CB_3=\CX_3=\CP=\CM=\CQ=\CN=\CC$, all edges are the canonical chiral gapless edges and two 0+1D gapless walls are the trivial walls. In this case, we have
\be \label{eq:iso-but-anomalous}
[m,m']_\CP \boxtimes_{\CB_2} [n,n']_\CQ = m'\otimes m^\ast \otimes n' \otimes n^\ast \simeq 
m' \otimes n' \otimes (m\otimes n)^\ast = [m\boxtimes_{\CX_2^\rev} n,m'\boxtimes_{\CX_2^\rev} n']. 
\ee
But in general, they are not isomorphic. Since $[m\boxtimes_{\CX_2^\rev} n,m'\boxtimes_{\CX_2^\rev} n']$ is the universal one, we always have the following commutative diagram: 
$$
\xymatrix{
& [m\boxtimes_{\CX_2^\rev} n,m'\boxtimes_{\CX_2^\rev} n'] \odot (m\boxtimes_{\CX_2^\rev} n) \ar[rd]^\ev & \\
([m,m']_\CP \boxtimes_{\CB_2} [n,n']_\CQ) \odot (m\boxtimes_{\CX_2} n) \ar[ur]^{\exists ! f\odot 1} \ar[r]^\simeq & \left( [m,m']_\CP\odot m \right)\boxtimes_{\CX_2^\rev} \left( [n,n']_\CQ\odot n \right) \ar[r]_-{\ev\boxtimes_{\CX_2^\rev} \ev} & m\boxtimes_{\CX_2^\rev} n. 
}
$$
The morphism 
$$
f: [m,m']_\CP \boxtimes_{\CB_2} [n,n']_\CQ \to [m\boxtimes_{\CX_2^\rev} n,m'\boxtimes_{\CX_2^\rev} n']
$$ 
is not an isomorphism in general. It means that naive fusion of observables on the two world lines of two 0d gapless wall (i.e. $[m,m']_\CP \boxtimes_{\CB_2} [n,n']_\CQ$) is not universal or at RG fixed point. It will flow to a RG fixed point, which is universal and defined by $[m\boxtimes_{\CX_2} n,m'\boxtimes_{\CX_2} n']$. In some sense, this morphism $f$ catches the information of the RG flow. Interestingly, even in the general cases, for a special class of edges excitations (or  excitations in the trivial wall), $f$ is an isomorphism (see \cite[Remark\,6.3]{kz4}).

From another point of view, that $f$ is not an isomorphism simply can be viewed as an indicator that there are certain anomaly, called {\it spatial fusion anomalies}. Indeed, 1d gapless edges, together with 0d walls between them, are anomalous 1d phases when the bulk $(\CC,c)$ is non-trivial. It is possible that this spatial fusion anomaly vanishes for some special anomalous phases as shown in the case discussed in the Eq.\,(\ref{eq:iso-but-anomalous}). But when the bulk phase $(\CC,c)$ is the trivial 2d topological order, the spatial fusion anomaly should definitely vanish. This is proved in \cite[Theorem\ 4.5]{kyz}. 

\begin{rem}
While spatial fusions are often anomalous, temporal fusions are often anomaly-free, i.e. $[m',m'']_\CP \boxtimes_{[m',m']_\CP} [m,m']_\CP \simeq [m,m'']_\CP$ for simple $m,m',m''\in\CM_i$ if all four internal homs are non-zero. 
\end{rem}

In general, the 1+1D chiral symmetries on two gapless walls $V_{12}$ and $V_{23}$ are potentially different, i.e. $V_{12}\neq V_{23}$. In this case, the spatial fusion of two walls causes the 1+1D chiral symmetries to break further down to a smaller VOA $V_{123}$. To compute the spatial fusion, we need further specify the fusion process. More precisely, we assume that this fusion is achieved in two steps: first breaking both 1+1D chiral symmetries $V_{12}$ and $V_{23}$ down to $V_{123}$ without changing $X$ and $Y$, then fusing according to (\ref{eq:fusion-0d-walls}). More explicitly, the first step gives: 
$$
(V_{12},X,{}^\CP\CM) \mapsto (V_{123}, X, {}^{\CP'}\CM') \quad\quad (V_{23},Y,{}^\CQ\CN) \mapsto (V_{123},Y,{}^{\CQ'}\CN'), 
$$ 
where $X$ and $Y$ should be viewed as their images in $(\Mod_{V_{123}})_{V_1|V_2}$ via two forgetful functors: 
$$
(\Mod_{V_{12}})_{V_1|V_2} \xrightarrow{\forget} (\Mod_{V_{123}})_{V_1|V_2}  \xleftarrow{\forget} (\Mod_{V_{23}})_{V_1|V_2},
$$ 
and $\CP'= ((\Mod_{V_{123}})_{V_1|V_2})_{X|X}$ and $\CQ'= ((\Mod_{V_{123}})_{V_1|V_2})_{Y|Y}$, and $\CM',\CN'$ are uniquely determined. The second step gives: 
\be \label{eq:fusion-0d-walls-2}
(V_{123}, X, {}^{\CP'}\CM') \boxtimes_{(V_2, {}^{\CB_2}\CX_2)} (V_{123},Y,{}^{\CQ'}\CN')
= ( V_{123}, X\boxtimes_{\CB_2} Y, {}^{(\CP'\boxtimes_{\CB_2}\CQ')}(\CM'\boxtimes_{\CX_2^\rev}\CN')).
\ee
where $X\boxtimes_{\CB_2} Y$ is naturally a symmetric separable $\dagger$-Frobenius algebra in $(\Mod_{V_{123}})_{V_1|V_3}$.

\begin{rem} \label{rem:abbreviation}
The spatial fusion formula (\ref{eq:fusion-0d-walls-2}) also suggests that it introduces very little confusion if we abbreviate the triple $(V_{12},X,{}^\CP\CM)$ to ${}^\CP\CM$ for simplicity unless there is a breaking of 1+1D chiral symmetries. 
\end{rem}

\subsection{Morita equivalence} \label{sec:morita}

The physical results in Section\,\ref{sec:0d-defects} and Section\,\ref{sec:fusion-anomaly} lead us to a representation theory of enriched monoidal categories as we will sketch in this subsection. This theory will be developed in details in \cite{kyz2}. 

\medskip
Let $\CA, \CB$ be UMTC's and $\CX, \CY$ two indecomposable UMFC's. Let ${}^{\CA}\CX$ and ${}^{\CB}\CY$ be the indecomposable enriched unitary multi-fusion categories obtained from the canonical construction. The time reversal of ${}^{\CB}\CY$ is defined by $({}^{\CB}\CY)^\rev:={}^{\overline{\CB}}\CY^\rev$. The Deligne tensor product 
$$
{}^{\CA}\CX \boxtimes {}^{\overline{\CB}}\CY^\rev := {}^{\CA\boxtimes\overline{\CB}} (\CX\boxtimes \CY^\rev)
$$
is again an indecomposable enriched unitary multi-fusion categories. 
We give the following working definition of modules over an indecomposable enriched unitary multi-fusion category first introduced in \cite{zheng}. 
\begin{defn} \label{def:module-enriched-monoidal-cat}
An enriched category ${}^\CP\CM$ obtained from canonical construction is called
\bnu
\item a left ${}^{\CA}\CX$-module if $\CP$ is a multi-fusion left $\CA$-module, 
and $\CM$ is a left $\CX^\rev\boxtimes_\CA\CP$-module such that the $\CP$-module structure on $\CM$ coincides with the following composed unitary monoidal functor
$$
\CP \hookrightarrow \CX^\rev \boxtimes \CP \xrightarrow{\boxtimes_\CA} \CX^\rev \boxtimes_\CA \CP \xrightarrow{\phi_\CM} \fun_\bh(\CM,\CM).
$$

\item a right ${}^{\CB}\CY$-module is a left $({}^\CB\CY)^\rev$-module; 

\item a ${}^{\CA}\CX$-${}^{\CB}\CY$-bimodule is a left $({}^\CB\CY)^\rev\boxtimes{}^{\CA}\CX$-module. 

\enu
The (left, right, bi-)bimodule ${}^\CP\CM$ is called closed if $\phi_\CM$ is also an equivalence. 
\end{defn}

\begin{rem} \label{rem:def-bimodule}
If ${}^{\CA}\CX$ and ${}^{\CB}\CY$ are two gapless edges of a 2d topological order, then ${}^\CP\CM$ describe a 0+1D gapless wall between two edges. Therefore, the following picture clarify the physical meaning of the notions in Definition\,\ref{def:module-enriched-monoidal-cat}: 
\be \label{pic:bimodule}
\raisebox{-30pt}{
  \begin{picture}(120,60)
   \put(0,10){\scalebox{1}{\includegraphics{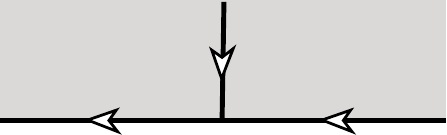}}}
   \put(0,10){
     \setlength{\unitlength}{.75pt}\put(0,0){
     \put(130,30)  {\scriptsize $\CB$}
     \put(130,-8) {\scriptsize $\CY$}
     \put(125,55) {\scriptsize ${}^\CB\CY$}
     \put(35,30) {\scriptsize $\CA$}
     \put(35,-8) {\scriptsize $\CX$}
     \put(30,55) {\scriptsize ${}^\CA\CX$}
     \put(93,27)   {\scriptsize $\CP$}
     \put(83,-5) {\scriptsize $\CM$}
     \put(75,55) {\scriptsize ${}^\CP\CM$}
     
     }\setlength{\unitlength}{1pt}}
  \end{picture}} 
\ee
Note that if ${}^\CP\CM$ is closed, then $\CP$ must be a closed multi-fusion $\CA$-$\CB$-bimodule, i.e. $\phi_\CP:\overline{\CA}\boxtimes\CB \to \FZ(\CP)$ is a braided equivalence. Note that we have also used the left-right convention in Remark\,\ref{rem:left-right-convention}.\footnote{Physically, $\CA$ acts on $\CP$ from left, but $\CX$ acts on $\CM$ from right. It seems that neither of the two left-right conventions is natural. This is due to the fact that we require $\CA$-acting on $\CX$ from left in our canonical construction, which makes the bulk and topological Wick rotation looks natural. If we only study the edge and ignore the bulk and topological Wick rotation, by requiring a right $\CA$-action on $\CX$ in a new ``canonical construction", we can flip the arrows in (\ref{pic:bimodule}) such that the left-right convention in Definition\,\ref{def:module-enriched-monoidal-cat} looks natural. This is what we will do in \cite{kyz2}.} 
\end{rem}

\begin{rem}
The representation theory of an enriched monoidal category will be developed in \cite{kyz2}. We briefly clarify the notion of a left ${}^{\CA}\CX$-module mathematically. All $\CA$-enriched categories form a 2-category $\mathbf{Cat}^\CA$. The monoidal functor $\otimes: \CA \times \CA \to \CA$ defines a pushforword 2-functor $\otimes_\ast: \mathbf{Cat}^{\CA \times \CA} \to \mathbf{Cat}^\CA$. As a consequence, $\mathbf{Cat}^\CA$ is a monoidal 2-category with the tensor product defined by 
$$
\mathbf{Cat}^\CA \times \mathbf{Cat}^\CA \xrightarrow{\times} \mathbf{Cat}^{\CA\times \CA} \xrightarrow{\otimes_\ast}  \mathbf{Cat}^\CA. 
$$ 
An $\CA$-enriched monoidal category ${}^\CA\CX$ is an algebra object in the monoidal 2-category $\mathbf{Cat}^{\CA}$. For a multi-fusion left $\CA$-module $\CP$, $\mathbf{Cat}^\CP$ is a naturally a left $\mathbf{Cat}^{\CA}$-module. The enriched category ${}^\CP\CM$ in Definition\,\ref{def:module-enriched-monoidal-cat} is precisely a left ${}^\CA\CX$-module in $\mathbf{Cat}^\CP$. 
\end{rem}

\begin{rem} \label{rem:op-enriched-module}
If ${}^\CP\CM$ is a ${}^{\CA}\CX$-${}^{\CB}\CY$-bimodule, then $({}^\CP\CM)^\op:={}^{\CP^\rev}(\CM^{\mathrm{op}})$ is automatically a ${}^{\CB}\CY$-${}^{\CA}\CX$-bimodule. 
\end{rem}


The following mathematical definition echoes with the physical fusion formula (\ref{eq:fusion-0d-walls}).
\begin{defn}[\cite{zheng}] \label{def:relative-tensor-product}
Let ${}^\CP\CM$ and ${}^\CQ\CN$ be a right ${}^{\CB}\CY$-module and a left ${}^{\CB}\CY$-module, respectively. We define a relative tensor product $\boxtimes_{{}^\CB\CY}$ as follows: 
$$
{}^\CP\CM \boxtimes_{{}^{\CB}\CY}  {}^\CQ\CN := {}^{(\CP\boxtimes_{\CB}\CQ)}(\CM\boxtimes_{\CY^\rev}\CN).  
$$
When $\CB=\bh,\CY=\bh$, it is just the Deligne tensor product $\boxtimes$, i.e. ${}^\CP\CM \boxtimes  {}^\CQ\CN := {}^{(\CP\boxtimes \CQ)}(\CM\boxtimes\CN)$.
\end{defn}

\begin{figure}
$$
\raisebox{-30pt}{
  \begin{picture}(120,95)
   \put(0,15){\scalebox{1}{\includegraphics{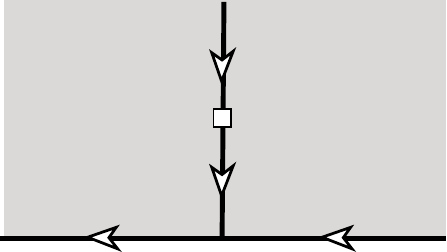}}}
   \put(0,15){
     \setlength{\unitlength}{.75pt}\put(0,0){
     \put(130,48)  {\scriptsize $\CB$}
     \put(130,-8) {\scriptsize $\CY$}
     \put(125,85) {\scriptsize ${}^\CB\CY$}
     \put(35,48) {\scriptsize $\CA$}
     \put(35,-8) {\scriptsize $\CX$}
     \put(30,85) {\scriptsize ${}^\CA\CX$}
     \put(93,27)   {\scriptsize $\CQ$}
     \put(83,-5) {\scriptsize $\CN$}
     \put(93,48) {\scriptsize $\CF$}
     \put(93,68) {\scriptsize $\CP$}
     
     }\setlength{\unitlength}{1pt}}
  \end{picture}} 
$$
\caption{This picture the physical intuition behind the notion of a spatial equivalence of bimodules (see Definition\,\ref{def:spatial-functor}).  
}
\label{fig:spatial-functor}
\end{figure}

There is a mathematical notion of a left ${}^{\CA}\CX$-module functor between two left ${}^{\CA}\CX$-modules ${}^\CP\CM$ and ${}^\CQ\CN$. It is just an enriched functor $F: {}^\CP\CM \to {}^\CQ\CN$, i.e. a 1-morphism in $\mathbf{Cat}^\CA$, such that $F$ intertwines the ${}^{\CA}\CX$-actions (see \cite{kyz2} for more details). It is called an ${}^{\CA}\CX$-module equivalence if $F$ is an enriched equivalence. In this case, we denote such an equivalence by ${}^\CP\CM \simeq {}^\CQ\CN$. 

It is, however, not enrough to describe spatial equivalences among 0+1D walls. We need a new notion of a module functor between enriched categories with different background categories. This new notion is given in Definition\,\ref{def:spatial-functor}, the physical intuition behind which is depicted in Figure\,\ref{fig:spatial-functor} (recall Figure\,\ref{fig:0d-defects-equivalence}). 
\begin{defn}[\cite{zheng}] \label{def:spatial-functor}
For ${}^\CA\CX$-${}^\CB\CY$-bimodules ${}^\CP\CM$ and ${}^\CQ\CN$, a spatial ${}^\CA\CX$-${}^\CB\CY$-bimodule functor from ${}^\CP\CM$ to ${}^\CQ\CN$ is a pair $(\CF,F)$, where $\CF$ is a closed left $\CP\boxtimes_{\overline{\CA}\boxtimes \CB} \CQ^\rev$-module, 
and $F: \CM \to \CF\boxtimes_\CQ \CN$ is a $\CX^\rev\boxtimes_\CA\CP\boxtimes_\CB\CY$-module functor. It is called a spatial equivalence if $F$ is an equivalence. We denote such a spatial equivalence by ${}^\CP\CM \ssimeq {}^\CQ\CN$. 
\end{defn}


\begin{expl}  \label{exam:fun-CC-H}
Recall Remark\,\ref{rem:gappable-excluded}, when we roll up a 1+1D anomaly-free RCFT to a cylinder $S^1\times \Rb^1$ then shrink $S^1$ to a point, we obtain a gapped 0+1D phase. Mathematically, by integrating the RCFT over this cylinder, we obtain a gappable gapless 0+1D phase $(\Cb,Y,\CSs)$, where the enriched category $\CSs$ is a ${}^\bh\bh$-${}^\bh\bh$-bimodule ${}^{\mathrm{Fun}_\bh(\CM,\CM)}\CM$ for a finite unitary category $\CM$. Note that $\CM$ is a $\mathrm{Fun}_\bh(\CM,\CM)$-$\bh$-bimodule. We have the following equivalences of bimodules: 
$$
F: \CM^{\mathrm{op}}\boxtimes_{\mathrm{Fun}_\bh(\CM,\CM)} \CM \simeq \bh
\quad\quad \mbox{and} \quad\quad 
G:\CM\boxtimes_\bh \CM^{\mathrm{op}} \simeq \mathrm{Fun}_\bh(\CM,\CM).
$$ 
It is clear that $\CM^\op$ is a closed $\bh\boxtimes_{\bh\boxtimes\bh}\mathrm{Fun}_\bh(\CM,\CM)^\rev$-bimodule. Therefore, 
$$
(\CM,F): {}^{\mathrm{Fun}_\bh(\CM,\CM)}\CM \xrightarrow{\ssimeq} {}^\bh\bh
$$ defines a spatial equivalence of ${}^\bh\bh$-${}^\bh\bh$-bimodules. This shows that the spatial equivalence is capable of describing how to gap out a 0+1D gapless phase as we claimed in Remark\,\ref{rem:gappable-excluded} and Remark\,\ref{rem:include-gappable}. We will discuss more general situations in Section\,\ref{sec:gappable-edges-morita}. 
\end{expl}

\begin{figure} 
$$
 \raisebox{-50pt}{
  \begin{picture}(185,100)
   \put(-10,0){\scalebox{0.5}{\includegraphics{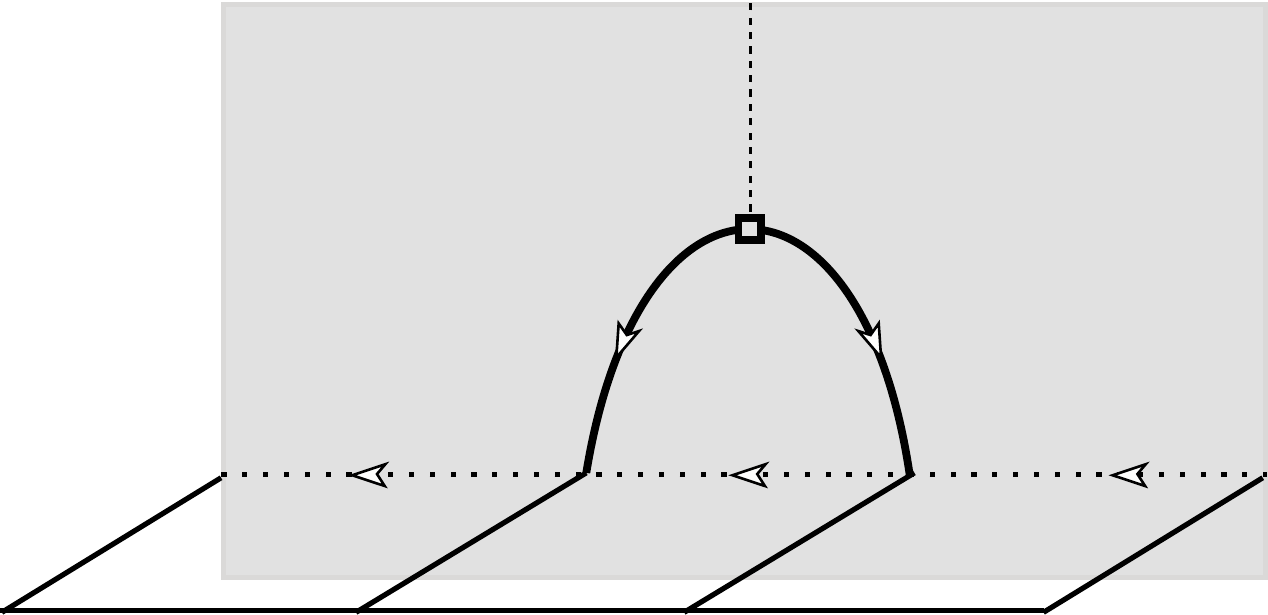}}}
   \put(-10,0){
     \setlength{\unitlength}{.75pt}\put(0,0){
      \put(22,6) {\footnotesize $(\CB_1,c)$}
      \put(180,6) {\footnotesize $(\CB_1,c)$}
      \put(115,122) {\footnotesize $(V_1,V_1,{}^{\CB_1}(\CP^\op)$)}
     \put(68,33)  {\footnotesize $\CB_1$}
     \put(140,15)  {\footnotesize $\CB_2$}
     \put(215,33)  {\footnotesize $\CB_1$}
     \put(120,42)  {\footnotesize $(V_2,{}^{\CB_2}\CB_2)$}
     \put(140,105)  {\footnotesize $\CB_1$}
     \put(137,60)  {\footnotesize $\CP^\op$}
     \put(45,100)  {\footnotesize $(V_1,{}^{\CB_1}\CB_1)$}
     
    
     \put(193,100) {\footnotesize $(V_1,{}^{\CB_1}\CB_1)$}
     \put(112,18) {\footnotesize $\CP$}
     \put(92,5) {\footnotesize $\CP$}
     \put(174,18) {\footnotesize $\CP^\op$}
     \put(155,5) {\footnotesize $\CP^\op$}
     
     \put(107,50)  {\footnotesize $\CP$}
     \put(173,50)  {\footnotesize $\CP^\rev$}
     
     \put(63,75)  {\footnotesize $\CB_1$}
     \put(205,75) {\footnotesize $\CB_1$}  
     }\setlength{\unitlength}{1pt}}  \end{picture}}
     \quad\quad
      \raisebox{-50pt}{
  \begin{picture}(150,100)
   \put(-20,0){\scalebox{0.5}{\includegraphics{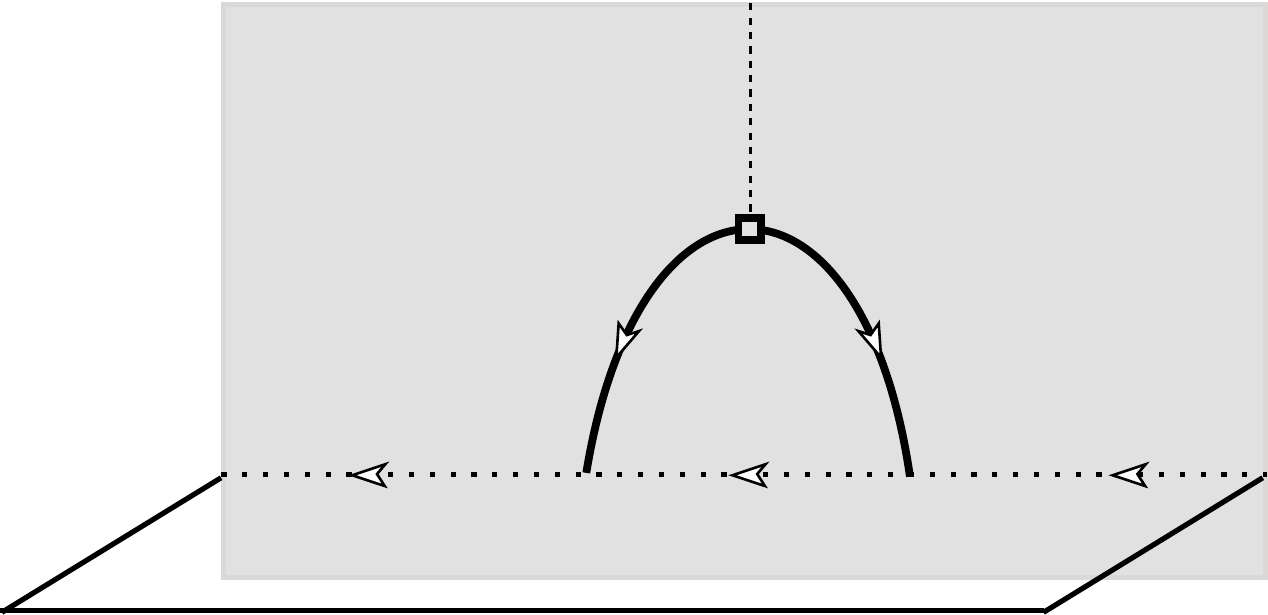}}}
   \put(-20,0){
     \setlength{\unitlength}{.75pt}\put(0,0){
     \put(22,6) {\footnotesize $(\CC,c)$}
     \put(68,33)  {\footnotesize $\CX_1$}
     \put(140,15)  {\footnotesize $\CX_2$}
     \put(215,33)  {\footnotesize $\CX_1$}
     \put(120,42)  {\footnotesize $(V_2,{}^{\CB_2}\CB_2)$}
     \put(140,105)  {\footnotesize $\CB_1$}
     \put(115,122) {\footnotesize $(V_1,V_1,{}^{\CB_1}\CX_1)$}
     \put(137,60)  {\footnotesize $\CP^\op$}
     \put(45,100)  {\footnotesize $(V_1,{}^{\CB_1}\CX_1)$}
     
    
     \put(195,100) {\footnotesize $(V_1,{}^{\CB_1}\CX_1)$}
     \put(109,18) {\footnotesize $\CM$}
     \put(172,18) {\footnotesize $\CM^\op$}
     
     \put(107,50)  {\footnotesize $\CP$}
     \put(173,50)  {\footnotesize $\CP^\rev$}
     
     \put(63,75)  {\footnotesize $\CB_1$}
     \put(205,75) {\footnotesize $\CB_1$}
     }\setlength{\unitlength}{1pt}}
  \end{picture}}
$$
$$
(a) \quad\quad\quad\quad\quad\quad\quad\quad\quad\quad\quad\quad\quad\quad\quad\quad
\quad\quad\quad\quad\quad\quad (b)
$$
\caption{Picture $(a)$ and $(b)$ illustrate the proofs in Example\,\ref{expl:AXBXA} and \ref{expl:edge-are-morita}, respectively. 
}
\label{fig:morita-proof}
\end{figure}

\begin{expl} \label{expl:AXBXA}
This example is illustrated in Figure\,\ref{fig:morita-proof} (a). Let $\CB_1,\CB_2$ be UMTC's and $\CP$ a closed multi-fusion $\CB_1$-$\CB_2$-bimodule. Then ${}^\CP\CP$ is a ${}^{\CB_1}\CB_1$-${}^{\CB_2}\CB_2$-bimodule and ${}^{\CP^\rev}\CP^{\mathrm{op}}$ is a ${}^{\CB_2}\CB_2$-${}^{\CB_1}\CB_1$-bimodule. Then we have the following ${}^{\CB_1}\CB_1$-${}^{\CB_1}\CB_1$-bimodule equivalences: 
$$
({}^{\CP^\rev}\CP^{\mathrm{op}}) \boxtimes_{{}^{\CB_2}\CB_2} {}^\CP\CP \simeq 
{}^{\CP^\rev\boxtimes_{\CB_2}\CP} (\CP^{\mathrm{op}}\boxtimes_{\CB_2^\rev}\CP) \simeq {}^{\fun_{\CB_1}(\CP,\CP)}\fun_{\CB_1}(\CP,\CP),
$$
where we have used the canonical monoidal equivalence $\CP\boxtimes_{\CB_2} \CP^\rev \simeq \fun_{\CB_1}(\CP,\CP)$ defined by $x\boxtimes_{\CB_2} y\mapsto x\otimes - \otimes y$ \cite[Corollary\,2.7]{zheng}. It is clear that $\CP^\op$ is a closed $\CB_1\boxtimes_{\FZ(\CB_1)} (\CP\boxtimes_{\CB_2}\CP^\rev)^\rev$-module. Therefore, $\CP^{\mathrm{op}}$, together with the canonical $\CB_1$-module equivalence: $F:\CP^\op \boxtimes_{\fun_{\CB_1}(\CP,\CP)} \fun_{\CB_1}(\CP,\CP) \simeq \CP^\op$, defines a spatial equivalence: 
$$
(\CP^{\mathrm{op}}, F): {}^{\fun_{\CB_1}(\CP,\CP)}\fun_{\CB_1}(\CP,\CP) \xrightarrow{\ssimeq} {}^{\CB_1}(\CP^\op). 
$$
Therefore, we obtain 
$$
{}^\CP\CP \boxtimes_{{}^{\CB_2}\CB_2} ({}^{\CP^\rev}\CP^{\mathrm{op}}) \ssimeq {}^{\CB_1}(\CP^\op).
$$
If $\CB_i=\Mod_{V_i}$ for VOA $V_i, i=1,2$, then $(V_1,V_1,{}^{\CB_1}(\CP^\op))$ defines a 0+1D gapless ``relative boundary'' of the 1d gapped wall $\CP\boxtimes_{\CB_2}\CP^\rev$ (see Figure\,\ref{fig:morita-proof} (a)). 
\void{
In the special case when $\CB_1=\bh$ and $\CB_2=\FZ(\CP)$, we obtain that 
$$
{}^\CP\CP \boxtimes_{({}^{\FZ(\CP)}\FZ(\CP))} ({}^\CP\CP)^\op \simeq {}^{\fun_\bh(\CP,\CP)}\fun_\bh(\CP,\CP) \ssimeq {}^\bh (\CP^\op). 
$$
Note that $(\Cb, \Cb, {}^\bh (\CP^\op))$ is a gapped 0d boundary of a 1d anomaly-free but unstable topological order $\fun_\bh(\CP,\CP)$. In other words, this fusion of two gapless 0d walls produces a gapped 0d wall under a spatial equivalence. 
}
\end{expl}

\begin{defn}
An ${}^\CA\CX$-${}^\CB\CY$-bimodule ${}^\CP\CM$ is called {\it spatially invertible} if there is a ${}^\CB\CY$-${}^\CA\CX$-bimodule ${}^\CQ\CN$ such that 
$$
{}^\CP\CM \boxtimes_{{}^\CB\CY} {}^\CQ\CN \ssimeq {}^\CA\CX, \quad\quad
\mbox{and} \quad\quad 
{}^\CQ\CN\boxtimes_{{}^\CA\CX}{}^\CP\CM \ssimeq {}^\CB\CY
$$
as bimodules. Two enriched multi-fusion category ${}^\CA\CX$ and ${}^\CB\CY$ are called {\it spatially Morita equivalent} if there exists a spatially invertible ${}^\CA\CX$-${}^\CB\CY$-bimodule. 
\end{defn}

Recall that if $\CY$ describes a 0d gapped wall between two gapped edges $\CS$ and $\CT$ (i.e. two UFC's), then $\CY$ is automatically an invertible $\CS$-$\CT$-bimodule with the inverse given by $\CY^\op$. We should expect that the similar phenomenon for a 0d wall between two chiral gapless edges. We explain this fact in the following example. 
\begin{expl} \label{expl:edge-are-morita}
Recall Example\,\ref{expl:fusion-two-opposite-walls}. As illustrated in Figure\,\ref{fig:morita-proof} (b), $\CP^\op$ is clearly a closed left $\CB_1\boxtimes_{\FZ(\CB_1)}(\CP\boxtimes_{\CB_2}\CP^\rev)^\rev$-module. By the property of factorization homology (recall Theorem\,\ref{thm:ai} and Example\,\ref{expl:ai}), we also have the following equivalence: 
$$
F: \CP^\op\boxtimes_{\fun_{\CB_1}(\CP,\CP)} (\CM\boxtimes_{\CX_2^\rev}\CM^\op) \xrightarrow{\simeq} \CX_1 
$$
as two left $\CX_1^\rev \boxtimes_{\CB_1}\CX_1$-modules. 
Therefore, we obtain the following spatial equivalence 
$$
{}^\CP\CM \boxtimes_{{}^{\CB_2}\CX_2} {}^{\CP^\rev}\CM^\op 
\simeq {}^{\fun_{\CB_1}(\CP,\CP)}(\CM\boxtimes_{\CX_2^\rev}\CM^\op) \ssimeq {}^{\CB_1}\CX_1.
$$
Similarly, one can also show that ${}^{\CP^\rev}\CM^\op \boxtimes_{{}^{\CB_1}\CX_1} {}^\CP\CM\ssimeq {}^{\CB_2}\CX_2$. In other words, the 
${}^{\CB_1}\CX_1$-${}^{\CB_2}\CX_2$-bimodule ${}^\CP\CM$ is spatially invertible, and defines a spatial Morita equivalence between ${}^{\CB_1}\CX_1$ and ${}^{\CB_2}\CX_2$. This implies that following spatial equivalence between 0+1D walls (recall (\ref{eq:fusion-0d-wall-inverse})): 
$$
(V, X\boxtimes_{\CB_2}X^\ast, {}^{\CP\boxtimes_{\CB_2} \CP^\rev}(\CM\boxtimes_{\CX_2^\rev}\CM^\op)) \ssimeq
(V_1,V_1,{}^{\CB_1}\CX_1). 
$$
If we discuss spatial equivalences, it is safe to abbreviate a triple $(V,X,{}^\CP\CM)$ to ${}^\CP\CM$ because the 1+1D and 0+1D chiral symmetries are not preserved under spatial equivalences. 
\end{expl}

Two 1d gapless edges are called spatially Morita equivalent if the associated enriched multi-fusion categories are spatially Morita equivalent. Then Example\,\ref{expl:edge-are-morita} gives the following physical theorem. 
\begin{pthm}
A 0+1D gapless wall between two 1d chiral gapless edges of the same 2d topological order defines a spatial Morita equivalence between these two chiral gapless edges. 
\end{pthm}

\section{Boundary-bulk relation II: chiral gapless edges} \label{sec:boundary-bulk}
In this section, we generalize the boundary-bulk relation for gapped edges to that for both gapped and chiral gapless edges.

\subsection{Bulk of a chiral gapless edge} \label{sec:bulk}

Given a chiral gapless edge $(V,{}^\CB\CX)$ of a bulk topological order $(\CC,c)$, how to understand those bulk excitations in terms of those on the edge? Let us first look at the gapped cases. When the edge is gapped, i.e. $V=\Cb, \CB=\bh$, a bulk topological excitation is precisely an edge excitation $x\in \CX$ that can be moved into the bulk. 
\bnu
\item An edge excitation that can be moved inside the bulk must be equipped with a half-braiding with all edge excitations in $\CX$. More explicitly, a bulk excitation can be realized by an edge excitation $x$, together with a family of isomorphisms
\be \label{eq:half-braiding}
x\otimes y \xrightarrow{\beta_{x,y}} y\otimes x, \quad\quad \forall y\in\CX,
\ee
such that the following diagrams
\be \label{diag:naturalness}
\raisebox{2em}{\xymatrix{
x\otimes y \ar[r]^{\beta_{x,y}} \ar[d]_{1f} & y\otimes x \ar[d]^{f1} \\
x\otimes z \ar[r]^{\beta_{x,z}} & z\otimes x 
}} \quad\quad\quad \forall f\in \hom_\CX(y,z)
\ee
are commutative. This family of isomorphisms $\beta_{x,-}=\{ \beta_{x,y} \}_{y\in\CX}$ defines a natural isomorphism $\beta_{x,-}: x\otimes - \to -\otimes x$, which is called a half-braiding. Therefore, the pair $(x,\beta_{x,-})$ defines a bulk excitation.

\item Moreover, morphisms (or instantons) between two bulk excitations $(x,\beta_{x,-})$ and $(y,\beta_{y,-})$ are precisely those morphisms (instantons) $f\in \hom_\CX(x,y)$ respecting the half-braiding, i.e. rending the following diagrams commutative:
\be \label{diag:hom-in-center}
\raisebox{2em}{\xymatrix{
x\otimes z \ar[r]^{\beta_{x,z}} \ar[d]_{f\otimes 1} & z\otimes x \ar[d]^{1\otimes f} \\
y\otimes z \ar[r]^{\beta_{y,z}} & z \otimes y\,  
}}
\quad\quad\quad \forall z\in\CX.
\ee
 
\enu
All such pairs form a category, which is precisely the Drinfeld center $\FZ(\CX)$ of $\CX$. The boundary-bulk relation says that $\CC\simeq \FZ(\CX)$ as UMTC's. We need generalize these arguments to the gapless edge $(V,{}^\CB\CX)$.

\begin{figure} 
$$
 \raisebox{-100pt}{
  \begin{picture}(70,150)
   \put(-80,8){\scalebox{0.6}{\includegraphics{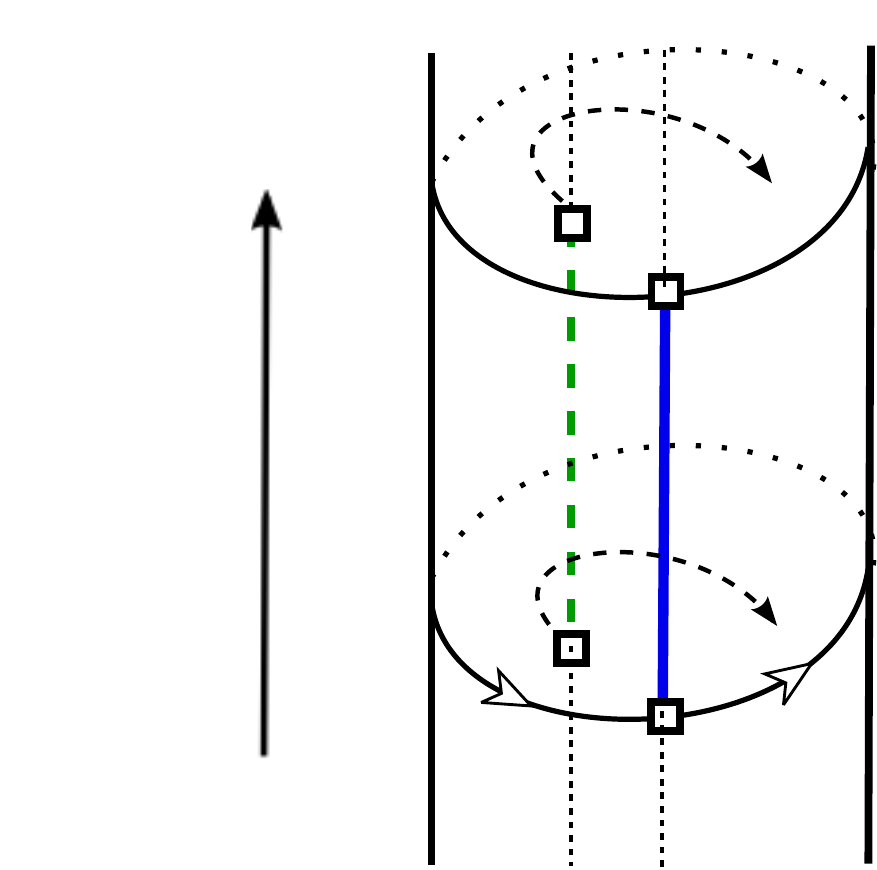}}}
   \put(-40,8){
     \setlength{\unitlength}{.75pt}\put(0,-75){
     \put(-5, 175) {$t$}
     \put(58,180)  {$\id_x$}
     \put(108,102)  {$y$}
     \put(108,200)  {$z$}
     \put(107, 182) {$[y,z]$}
     \put(85,123)  {$x$}
     \put(85,225) {$x$}
     }\setlength{\unitlength}{1pt}}
  \end{picture}}
\quad\quad\quad\quad\quad \quad\quad\quad\quad
 \raisebox{-100pt}{
  \begin{picture}(100,150)
   \put(-60,8){\scalebox{0.6}{\includegraphics{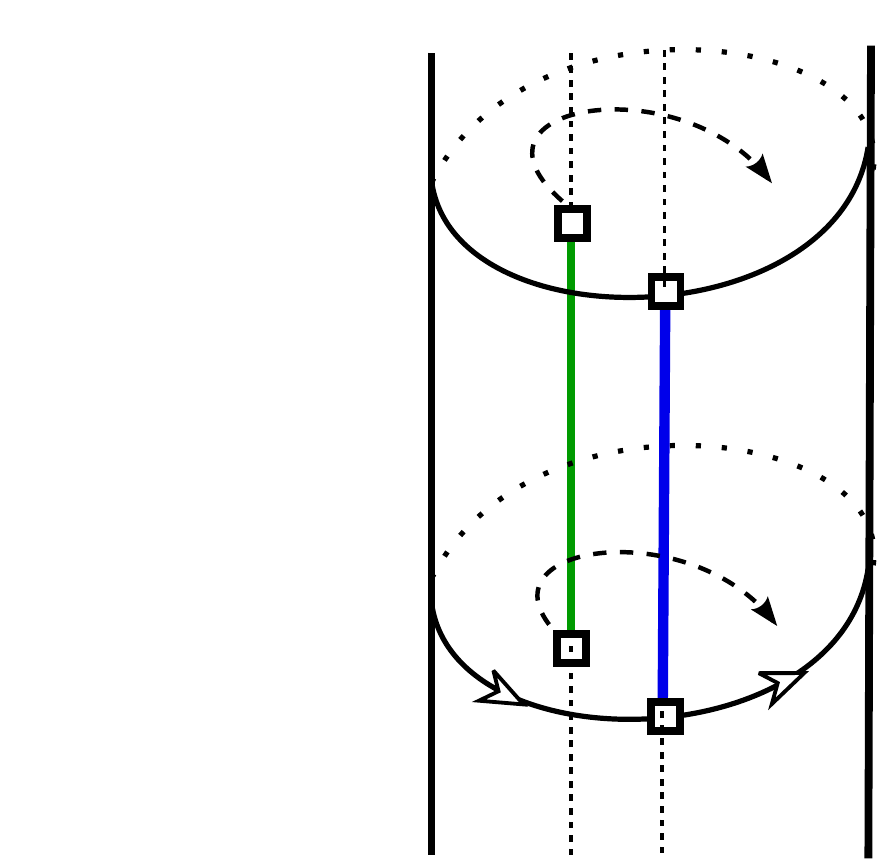}}}
   \put(0,8){
     \setlength{\unitlength}{.75pt}\put(-40,-75){
     \put(80,175)  {$s$}
     \put(120,97)  {$z$}
     \put(123,198)  {$z$}
     \put(121,180) {$[z,z]$}
     \put(79,118)  {$x$}
     \put(79,220) {$y$}
     }\setlength{\unitlength}{1pt}}
  \end{picture}}  
$$
$$
(a) \quad\quad\quad\quad \quad \quad\quad\quad\quad \quad\quad \quad
\quad\quad\quad\quad  (b)
$$
\caption{These two pictures depicts observables on the world line supported on a topological excitation in the bulk can be half-braided and fused with those on the world line supported on edge. Picture (a) illustrates the meaning of a half-braiding, and the label $\id_x$ of the green dotted line represents the canonical morphism $\id_x: \one \to [x,x]_\CBs$ and the vacuum state in the boundary CFT $[x,x]_\CB$; Picture (b) illustrate instantons (labeled by $s$) between $x$ and $y$ in the bulk and its compatibility with the half-braidings. 
}
\label{fig:hb+fusion}
\end{figure}

\medskip
Now we consider a chiral gapless edge as depicted in Figure\,\ref{fig:hb+fusion} (a), where $x$ is a bulk topological excitation and $y,z$ are two edge excitations. We should expect again that a bulk excitation can be realized by an edge excitation, together with a ``half-braiding'', a notion which will be made precise below. 
\bnu
\item An edge excitation $x\in\CX$ can be moved into the bulk if it is equipped with a half-braiding, which should consist of the following isomorphisms in $\CX$: 
\be \label{eq:beta}
\beta_{x,y}: x\otimes y \xrightarrow{\simeq} y\otimes x, \quad\quad\quad \forall y\in\CX.
\ee  
Moreover, they should satisfy a similar naturalness condition as in (\ref{diag:naturalness}). Namely, $\beta_{x,-}: x\otimes - \to -\otimes x$ should be a natural isomorphism between two endo-functor of $\CX$. But this condition is not enough because $\hom_\CX(y,z)$ contains only the vacuum channels of the whole physical hom space $[y,z]_\CB$.  

Note that a half-braiding is an adiabatic process of moving the bulk excitation $x$ around an edge excitation $y\in\CX$. This move automatically moves all observables on the world line supported on $x$. What observables could live on
this world line in the bulk? It has to be a subspace of the boundary CFT $[x,x]_\CB$. If this subspace is zero, then it is reasonable to say that $x$ is not equipped with any half-braiding. The minimal requirement for a non-zero edge excitation $x$ to move into the bulk is that the vacuum state in the boundary CFT $[x,x]_\CB$ survives on the world line in the bulk. This vacuum state is characterized by the canonical morphism $\id_x: \one_\CB \to [x,x]_\CB$ under the assumption that the chiral symmetry $V=\one_\CB$ is preserved. This vacuum state can be fused into the space of observables on the world line supported on the edge, say $[y,z]_\CB$, along a path from the bulk to the edge. As illustrated in Figure\,\ref{fig:hb+fusion} (a), it is clear that this fusion should be path independent. Namely, we can fuse it into $[y,z]_\CB$ from left or first half-braid it to the right then fuse it from right without making any difference. This leads to the following commutative diagram: 
\be \label{diag:half-braiding-0}
\raisebox{2em}{\xymatrix@!C=20ex{
[y,z]_\CB \ar[r]^-{1 \otimes \id_x} \ar[d]_{\id_x \otimes 1} & [y,z]_\CB \otimes [x,x]_\CB \ar[r]^\otimes &
[y\otimes x, z\otimes x]_\CB  \ar[d]^{-\circ \beta_{x,y}} \\
[x,x]_\CB \otimes [y,z]_\CB \ar[r]^\otimes & [x\otimes y, x\otimes z]_\CB \ar[r]^{\beta_{x,z}\circ -}  & [x\otimes y, z\otimes x]_\CB\, .
}}
\ee
The data (\ref{eq:beta}) and the condition (\ref{diag:half-braiding-0}) give the precisely meaning of a ``half-braiding'' for a gapless edge. Therefore, such a pair $(x,\beta_{x,-})$ should define a bulk excitation.

\item What about the instantons between two such bulk excitations, say $(x,\beta_{x,-})$ and $(y,\beta_{y,-})$? Consider the situation depicted in Figure\,\ref{fig:hb+fusion} (b), where $z$ is an edge excitation. Not all observables in $[x,y]$ are allowed to live in the bulk. We denote the maximal sub-object of $[x,y]_\CB$ that is allowed to live in the bulk by $\iota: s\hookrightarrow [x,y]_\CB$. Then fusing $s$ with $[z,z]_\CB$ from left should not be different from first half-braiding it with $[z,z]_\CB$ then fusing it with $[z,z]_\CB$ from right. As a consequence, we obtain the following commutative diagram: 
\be \label{diag:hb+fusion}
\raisebox{2em}{\xymatrix@!C=20ex{
s  \ar[d]_{\iota\, \otimes \id_z} \ar[r]^-{\id_z \otimes\, \iota} & [z,z]_\CB \otimes [x,y]_\CB \ar[r]^{\otimes} & [z\otimes x, z\otimes y]_\CB \ar[d]^{-\circ \beta_{x,z}} \\
[x,y]_\CB\otimes [z,z]_\CB \ar[r]^\otimes & [x\otimes z, y\otimes z]_\CB \ar[r]^{\beta_{y,z}\circ -} & [x\otimes z, z\otimes y]_\CB\, . 
}}
\ee
\enu

\begin{expl}
In the case of canonical gapless edge, i.e. $\CC=\CB,{}^\CB\CX={}^\CB\CB$ and $[y,z]_\CB:=z\otimes y^\ast$, by restricting to the case $y=\one$, the commutative diagram (\ref{diag:half-braiding-0}) implies immediately, $\beta_{x,-} = c_{x,-}$, where $c_{x,-}: x\otimes - \to - \otimes x$ is the braiding of UMTC $\CB$. Importantly, this already means that $(x, c_{-,x}^{-1})$ for $x\in \CBs$ are not allowed to live in the bulk! In other words, by promoting $\hom_\CB(y,z)$ to $[y,z]_\CB$, it chops off the $\overline{\CB}$-factor in $\FZ(\CB) =\CB\boxtimes \overline{\CB}$ entirely. By spelling out the condition (\ref{diag:hb+fusion}) explicitly in this case, we see immediately that $s$ should be symmetric to all $z\in \CB$. Since the braidings in $\CB$ are non-degenerate, it means that $s$ can only be a direct sum of $\one$, or equivalently, $s\in \bh$. In other words, $s$ can be identified with $\hom_{\CB}(\one,[x,y]) \simeq \hom_\CB(x,y)$. Therefore, we have recovered the bulk UMTC $\CB$ as the bulk of the canonical chiral gapless edge $(V,{}^\CB\CB)$. 
\end{expl}

\subsection{Bulk is the center of the edge} \label{sec:drinfeld-center}

In this subsection, we translate the data (\ref{eq:beta}) and conditions (\ref{diag:half-braiding-0}),(\ref{diag:hb+fusion}) into mathematical notions of half-braidings and the center of an enriched monoidal category first introduced in \cite{kz2}. 

\medskip
Let $\CB$ be a braided multi-fusion category, and let $\CXs$ be a $\CB$-enriched multi-fusion category. We denote the underlying category of $\CXs$ by $\CX$. The tensor product in $\CXs$ is an enriched functor $\otimes:\CXs \times \CXs \to \CXs$. As a consequence, for $x\in\CXs$, both $x\otimes -, -\otimes x: \CXs \to \CXs$ are enriched functors. Using this language, the data (\ref{eq:beta}) and condition (\ref{diag:half-braiding-0}) can be translated to the following mathematical definition of a half-braiding for an enriched monoidal category. 
\begin{defn} \label{def:half-braiding}
A {\em half-braiding} for an object $x\in\CXs$ is an enriched natural isomorphism 
$$
\beta_x:x\otimes-\to-\otimes x
$$ 
between two enriched endo-functors of $\CXs$ such that it defines a half-braiding in the underlying monoidal category $\CX$, and the following diagram: 
\be \label{diag:half-braiding}
\raisebox{2em}{\xymatrix{
\hom_\CXs(y,z) \ar[r]^-{1 \otimes \id_x} \ar[d]_{\id_x \otimes 1}  & \hom_\CXs(y,z) \otimes \hom_\CXs(x,x) \ar[r]^-{\otimes} &  \hom_\CXs(z\otimes x, y\otimes x) \ar[d]^{-\circ \beta_{x,y}} \\
\hom_\CXs(x,x) \otimes \hom_\CXs(y,z) \ar[r]^-\otimes & \hom_\CXs(x\otimes y, x\otimes z) \ar[r]^{\beta_{x,z} \circ -} & 
\hom_\CXs(x\otimes y, z\otimes x). 
}}
\ee
is commutative for $y,z\in \CX$. 
\end{defn}

Similarly, using (\ref{diag:hb+fusion}), we obtain the definition of the center of $\CXs$.  
\begin{defn} \label{def:drinfeld-center}
The {\em center} of $\CXs$ is a category $\FZ(\CXs)$ enriched in $\CB$ defined as follows:
\begin{itemize}
  \item an object is a pair $(x,\beta_{x,-})$, where $x\in\CX$ and $\beta_{x,-}$ is a half-braiding for $x$;
  \item $\hom_{\FZ(\CC^\sharp)}((x,\beta_x),(y,\beta_y))$ is the maximal subobject $\iota:s\hookrightarrow\hom_{\CXs}(x,y)$ rendering the following diagram commutative for any $z\in\CX$:
\be \label{diag:hb+fusion-2}
\raisebox{2em}{\xymatrix@!C=22ex{
  s \ar[r]^-{\id_z\otimes\,\iota} \ar[d]_{\iota\,\otimes\id_z} & \hom_\CXs(z,z)\otimes\hom_\CXs(x,y) \ar[r]^-{\otimes} & \hom_\CXs(z\otimes x,z\otimes y) \ar[d]^{-\circ \beta_{x,z}} \\
  \hom_\CXs(x,y)\otimes\hom_\CXs(z,z) \ar[r]^-{\otimes} & \hom_\CXs(x\otimes z,y\otimes z) \ar[r]^-{\beta_{y,z}\circ-} & \hom_\CXs(x\otimes z,z\otimes y); \\
}}
\ee
  \item the identity morphisms and the composition maps $\circ$ are induced from those in $\CXs$. 
\end{itemize}
\end{defn}

\begin{rem}
The center $\FZ(\CXs)$ has an obvious monoidal structure induced from that of $\CXs$ and that of the ordinary Drinfeld center $\FZ(\CX)$. The underlying category of $\FZ(\CXs)$ is a full subcategory of $\FZ(\CX)$ \cite[Proposition\,4.3]{kz2}.
\end{rem}

Let $\CB$ be a UMTC and $\CX$ be a left fusion $\CB$-module. We denote the centralizer of the image of $\overline{\CB}$ by $\phi_\CX(\overline{\CB})'|_{\FZ(\CX)}$, which is an $\bh$-enriched category but can also be viewed as a $\CB$-enriched monoidal category by identifying an object $a\in\bh$ with the object $a\otimes \one_\CB$ in $\CB$. 
\begin{thm}[\cite{kz2}] \label{thm:KZ2}
We have $\FZ({}^\CB\CX) \simeq \phi_\CX(\overline{\CB})'|_{\FZ(\CX)}$ as $\CB$-enriched braided monoidal categories. 
\end{thm}

As a consequence of above mathematical theorem, given a chiral gapless edge $(V,{}^\CB\CX)$ of a 2d topological order $(\CC,c)$, we have the following boundary-bulk relation: 
\be \label{eq:bulk=center}
\FZ({}^\CB\CX) \simeq \overline{\CB}'|_{\FZ(\CX)} \simeq \CC. 
\ee
In other words, the UFC $\CX$ describes a gapped wall between two 2d topological orders $(\CB,c)$ and $(\CC,c)$. 
It also means that all chiral gapless edges are obtained from a topological Wick rotation as illustrated by the following pictures:
\be \label{pic:fusing-edge-wall}
\raisebox{-30pt}{
  \begin{picture}(120,75)
   \put(0,15){\scalebox{0.5}{\includegraphics{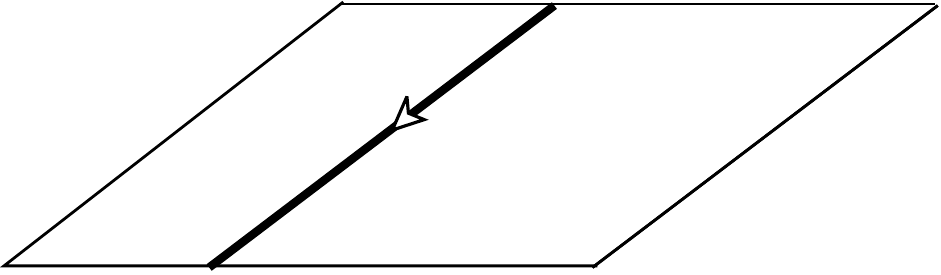}}}
   \put(0,15){
     \setlength{\unitlength}{.75pt}\put(0,0){
     \put(140,40)  {\scriptsize $ (\CC,c) $}
     \put(15,7) {\scriptsize $(\CB,c)$}
     \put(85,27)   {\scriptsize $\CX$}
     }\setlength{\unitlength}{1pt}}
  \end{picture}} 
\quad \xrightarrow{\mbox{\footnotesize topological wick rotation}} \quad 
\raisebox{-30pt}{
  \begin{picture}(100,75)
   \put(0,10){\scalebox{0.5}{\includegraphics{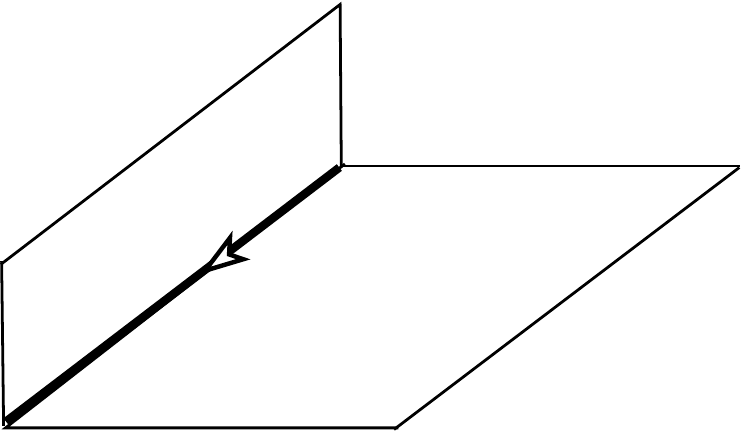}}}
   \put(0,10){
     \setlength{\unitlength}{.75pt}\put(0,0){
     \put(100,40)  {\scriptsize $ (\CC,c) $}
     \put(25,43) {\scriptsize $(\CB,c)$}
     \put(70, 70)  {\scriptsize $(V,{}^\CB\CX)$}
     \put(30,15)   {\scriptsize $\CX$}
     }\setlength{\unitlength}{1pt}}
  \end{picture}}
\ee
which was explained in details in \cite[Section\,5.2]{kz4}. 

\begin{rem}
The isomorphisms in (\ref{eq:bulk=center}) simply prove a special case of a general ``holographic principle'' or boundary-bulk relation: the bulk is the center of the boundary for topological orders in all dimensions regardless the boundary is gapped or gapless \cite{kong-wen-zheng-2}. Note that the boundary-bulk relation (\ref{eq:bulk=center}) automatically includes gapped edges (i.e. $V=\Cb,\CB=\bh$) as special cases. It turns out that it also holds for all non-chiral gapless edge as we will see later. 
\end{rem}

\begin{thm}[\cite{zheng}] \label{thm:morita}
Two indecomposable enriched unitary multi-fusion categories ${}^\CA\CX$ and ${}^\CB\CY$ are spatially Morita equivalent if and only if $\FZ({}^\CA\CX)\simeq \FZ({}^\CB\CY)$. 
\end{thm}

The physical meaning of above mathematical theorem can be reformulated as the following physical theorem. 
\begin{pthm}
Two 1+1D gapped or chiral gapless edges share the same bulk if and only if the associated enriched unitary multi-fusion categories are spatially Morita equivalent. 
\end{pthm}

\subsection{Boundary-bulk relation for gapless edges}  \label{sec:Z1}

\begin{figure} 
$$
 \raisebox{-50pt}{
  \begin{picture}(100,120)
   \put(-130,10){\scalebox{0.8}{\includegraphics{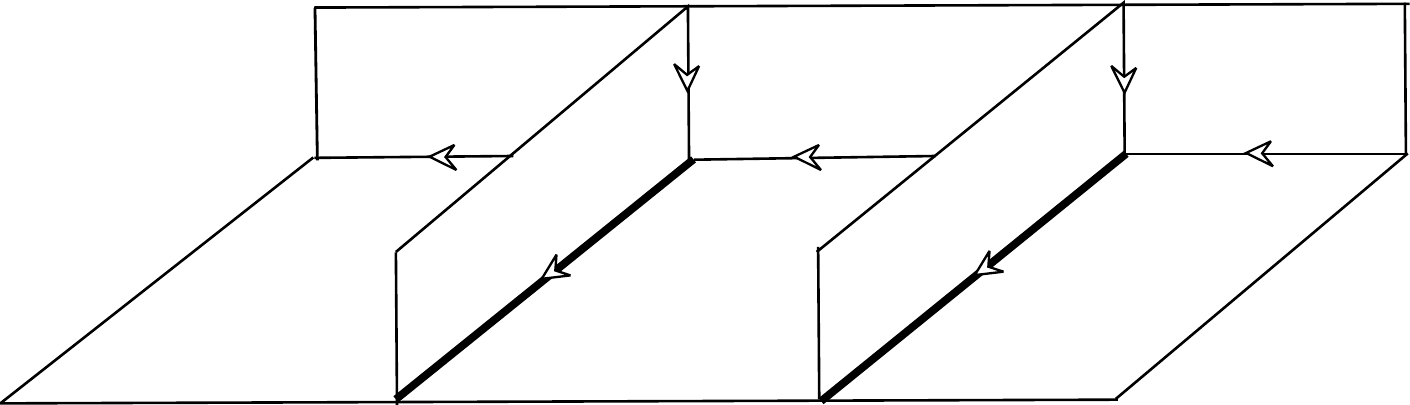}}}
   \put(-123,5){
     \setlength{\unitlength}{.75pt}\put(0,0){
     \put(5,10)  {\footnotesize $ (\CC_3,c_1+c_2+c_3) $}
     \put(155,10) {\footnotesize $(\CC_2,c_1+c_2)$}
     \put(305,10) {\footnotesize $(\CC_1,c_1)$}
     \put(127, 60) {\footnotesize $(V_5,{}^{\CB_5}\CX_5)$}
     \put(160, 82)  {\footnotesize $(\CB_5,c_3)$}
     \put(163,36)   {\footnotesize $\CX_5$}
     \put(296,36)   {\footnotesize $\CX_4$}
     \put(260, 60) {\footnotesize $(V_4,{}^{\CB_4}\CX_4)$}
     \put(295, 82)  {\footnotesize $(\CB_4,c_2)$}
     \put(400,73)  {\footnotesize $\CX_1$}
     \put(390,110)  {\footnotesize $(\CB_1,c_1)$}
     \put(390,137)  {\footnotesize $(V_1,{}^{\CB_1}\CX_1)$}
     \put(245,70)  {\footnotesize $\CX_2$}
     \put(240,110) {\footnotesize $(\CB_2,c_1+c_2)$}
     \put(245,137){\footnotesize $(V_2,{}^{\CB_2}\CX_2)$}
     \put(100,73)  {\footnotesize $\CX_3$}
     \put(90,110) {\footnotesize $(\CB_3,c_1+c_2+c_3)$}
     \put(90,137) {\footnotesize $(V_3,{}^{\CB_3}\CX_3)$}
     \put(207,72) {\footnotesize $\CT_0$}
     \put(209,105)  {\footnotesize $\CT$}
     \put(341,73) {\footnotesize $\CS_0$}
     \put(344,105)  {\footnotesize $\CS$}   
     \put(190,135)  {\footnotesize ${}^\CT\CT_0$}  
     \put(330,135)  {\footnotesize ${}^\CS\CS_0$} 
     }\setlength{\unitlength}{1pt}}
  \end{picture}}
$$
\caption{This picture depicts three 2d topological orders $(\CC_i,\sum_{k=1}^i c_k)$ for $i=1,2,3$, three 1d gapless edges and two gapless walls $(V_j,{}^{\CB_j}\CX_j)$ for $j=1,2,3,4,5$ and two 0d gapless defects $(V_{142},{}^{\CS}\CS_0)$ and $(V_{253}, {}^{\CT}\CT_0)$. 
}
\label{fig:bb-duality-gapless}
\end{figure}

In this subsection, we consider more general 0d defects up to spatial equivalences. Consider the physical configuration depicted in Figure\,\ref{fig:bb-duality-gapless}. There are three 2d topological orders $(\CC_1,c_2), (\CC_2,c_1+c_2), (\CC_3,c_1+c_2+c_3)$, which have chiral gapless edges $(V_i, {}^{\CB_i}\CX_i)$ for $i=1,2,3$, respectively, and are separated by two chiral gapless walls $(V_i,{}^{\CB_i}\CX_i)$ for $i=4,5$. Moreover, the normal directions of the two vertical rectangles labeled by $(\CB_4,c_2)$ and $(\CB_5,c_3)$ are pointing toward right. By our convention, we have unitary braided monoidal equivalences: 
$$
\overline{\CC_2} \boxtimes \overline{\CB_4} \boxtimes \CC_1 \simeq \FZ(\CX_4), 
\quad\quad
\overline{\CC_3} \boxtimes \overline{\CB_5} \boxtimes \CC_2 \simeq \FZ(\CX_5).
$$
There are two 0+1D gapless defects junctions given by ${}^{\CS}\CS_0$ and ${}^{\CT}\CT_0$. We have ignored the 1+1D and 0+1D chiral symmetries because we only care about the spatial equivalence classes here. These 0+1D defects are uniquely determined by their neighborhoods. We explain this fact in details below. 
\bnu
\item $\CS$ is a closed multi-fusion $(\CB_2\boxtimes\CB_4)$-$\CB_1$-bimodule, and $\CT$ is a closed multi-fusion $(\CB_3\boxtimes\CB_5)$-$\CB_2$-bimodule, i.e. UMFC's equipped with unitary braided monoidal equivalences: 
\be
\phi_\CS: \overline{\CB_2} \boxtimes \overline{\CB_4} \boxtimes \CB_1 \xrightarrow{\simeq} \FZ(\CS), \quad\quad
\phi_\CT: \overline{\CB_3} \boxtimes \overline{\CB_5} \boxtimes \CB_2 \xrightarrow{\simeq} \FZ(\CT). \label{eq:negative-orientation}
\ee

\item $\CS_0$ is a closed left
$\CX_2^\rev \boxtimes_{\CB_2\boxtimes \overline{\CC_2}} (\CX_4^\rev \boxtimes_{\CB_4} \CS) \boxtimes_{\overline{\CC_1}\boxtimes \CB_1} \CX_1$-module;  \\
$\CT_0$ is a closed left
$\CX_3^\rev \boxtimes_{\CB_3\boxtimes \overline{\CC_3}} (\CX_5^\rev \boxtimes_{\CB_5} \CT) \boxtimes_{\overline{\CC_2}\boxtimes \CB_2} \CX_2$-module. In particular, we have the following unitary monoidal equivalences:  
\begin{align*}
&\phi_{\CS_0}: \CX_2^\rev \boxtimes_{\CB_2\boxtimes \overline{\CC_2}} (\CX_4^\rev \boxtimes_{\CB_4} \CS) \boxtimes_{\overline{\CC_1}\boxtimes \CB_1} \CX_1 \xrightarrow{\simeq} \fun_\bh(\CS_0,\CS_0),  \\
&\phi_{\CT_0}: \CX_3^\rev \boxtimes_{\CB_3\boxtimes \overline{\CC_3}} (\CX_5^\rev \boxtimes_{\CB_5} \CT) \boxtimes_{\overline{\CC_2}\boxtimes \CB_2} \CX_2 \xrightarrow{\simeq} \fun_\bh(\CT_0,\CT_0).
\end{align*}

\item Then enriched categories ${}^{\CS}\CS_0$ and ${}^{\CT}\CT_0$ are determined by the left $\CS$-module structure on $\CS_0$ and the left $\CT$-module structure on $\CT_0$, respectively, as follows: 
\begin{align*}
&\CS \to \CX_2^\rev \boxtimes_{\CB_2\boxtimes \overline{\CC_2}} (\CX_4^\rev \boxtimes_{\CB_4} \CS) \boxtimes_{\overline{\CC_1}\boxtimes \CB_1} \CX_1 \xrightarrow[\simeq]{\phi_{\CS_0}} \fun_\bh(\CS_0,\CS_0),  \\
&\CT \to \CX_3^\rev \boxtimes_{\CB_3\boxtimes \overline{\CC_3}} (\CX_5^\rev \boxtimes_{\CB_5} \CT) \boxtimes_{\overline{\CC_2}\boxtimes \CB_2} \CX_2 \xrightarrow[\simeq]{\phi_{\CT_0}} \fun_\bh(\CT_0,\CT_0).
\end{align*}
\enu

\begin{rem}
There are many different ways to see that $\CS_0$ and $\CT_0$ are uniquely determined by its neighborhood. More precisely, by Theorem\,\ref{thm:KZ}, we have the following different but equivalent ways of characterize $\CS_0$ and $\CT_0$ uniquely (up to equivalences)
\begin{itemize}
\item by the following unitary monoidal equivalences, respectively, 
$$
\CX_4^\rev \boxtimes_{\CB_4} \CS \xrightarrow{\simeq} \fun_{\CX_1|\CX_2}(\CS_0,\CS_0), 
\quad\quad 
\CX_5^\rev \boxtimes_{\CB_5} \CT \xrightarrow{\simeq} \fun_{\CX_2|\CX_3}(\CT_0,\CT_0); 
$$
\item by the following unitary monoidal equivalences, respectively, 
$$
\CX_4^\rev \boxtimes\CB_1 \xrightarrow{\simeq} \fun_{\CX_1|\CS^\rev\boxtimes_{\CB_2}\CX_2}(\CS_0,\CS_0), 
\quad\quad 
\CX_5^\rev \boxtimes \CB_2 \xrightarrow{\simeq} \fun_{\CX_2|\CT^\rev\boxtimes_{\CB_3}\CX_3}(\CT_0,\CT_0); 
$$
\item by the following unitary monoidal equivalences, respectively, 
\be \label{eq:cx-45}
\CX_4^\rev \xrightarrow{\simeq} \fun_{\CX_2^\rev\boxtimes_{\CB_2}\CS\boxtimes_{\CB_1}\CX_1}(\CS_0,\CS_0), 
\quad\quad 
\CX_5^\rev \xrightarrow{\simeq} \fun_{\CX_3^\rev\boxtimes_{\CB_3}\CT\boxtimes_{\CB_2}\CX_2}(\CT_0,\CT_0). 
\ee
\end{itemize}
\end{rem}

\begin{rem} \label{rem:two-functors-coincide}
The unitary braided monoidal functor $\phi_{\CX_4}: \overline{\CB_4} \to \FZ(\CX_4)$ that defines the enriched multi-fusion category ${}^{\CB_4}\CX_4$ is isomorphic to the following functor: 
$$
\overline{\CB_4} \hookrightarrow \overline{\CC_2} \boxtimes \overline{\CB_4} \boxtimes \CC_1 \xrightarrow{\simeq} 
\overline{\FZ(\CX_2^\rev\boxtimes_{\CB_2}\CS\boxtimes_{\CB_1}\CX_1)} \xrightarrow{\simeq} \FZ(\CX_4), 
$$
where the first ``$\simeq$'' was explained in the proof of \cite[Theorem\,3.3.6.]{kz1} and the second ``$\simeq$'' is determined by the invertible $(\CX_2^\rev\boxtimes_{\CB_2}\CS\boxtimes_{\CB_1}\CX_1)$-$\CX_4$-bimodule $\CS_0$ (recall (\ref{eq:cx-45})) \cite{eno2009}. 
\end{rem}

Conversely, one can also viewed the 1d gapless wall $(V_4,{}^{\CB_4}\CX_4)$ as the 1d  ``relative bulk'' of the 0+1D wall ${}^\CS\CS_0$ on the edge. In this setting, we obtain a generalization the unique-bulk principle. More precisely, by assuming all the data on the edge (not the data in the bulk), we will show that the 1d ``relative bulk'' $(V_4,{}^{\CB_4}\CX_4)$ is uniquely determined by 0d wall ${}^\CS\CS_0$ on the edge.

\begin{defn}
The $\FZ^{(1)}$-center of the ${}^{\CB_1}\CX_1$-${}^{\CB_1}\CX_1$-bimodule ${}^\CS\CS_0$, denoted by $\FZ^{(1)}({}^\CS\CS_0)$, 
is an enriched unitary multi-fusion category 
$$
\FZ^{(1)}({}^\CS\CS_0):= {}^{\FZ^{(1)}_2({}^\CS\CS_0)} \FZ^{(1)}_1({}^\CS\CS_0),
$$
which is defined by a triple $(\FZ^{(1)}_2({}^\CS\CS_0), \FZ^{(1)}_1({}^\CS\CS_0), F)$ via the canonical construction, and
\bnu

\item $\FZ^{(1)}_2({}^\CS\CS_0)$ is the UMTC defined by 
\be \label{eq:Z1-background-cat}
\FZ^{(1)}_2({}^\CS\CS_0):=(\CB_2\boxtimes\overline{\CB_1})'|_{\overline{\FZ(\CS)}};
\ee

\item $\FZ^{(1)}_1({}^\CS\CS_0)$ is a UFC defined by 
\be \label{eq:Z1-underlying-cat}
\FZ^{(1)}_1({}^\CS\CS_0) := \fun_{\CX_2^\rev\boxtimes_{\CB_2}\CS\boxtimes_{\CB_1}\CX_1}(\CS_0,\CS_0)^\rev; 
\ee

\item $F: \overline{\FZ^{(1)}_2({}^\CS\CS_0)} \to \FZ(\FZ^{(1)}_1({}^\CS\CS_0))$ is a unitary braided monoidal functor defined by
\begin{align}
&\overline{\FZ^{(1)}_2({}^\CS\CS_0)} \hookrightarrow \overline{\CB_2} \boxtimes \overline{\FZ^{(1)}_2({}^\CS\CS_0)} \boxtimes  \CB_1 \xrightarrow{\simeq} \FZ(\CX_2^\rev\boxtimes_{\CB_2}\CS\boxtimes_{\CB_1}\CX_1) \xrightarrow{\simeq} \FZ(\FZ^{(1)}_1({}^\CS\CS_0)), \nonumber
\end{align}
where 
the second ``$\simeq$'' is determined by the  invertible $(\CX_2^\rev\boxtimes_{\CB_2}\CS\boxtimes_{\CB_1}\CX_1)$-$\FZ^{(1)}_0({}^\CS\CS_0)^\rev$-bimodule $\CS_0$ (recall (\ref{eq:Z1-underlying-cat})) \cite{eno2009}. 
\enu
\end{defn}

Then we can see that ${}^{\CB_4}\CX_4$ can be determined by ${}^\CS\CS_0$ as the $\FZ^{(1)}$-center $\FZ^{(1)}({}^\CS\CS_0)$. Is the chiral symmetry $V_4$ also determined by ${}^\CS\CS_0$? Yes, indeed. Recall Remark\,\ref{rem:add-information-to-P}, we have $\CS = ((\Mod_{V_{1|24}})_{V_1|V_2\otimes_\Cb V_4})_{X|X}$. The relation between the 1+1D chiral symmetry $V_{1|24}$ and the 0+1D chiral symmetry $X$ is given in Diagram (\ref{diag:relation-V-X}). The VOA $V_4$ can be recovered as the commutant of $V_2$ in $V_2\otimes_\Cb V_4$. For convenience, we can also denote $(V_4,{}^{\CB_4}\CX_4)$ by $\FZ^{(1)}({}^\CS\CS_0)$ and refer to it as the $\FZ^{(1)}$-center of ${}^\CS\CS_0$.

\begin{figure}[tb]
 \begin{picture}(150, 100)
   \put(100,10){\scalebox{2}{\includegraphics{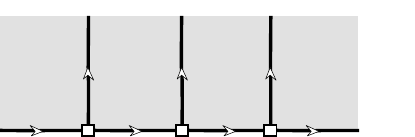}}}
   \put(60,-55){
     \setlength{\unitlength}{.75pt}\put(-18,-19){
     \put(95, 97)       {\scriptsize ${}^{\CB_1}\CX_1$}
     \put(170, 97)       {\scriptsize ${}^{\CB_2}\CX_2$}
     \put(242, 97)       {\scriptsize ${}^{\CB_3}\CX_3$}
     \put(310, 97)       {\scriptsize ${}^{\CB_4}\CX_4$}
     \put(133,95)      {\scriptsize ${}^{\CS}\CS_0$}
     \put(208,95)      {\scriptsize ${}^{\CT}\CT_0$}
     \put(278,95)      {\scriptsize ${}^{\CU}\CU_0$}
     \put(85, 160)    {\scriptsize $\Z({}^{\CB_1}\CX_1)$}
     \put(165, 160)    {\scriptsize $\Z({}^{\CB_2}\CX_2)$}
     \put(240, 160)    {\scriptsize $\Z({}^{\CB_3}\CX_3)$}
     \put(310, 160)    {\scriptsize $\Z({}^{\CB_4}\CX_4)$}
     \put(130,207)     {\scriptsize $\Z^{(1)}({}^{\CS}\CS_0)$}
     \put(200,207)     {\scriptsize $\Z^{(1)}({}^{\CT}\CT_0)$}
     \put(275,207)     {\scriptsize $\Z^{(1)}({}^{\CU}\CU_0)$}
     }\setlength{\unitlength}{1pt}}
  \end{picture}
\caption{The picture depicts the complete boundary-bulk relation, which can be summarized mathematically as fully faithful functor. The arrows indicate the orientation of the edges or walls and the order of tensor product of topological excitations on the edges or walls. 
}
\label{fig:bbr-gapless}
\end{figure}

\medskip
As a consequence, we obtain a generalization of unique-bulk principle for 2d topological orders with gapped and chiral gapless edges as illustrated in Figure\,\ref{fig:bbr-gapless}. Again this relation can be stated as the functoriality of the center. We will make it precisely in Section\,\ref{sec:main-thm}.

\subsection{Center functor is an monoidal equivalence} \label{sec:main-thm}

In this subsection, we obtain a mathematical theorem inspired from the boundary-bulk relation of 2d topological orders with gapped and chiral gapless edges. 

\begin{defn} \label{def:en-monoidal-bimodule}
Let $\CC$ and $\CD$ be UMTC's. A closed enriched multi-fusion $\CD$-$\CC$ bimodule is an indecomposable enriched unitary multi-fusion category ${}^\CB\CX$, together with a unitary braided monoidal equivalence $\phi: \overline{\CD} \boxtimes \overline{\CB} \boxtimes \CC \xrightarrow{\simeq} \FZ(\CX)$, such that ${}^\CB\CX$ is obtained from the canonical construction with the left multi-fusion $\CB$-module structure on $\CX$ defined by $\overline{\CB} \hookrightarrow \overline{\CD} \boxtimes \overline{\CB} \boxtimes \CC \xrightarrow{\simeq} \FZ(\CX)$.
\end{defn}

\begin{defn}
Two such closed enriched multi-fusion $\CD$-$\CC$-bimodules ${}^\CB\CX$ and ${}^{\CB'}\CX'$ are called equivalent if there are a braided monoidal equivalence $f: \CB \to \CB'$ and a monoidal equivalence $g: \CX \to \CX'$ such that the following diagram
$$
\xymatrix{
\overline{\CD} \boxtimes \overline{\CB} \boxtimes \CC \ar[r]^-\phi \ar[d]_{1f1} & \FZ(\CX) \ar[r]^\forget & \CX \ar[d]^g \\
\overline{\CD} \boxtimes \overline{\CB'} \boxtimes \CC \ar[r]^-{\phi'} & \FZ(\CX') \ar[r]^\forget & \CX'
}
$$
is commutative and $g$ defines a multi-fusion $(\CD\boxtimes\CB)$-$\CC$-bimodule equivalence between $\CX$ and $\CX'$ (recall Definition\,\ref{def:monoidal-module-map}).  
\end{defn}

\begin{lem} \label{lem:monoidal-relative-tensor-product}
Let $\CC,\CD,\CE$ be UMTC's. Let ${}^\CA\CX$ and ${}^\CB\CY$ be a closed enriched multi-fusion $\CE$-$\CD$-bimodule and a closed enriched multi-fusion $\CD$-$\CC$-bimodule, respectively. The following relative tensor product 
$$
{}^\CA\CX \boxtimes_{\CD} {}^\CB\CY := {}^{\CA\boxtimes\CB} (\CX\boxtimes_\CD\CY)
$$
is well-defined and is a closed enriched multi-fusion $\CE$-$\CC$-bimodule by Theorem\,\ref{thm:KZ}. Its physical meaning is illustrated in Figure\,\ref{fig:fusion-walls}. 
\end{lem}

\medskip
We introduce two categories $\iemfc$ and $\nbfc$ as follows (recall Remark\,\ref{rem:def-bimodule}): 
\begin{itemize}
\item $\iemfc$: Objects are indecomposable enriched unitary multi-fusion categories ${}^\CA\CX$; morphisms in $\hom_\iemfc({}^\CA\CX,{}^\CB\CY)$ are the spatial equivalence classes of ${}^\CB\CY$-${}^\CA\CX$-bimodules (recall Definition\,\ref{def:module-enriched-monoidal-cat} and \ref{def:spatial-functor}), the background category of which, as multi-fusion categories, are indecomposable; the identity morphism in $\hom_\iemfc({}^\CA\CX,{}^\CA\CX)$ is the trivial bimodule ${}^\CA\CX$; the composition map is defined by the relative tensor product of bimodules (recall Definition\,\ref{def:relative-tensor-product}). 

\item $\CU\CM\CT^{\mathrm{en-cl}}$: Objects are UMTC's $\CC,\CD,\cdots$; morphisms in $\hom_{\CU\CM\CT^{\mathrm{en-cl}}}(\CC,\CD)$ are the equivalence classes of closed enriched multi-fusion $\CD$-$\CC$-bimodules (recall Definition\,\ref{def:en-monoidal-bimodule}); the identity morphism from $\CC$ to $\CC$ is given by $\CC$. The composition map is defined by the relative tensor product of bimodules (see Lemma\,\ref{lem:monoidal-relative-tensor-product}). 

\end{itemize}
Both categories are symmetric monoidal with the tensor product defined by the Deligne tensor product $\boxtimes$ (recall Definition\,\ref{def:relative-tensor-product}).

\medskip
The boundary-bulk relation of 2d topological orders with gapped and chiral gapless edges can be stated as the following mathematical theorem. 
\begin{thm} \label{thm:ff-functor}
The functor $\FZ: \iemfc \to \CU\CM\CT^{\mathrm{en-cl}}$, which is defined by 
$$
{}^\CA\CX \mapsto \FZ({}^\CA\CX) \quad\quad \mbox{and} \quad\quad
\hom_\iemfc({}^\CA\CX,{}^\CB\CY) \ni {}^\CP\CM \mapsto \FZ^{(1)}({}^\CP\CM), 
$$
is a well-defined symmetric monoidal equivalence. 
\end{thm}
\pf
The essential surjectivity follows from $\FZ({}^\CC\CC)\simeq\CC$ for any UMTC $\CC$ \cite[Corollary\ 4.9]{kz2}. The fully faithfulness follows from that of Drinfeld center Theorem\,\ref{thm:KZ} and the definition of a spatial equivalence. The symmetric monoidalness is obvious. 
\epf

\begin{rem}
We conjecture that the complete boundary-bulk relation for $n$d topological orders with gapped/gapless boundaries and higher codimensional gapped/gapless defects on the boundary can also be stated as a symmetric monoidal equivalence of higher monoidal categories. This generalizes a conjecture proposed in \cite{kong-wen-zheng-1} for $n$d topological orders with only gapped boundaries and gapped higher codimensional defects on the boundary. 
\end{rem}

\section{Non-chiral gapless edges}  \label{sec:class-gapless-edge}

In this section, we develop the mathematical theory of non-chiral gapless edges. 


\subsection{A construction of non-chiral gapless edge}  \label{sec:non-chiral-walls}

In this subsection, we construction a non-chiral gapless edge from chiral gapless edges.

\begin{figure} 
$$
 \raisebox{-50pt}{
  \begin{picture}(170,150)
   \put(-50,0){\scalebox{0.65}{\includegraphics{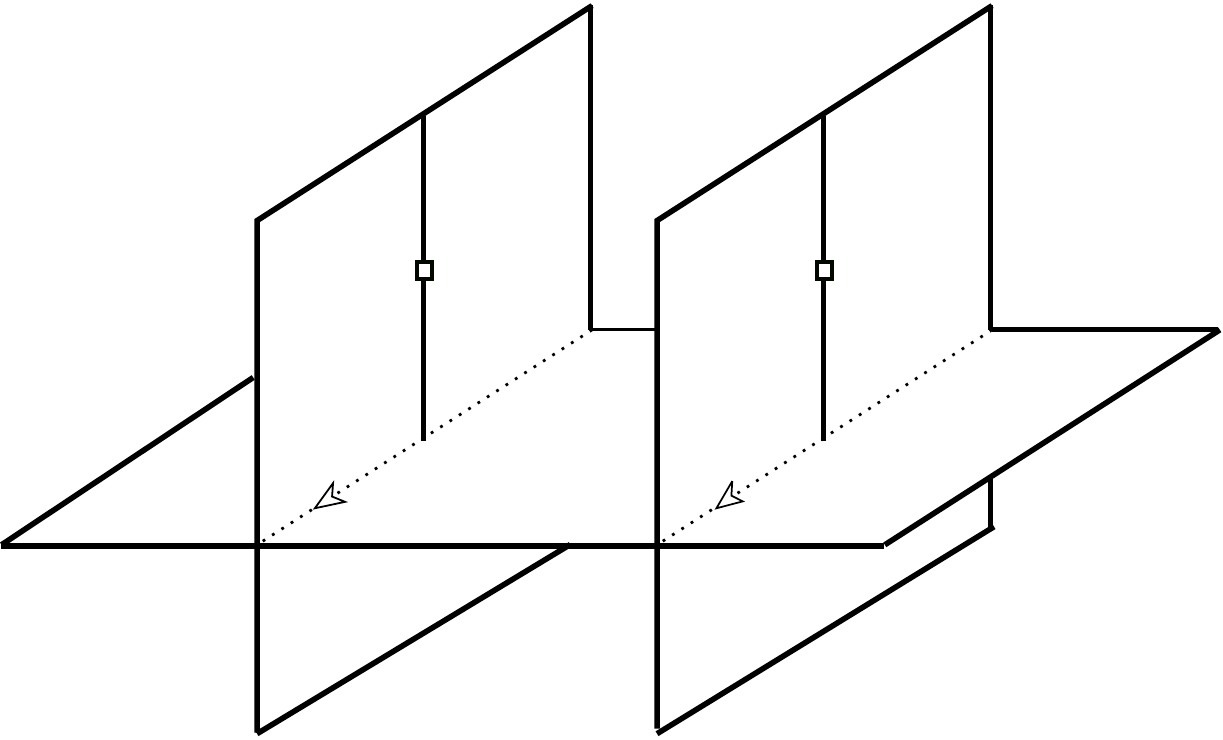}}}
   \put(-50,0){
     \setlength{\unitlength}{.75pt}\put(0,0){
   
     \put(253,165)  {\scriptsize $(V_\CB,{}^\CB\CY)$}
     \put(153,165)    {\scriptsize $(V_\CA,{}^\CA\CX)$}
     
     \put(125,90) {\scriptsize $\CX$}
     \put(115,130) {\scriptsize $(\CA,c_2)$}
     \put(227,90) {\scriptsize $\CY$}
     \put(215,130) {\scriptsize $(\CB,c_3)$}
     
     \put(110,72)  {\scriptsize $x\in \CX$}
     \put(210,72)  {\scriptsize $y\in \CY$}
     \put(110,152) {\scriptsize $x'\in \CX$}
     \put(210,152) {\scriptsize $y' \in \CY$}
     
     \put(26,55)  {\scriptsize $(\CC,c_1)$}
     \put(100,55) {\scriptsize $(\CD,c_1+c_2)$}
     \put(270,108) {\scriptsize $(\CE,c_1+c_2+c_3)$}
     \put(73,115)   {\scriptsize $[x,x']_\CA$}
     \put(175,115)   {\scriptsize $[y,y']_\CB$}
     
     }\setlength{\unitlength}{1pt}}
  \end{picture}}
 \quad\quad
   \raisebox{-50pt}{
  \begin{picture}(100,150)
   \put(0,0){\scalebox{0.65}{\includegraphics{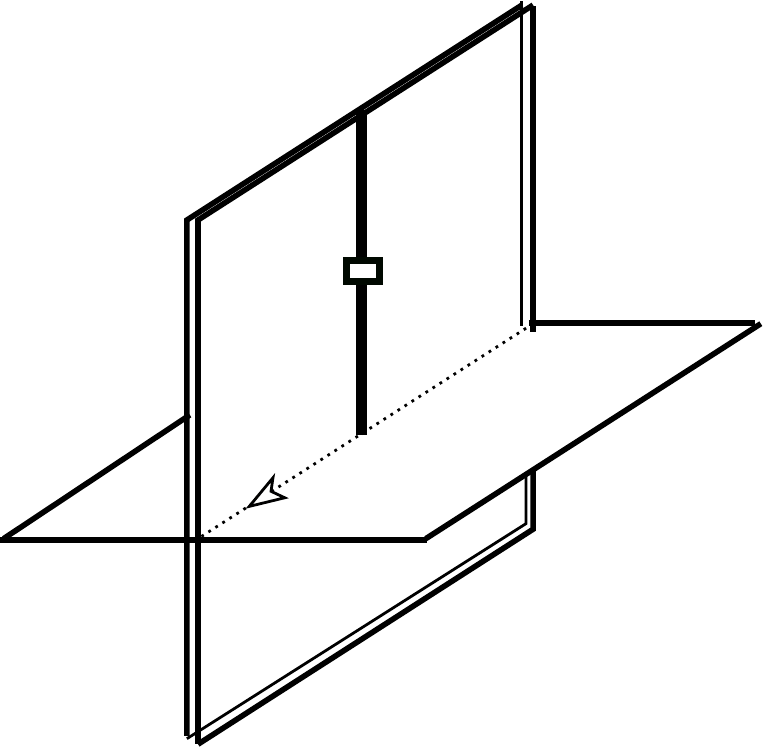}}}
   \put(0,0){
     \setlength{\unitlength}{.75pt}\put(0,0){
    
    \put(20,58)  {\scriptsize $(\CC,c_1)$}
     \put(150,113) {\scriptsize $(\CE,c_1+c_2+c_3)$}
     
     \put(100,95) {\scriptsize $\CX\boxtimes_\CD\CY$}
     \put(100,130) {\scriptsize $\CA\boxtimes\CB$}
     
     \put(95,75) {\scriptsize $xy$}
     \put(95,153) {\scriptsize $x'y'$}
     
     \put(51,110)   {\scriptsize $[xy,x'y']$}
     \put(-5,177) {\scriptsize $(V_\CA\otimes_\Cb V_\CB, {}^{\CA\boxtimes\CB}(\CM\boxtimes_\CD\CN)$}
     
     }\setlength{\unitlength}{1pt}}
  \end{picture}}  
$$
$$
(a) \quad\quad\quad\quad\quad\quad\quad\quad\quad\quad\quad\quad\quad\quad\quad\quad \quad\quad\quad\quad
(b)
$$
\caption{The picture (a) depicts two chiral gapless walls $(V_\CA, {}^\CA\CM)$ and $(V_\CB, {}^\CB\CN)$. The vertical direction is the direction of time. The picture (b) depicts the new 1d wall obtained after the fusion, where $xp:=x\boxtimes_\CD p, yp:=y\boxtimes_\CD q \in \CM\boxtimes_\CD\CN$. The arrows on the dotted lines are the orientation of the wall. It determines the order of the fusion product of wall excitations. 
}
\label{fig:fusion-walls}
\end{figure}

\medskip
First, we recall a useful fusion formula of 1d chiral gapless walls in \cite{kz4}. 
We illustrate two 1d chiral gapless walls before the fusion in Figure\,\ref{fig:fusion-walls} (a) and after the fusion in Figure\,\ref{fig:fusion-walls} (b). More precisely, $\CA,\CB,\CC,\CD,\CE$ are UMTC's, and $\CX$ is a closed fusion $(\CC\boxtimes\CA)$-$\CD$-bimodule, and $\CY$ is a closed fusion $(\CD\boxtimes\CB)$-$\CE$-bimodule\footnote{Our convention is that the fictional bulk phase $\CA$ (or $\CB$) sits on the left side of the oriented wall (recall Remark\,\ref{rem:left-right-convention}).}. The vertical direction is the direction of time. Two vertical planes depict the 1+1D world sheets (or fictional bulk phases) of two chiral gapless walls $(V_\CA, {}^\CA\CX)$ and $(V_\CB, {}^\CB\CY)$. Two VOA's $V_\CA$ and $V_\CB$ have central charge $c_2$ and $c_3$, respectively. The spatial fusion of these two walls can be computed by the following formula:
\be \label{eq:fusing-walls-1}
(V_\CA, {}^\CA\CX) \boxtimes_{(\CD, c_1+c_2)} (V_\CB, {}^\CB\CY) 
= (V_\CA\otimes_\Cb V_\CB, {}^{\CA\boxtimes \CB} (\CX\boxtimes_\CD\CY)).  
\ee
which was explained in details in Section\,6.3 in \cite{kz4}.

\begin{figure}
$$
 \raisebox{-50pt}{
  \begin{picture}(105,140)
   \put(-50,0){\scalebox{0.65}{\includegraphics{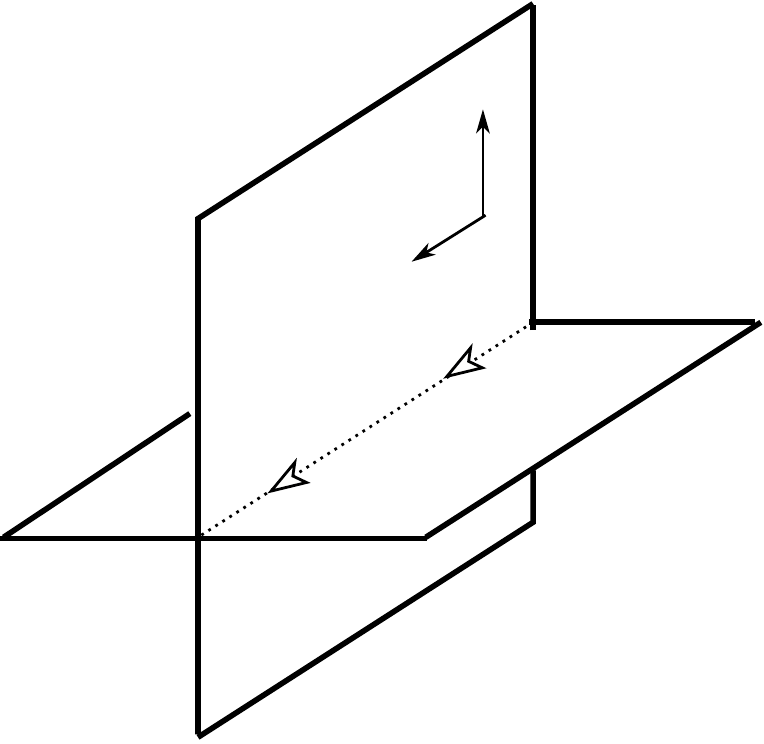}}}
   \put(-50,0){
     \setlength{\unitlength}{.75pt}\put(0,0){
   
     \put(53,165)    {\scriptsize $(V,{}^\CA\CX)$}
     
     \put(90,78) {\scriptsize $\CX$}
     \put(60,120) {\scriptsize $(\CA,c_2)$}
     
     \put(105,115) {\scriptsize $x$}
     \put(113,152) {\scriptsize $t$}
          
     \put(21,56)  {\scriptsize $(\CC,c_1)$}
     \put(160,110) {\scriptsize $(\CD,c_1+c_2)$}
     
     \put(155,145) {\scriptsize $z=t+ix$}
     
     }\setlength{\unitlength}{1pt}}
  \end{picture}}
 \quad \Longleftrightarrow \quad\quad
   \raisebox{-50pt}{
  \begin{picture}(100,140)
   \put(0,0){\scalebox{0.65}{\includegraphics{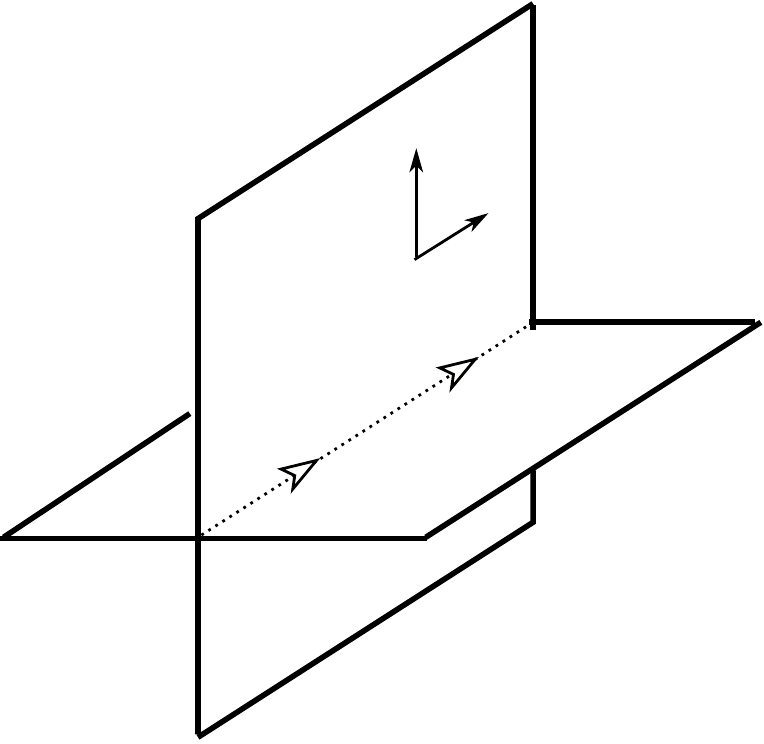}}}
   \put(0,0){
     \setlength{\unitlength}{.75pt}\put(0,0){
    
    \put(46,165)    {\scriptsize $(\overline{V},{}^{\overline{\CA}}\CX^\rev)$}
     
     \put(89,78) {\scriptsize $\CX^\rev$}
     \put(60,120) {\scriptsize $(\overline{\CA},-c_2)$}
     
     \put(119,122) {\scriptsize $x$}
     \put(108,146) {\scriptsize $t$}
          
     \put(21,56)  {\scriptsize $(\CC,c_1)$}
     \put(160,110) {\scriptsize $(\CD,c_1+c_2)$}     
     
     \put(155,145) {\scriptsize $\bar{z}=t-ix$}
     
     }\setlength{\unitlength}{1pt}}
  \end{picture}}  
$$
$$
(a) \quad\quad\quad\quad\quad\quad\quad \quad\quad\quad\quad\quad\quad\quad\quad\quad \quad\quad \quad
(b)
$$
\caption{These two picture depict two physically equivalent 1d chiral gapless walls, which are equipped with the opposite orientations. 
}
\label{fig:flip-orientation}
\end{figure}

\medskip
Secondly, notice that flipping orientation is associated to changing chirality. Figure\,\ref{fig:flip-orientation} (a) depicts a chiral gapless wall $(V,{}^\CA\CX)$ with a chosen orientation, which is indicated by the complex coordinate $z=t+ix$ on the 1+1D world sheet, or equivalently, by the orientation of the spatial dimension (i.e. $x$-axis or the arrows on the dotted line) because the orientation of time is fixed, or equivalently, by the normal direction of the world sheet (pointing towards right in this case). The underlying category $\CX$ of ${}^\CA\CX$ is the category of topological wall excitations, and the order of the fusion product in $\CX$ is determined by the orientation of the wall. The chiral central charge of $V_\CA$ is $c_2$, and that of $V_\CB$ is $c_3$. 

Without altering the physics, we can flip the orientation of this wall (i.e. flipping the direction of $x$-axis) and, at the same time, change all the data according to Figure\,\ref{fig:flip-orientation} (b). As a consequence, a point at $z$ in the old coordinate becomes $\bar{z}$ in the new coordinate; a chiral field $\psi(z)$ in $V$ becomes an anti-chiral field $\psi(\bar{z})$ in $\overline{V}$; The chiral central charge $c_2$ of $V$ becomes the anti-chiral central charge $c_2$ of $\overline{V}$, or equivalently, the chiral central charge $-c_2$ of $\overline{V}$; $\CX$ becomes $\CX^\rev$. In summary, we will say that a gapless wall defined by $(V,{}^\CA\CX)$ with a given orientation is entirely same as the one defined by $(\overline{V},{}^{\overline{\CA}}\CX^\rev)$ but with the opposite orientation.

\medskip
Thirdly, we start with two parallel and adjacent gapless walls with the opposite orientations, then we change the orientation of one of the walls and the data on the wall according to Figure\,\ref{fig:flip-orientation}, at last, we apply the formula (\ref{eq:fusing-walls-1}). This produces a non-chiral gapless wall or edge. We give some examples below.

\begin{expl} \label{expl:non-chiral-edge-1}
We start with a bulk phase $(\CC,c)$ with a chiral gapless edge $(V,{}^\CB\CX)$, then flip the arrow of a right semicircle of the edge, then folding the disk as illustrated in the following pictures: 
\be \label{fig:flip+folding}
 \raisebox{-30pt}{
  \begin{picture}(90,80)
   \put(10,0){\scalebox{0.45}{\includegraphics{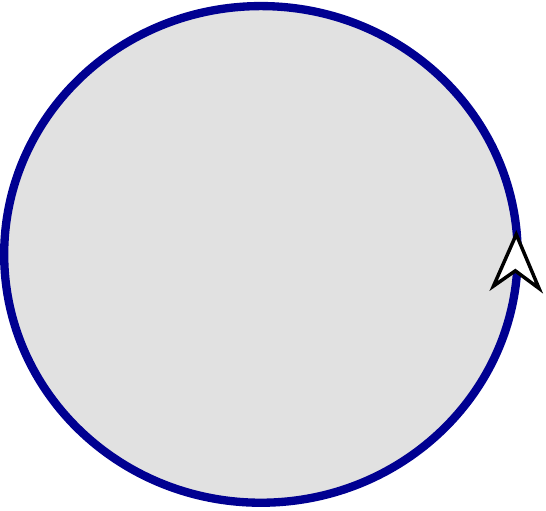}}}
   \put(10,0){
     \setlength{\unitlength}{.75pt}\put(0,0){
          
     \put(33,45)  {\scriptsize $(\CC,c)$}
     \put(-23,75) {\scriptsize $(V,{}^\CB\CX)$}
     
     }\setlength{\unitlength}{1pt}}
  \end{picture}}
\quad\quad\quad\quad
 \raisebox{-30pt}{
  \begin{picture}(90,80)
   \put(0,0){\scalebox{0.45}{\includegraphics{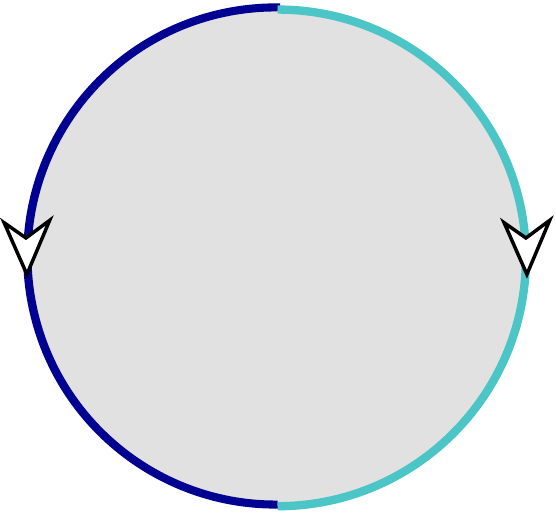}}}
   \put(0,0){
     \setlength{\unitlength}{.75pt}\put(0,0){
   
     \put(33,45)  {\scriptsize $(\CC,c)$}
     \put(-23,75) {\scriptsize $(V,{}^\CB\CX)$}
     \put(90,20)  {\scriptsize $(\overline{V},{}^{\overline{\CB}}\CX^\rev)$}
     
     }\setlength{\unitlength}{1pt}}
  \end{picture}}
\quad\quad\quad\quad\quad
 \raisebox{-30pt}{
  \begin{picture}(60,80)
   \put(0,0){\scalebox{0.45}{\includegraphics{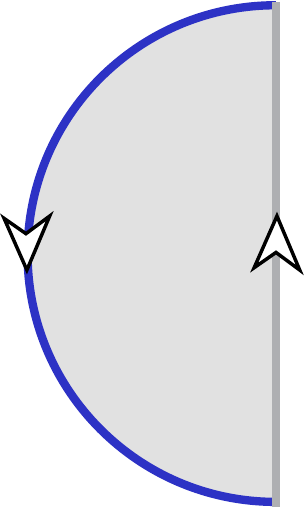}}}
   \put(0,0){
     \setlength{\unitlength}{.75pt}\put(0,0){
     
     \put(52,20) {\scriptsize $(\Cb,{}^\bh\CC)=\CC$}
     \put(10,45)  {\scriptsize $(\FZ(\CC),0)$}
     \put(-80,88) {\scriptsize $(V\otimes_\Cb\overline{V},{}^{\CB\boxtimes \overline{\CB}}(\CX\boxtimes \CX^\rev))$}
     
     }\setlength{\unitlength}{1pt}}
  \end{picture}}
\ee
One can see that, in the third picture, there are two edges of the same 2d bulk $(\FZ(\CC),c)$:
\bnu
\item One is a non-chiral gapless edge given by $(V\otimes_\Cb\overline{V},{}^{\CB\boxtimes \overline{\CB}}(\CX\boxtimes \CX^\rev))$; 
\item the other one is a gapped edge given by $\CC$ viewed as a UFC. Note that the boundary-bulk relation still holds, i.e. 
\be \label{eq:same-center}
\FZ({}^{\CB\boxtimes \overline{\CB}}(\CX\boxtimes \CX^\rev))\simeq \FZ(\CC) \simeq \CC \boxtimes \overline{\CC}, 
\ee
and this non-chiral gapless edge is clearly gappable.

\item Interestingly, this process also creates two 0d gapless walls between these two different edges in the third picture. It is clear that this 0d wall can be described by $(V,{}^\CB\CX)$, in which ${}^\CB\CX$ is the trivial ${}^\CB\CX$-${}^\CB\CX$-bimodule (see the first picture in (\ref{fig:flip+folding})), or equivalently, a ${}^{\CB\boxtimes \overline{\CB}}(\CX\boxtimes \CX^\rev)$-$\CC$-bimodule (see the third picture in (\ref{fig:flip+folding})). By Theorem\,\ref{thm:morita} and (\ref{eq:same-center}), we see that ${}^{\CB\boxtimes \overline{\CB}}(\CX\boxtimes \CX^\rev)$ and  ${}^\bh\CC$ are spatial Morita equivalent with the spatially invertible bimodule given by ${}^\CB\CX$. We will study this type of 0d walls in Section\,\ref{sec:non-chiral-edge-chiral-wall}. 
\enu 
\end{expl}

In general, let $V_L$ and $V_R$ be unitary rational VOA's with central charge $c_L$ and $c_R$, respectively, such that $\Mod_{V_L}$ and $\Mod_{V_R}$ are UMTC's. The following pair   
$$
\left( V_L\otimes_\Cb\overline{V}_R,{}^{\Mod_{V_L}\boxtimes \overline{\Mod_{V_R}}}(\Mod_{V_L}\boxtimes \Mod_{V_R}^\rev)\right),
$$ 
which defines a so-called {\it the canonical non-chiral gapless edge} of $(\Mod_{V_L}\boxtimes\overline{\Mod_{V_R}},c_L-c_R)$. We will call $V_L$ the chiral symmetry, $V_R$ {\it the anti-chiral symmetry}, and $V_L\otimes_\Cb\overline{V}_R$ {\it the non-chiral symmetry}. When $V_L\nsimeq V_R$, the non-chiral gapless edge will be called {\it hieterotic}.

\subsection{Classification of non-chiral gapless edges} \label{sec:classification-non-chiral-edges}

It turns out that $V_L\otimes_\Cb\overline{V}_R$ is not the most general non-chiral symmetry. The algebraic structure on $V_L\otimes_\Cb\overline{V}_R$ is not a VOA but a so-called full field algebra of central charges $(c_L,c_R)$, where $c_L$ (resp. $c_R$) is called the chiral (resp. anti-chiral) central charge, or just a full field algebra for simplicity \cite{ffa,kong-ffa}. Let $V_L$ and $V_R$ be two (unitary) rational VOA's of central charges $c_L$ and $c_R$, respectively. We will be interested in the so-called full field algebras over $V_L\otimes_\Cb \overline{V}_R$, which is a certain full field algebra of central charge $(c_L,c_R)$ containing $V_L\otimes_\Cb \overline{V}_R$ as a subalgebra (see \cite[Definition\,1.17]{ffa} and texts below \cite[Proposition\,1.21]{ffa}). The following theorem is a partial result proved in \cite[Theorem\,4.15]{kong-ffa}. 
\begin{thm}
 A full field algebra (of central charges $(c_L,c_R)$) over $V_L\otimes_\Cb \overline{V}_R$ is equivalent to a commutative algebra in $\Mod_{V_L}\boxtimes \overline{\Mod_{V_R}}$.  
\end{thm}

In this work, by a non-chiral symmetry, we mean a unitary rational full field algebra. We provide a working definition of this notion below. 
\begin{defn}
A full field algebra $W$ of central charges $(c_L,c_R)$ is called unitary rational if 
there exist two unitary rational VOA's $V_L$ and $V_R$ of central charges $c_L$ and $c_R$, respectively, such that $\Mod_{V_L}$ and $\Mod_{V_R}$ are UMTC's, and $W$ is a full field algebra over $V_L\otimes_\Cb \overline{V}_R$, and, as a commutative algebra in $\Mod_{V_L}\boxtimes \overline{\Mod_{V_R}}$, it is connected and separable (i.e. condensable \cite{anyon}). 
\end{defn}

It is possible to give a direct definition of the notion of a module over a full field algebra $W$ such that the category $\Mod_W$ of $W$-modules is equivalent to $(\Mod_{V_L}\boxtimes \overline{\Mod_{V_R}})_W^0$, which denotes the category of local $W$-modules in $\Mod_{V_L}\boxtimes \overline{\Mod_{V_R}}$. For the purpose of this work, we can simply set $\Mod_W:=(\Mod_{V_L}\boxtimes \overline{\Mod_{V_R}})_W^0$. This definition is independent of the choices of $V_L$ and $V_R$.

\begin{rem}
A condensable algebra in $\Mod_{V_L}\boxtimes \overline{\Mod_{V_R}}$ is automatically equipped with a canonical structure of a simple special symmetric $\dagger$-Frobenius algebra ($\dagger$-SSSFA) in $\Mod_{V_L}\boxtimes \overline{\Mod_{V_R}}$ (see for example \cite{anyon}). 
\end{rem}

Let $\CB:=\Mod_W$ for a unitary rational full field algebra $W$ of central charge $(c_L,c_R)$. Then $(\CB,c_L-c_R)$ defines a 2d topological order, and $(W,{}^\CB\CB)$ defines the canonical non-chiral gapless edge of $(\CB, c_L-c_R)$ with a non-chiral symmetry $W$. 

By fusing canonical non-chiral gapless edges with some gapped walls, we obtain more general non-chiral gapless edges. Let $\CX$ by a gapped wall between two 2d topological orders $(\CB,c_L-c_R)$ and $(\CC,c_L-c_R)$.  Then the following fusion formula: 
$$
\left( W,{}^\CB\CX \right)
= \left( W,{}^\CB\CB \right) \boxtimes_{(\CB,\, c_L-c_R)} \left( \Cb,{}^\bh\CX \right) 
$$
defines a non-chiral gapless edge of $(\CC,c_L-c_R)$. All of these non-chiral gapless edges can also be obtained from topological Wick rotations. 

\medskip
Sometimes, a non-chiral gapless edge can be gapped out. In this case, its bulk is a non-chiral 2d topological order. Non-chiral gapless edges of a non-chiral 2d topological order are always gappable. We gives some non-trivial examples.

\begin{expl} 
\label{expl:ising2+toric}
Let $\ising$ be the Ising UMTC given by $\Mod_{V_\ising}$, where $V_\ising$ is the well known Ising VOA with the central charge $c=\frac{1}{2}$. It has three simple objects $\one, \psi, \sigma$ with the fusion rule given by $\psi\otimes\psi=\one$, $\psi\otimes\sigma=\sigma$ and $\sigma\otimes \sigma=\one\oplus\psi$. We have $\FZ(\ising)\simeq \ising \boxtimes \overline{\ising}$. Let $\toric$ be the UMTC describing the $\Zb_2$ 2d topological order. It has four simple objects $1,e,m,f$ with the fusion rule given by $e\otimes e=m\otimes m = f\otimes f=1$ and $m\otimes e=f$.  It is known that $\toric=\FZ(\mathrm{Rep}(\mathbb{Z}_2))$, where $\mathrm{Rep}(\mathbb{Z}_2)$ is the category of finite dimensional representations of the group $\mathbb{Z}_2$. The Lagrangian algebra $B=\one\boxtimes\one \oplus \psi\boxtimes\psi \oplus \sigma\boxtimes\sigma$ in $\FZ(\ising)$ has a subalgebra 
\be \label{eq:def-A}
W = \one\boxtimes\one \oplus \psi\boxtimes\psi,
\ee
which is also condensable. By condensing $W$, we obtain precisely the $\Zb_2$ 2d topological order, i.e. $\FZ(\ising)_W^0\simeq \toric$ \cite{bs,cjkyz}. The UFC $(\FZ(\ising))_W$ describes a gapped wall between $(\FZ(\ising),0)$ and $(\toric,0)$. By fusing this gapped wall with the canonical non-chiral gapless edge of $\FZ(\ising)$, we obtain a non-trivial non-chiral gapless edge of the toric code phase: 
\be \label{eq:toric}
\left( V_\ising\otimes_\Cb \overline{V}_\ising, {}^{\FZ(\ising)}\FZ(\ising)\right) \boxtimes_{(\FZ(\ising),0)} (\FZ(\ising))_W = 
\left( V_\ising\otimes_\Cb \overline{V}_\ising, {}^{\FZ(\ising)}(\FZ(\ising))_W \right). 
\ee
By \cite[Theorem\, 3.3.6]{kz1}, we see that the boundary-bulk relation still holds, i.e. 
$$
\FZ({}^{\FZ(\ising)}(\FZ(\ising))_W) \simeq \toric.
$$ 
It is also worth pointing out that the partition function of the non-chiral gapless edge (i.e. that of $M_{\one,\one}$) is given by $|\chi_0(\tau)|^2 + |\chi_{\frac{1}{2}}(\tau)|^2$, which is not modular invariant because the the edge is anomalous as a gapless 1d phase. It is perhaps the first time to find a physical meaning of non-modular-invariant partition functions. 
\end{expl}

\begin{expl} \label{expl:canonical-edge-toric}
$W$ defined by (\ref{eq:def-A}) is automatically a unitary rational full field algebra over $V_\ising\otimes_\Cb\overline{V}_\ising$. We have $\Mod_W:=\FZ(\ising)_W^0\simeq \toric$. Therefore, 
\be \label{eq:toric-canonical-edge}
(W,{}^{\Mod_W}\toric)
\ee 
defines a canonical non-chiral gapless edge of the $\Zb_2$ 2d topological order $(\toric,0)$. In this case, the partition function of the non-chiral gapless edge (i.e. that of $M_{\one,\one}$) is again given by $|\chi_0(\tau)|^2 + |\chi_{\frac{1}{2}}(\tau)|^2$. But (\ref{eq:toric-canonical-edge}) and (\ref{eq:toric}) describe different edges because the sets of topological edge excitations (or equivalently, the non-chiral symmetries) are different. 
\end{expl}

\begin{expl} \label{expl:edge-toric-III}
Recall that there are two gapped edges of $\Zb_2$ topological orders described by two UFC's $\rep(\Zb_2)$ and $\mathrm{Vec}_{\Zb_2}$, corresponding to condensing $m$-particles and $e$-particles, respectively. Recall that 
\be \label{eq:def-B}
B:=\one\boxtimes\one \oplus \psi\boxtimes\psi \oplus \sigma\boxtimes\sigma
\ee 
is a Lagrangian algebra in $\FZ(\ising)$. It is a modular invariant full field algebra extending $V_\ising\boxtimes \overline{V}_\ising$. We have $\Mod_B:= \FZ(\ising)_B^0 \simeq \bh$. Then we obtain two new non-chiral gapless edges of $(\toric,0)$ defined by 
$(B,{}^{\Mod_B}\rep(\Zb_2))$ and $(B, {}^{\Mod_B}\vect_{\Zb_2})$. Both of them can be factorized as follows: 
\begin{align}
(B,{}^{\Mod_B}\rep(\Zb_2)) &\simeq (B,{}^{\Mod_B}\bh) \boxtimes (\Cb,{}^\bh\rep(\Zb_2)) = (B,{}^{\Mod_B}\bh) \boxtimes \rep(\Zb_2), \label{eq:BH-repZ2} \\
(B, {}^{\Mod_B}\vect_{\Zb_2}) &\simeq (B,{}^{\Mod_B}\bh) \boxtimes (\Cb,{}^\bh\vect_{\Zb_2}) = (B,{}^{\Mod_B}\bh) \boxtimes \vect_{\Zb_2}.\label{eq:BH-vecZ2}
\end{align}
Note that $(B,{}^{\Mod_B}\bh)$ is a non-chiral gapless edge of the 2d trivial phase because $\FZ({}^{\Mod_B}\bh)=\FZ(\bh)=\bh$, thus provides a mathematical description of an anomaly-free 1d gapless phase. Therefore, both gapless edges $(B,{}^{\Mod_B}\rep(\Zb_2))$ and $(B, {}^{\Mod_B}\vect_{\Zb_2})$ are obtained by stacking the anomaly-free 1d gapless phase $(B,{}^{\Mod_B}\bh)$ with the gapped edges of $\Zb_2$ topological orders. 
\end{expl}

\begin{expl}
By a topological Wick rotation, we obtain an edge of the trivial 2d topological order, or equivalently, an anomaly-free 1d gapless phase,  $(V_\ising\otimes_\Cb\overline{V}_\ising, {}^{\FZ(\ising)}\ising)$. It is different from $(B,{}^{\Mod_B}\bh)$ in Example\,\ref{expl:edge-toric-III} in their non-chiral symmetries and the categories of topological edge excitations. One can obtain the first one from the second one via a 1d phase transition, which breaks the non-chiral symmetry from $B$ to $V_\ising\otimes_\Cb\overline{V}_\ising$.  
\end{expl}


Similar to chiral gapless edges, we would like to propose that all non-chiral gapless edges of a 2d topological order can be obtained by fusing canonical non-chiral gapless edges with gapped walls, or equivalently, by topological Wick rotations. Moreover, we expect that different pairs $(A,{}^\CB\CX)$ describe different non-chiral gapless edges (see Section\,\ref{sec:edge-tpt} for more discussion). As a consequence, we obtain the following classification result stated as a physical Theorem. 
\begin{pthm} \label{pthm:non-chiral-edge}
Non-chiral gapless edges of a 2d topological order $(\CC,c)$ are mathematically described and classified by pairs $(W, {}^\CB\CX)$, where 
\begin{itemize}
\item $W$ is the non-chiral symmetry, which is a unitary rational full field algebra with chiral central charge $c_L$ and the anti-chiral central charge $c_R$ such that $c=c_L-c_R$. 
\item ${}^\CB\CX$ is the enriched monoidal category defined by the pair $(\CB, \CX)$ via the canonical construction, where $\CB:=\Mod_W$ and $\CX$ is a closed fusion $\CB$-$\CC$-bimodule. 
\end{itemize}
For the convenience of numerical computation, one can replace the pair $(W, {}^\CB\CX)$ by a new pair $(W, A)$, where $A$ is a Lagrangian algebra in $\overline{\CB} \boxtimes \CC$. 
\end{pthm}

\begin{rem}
The mathematical classification of non-chiral gapless walls is automatic by the folding trick. 
\end{rem}

\begin{rem} \label{remark:non-chiral-edge}
Theorem$^{\mathrm{ph}}$\,\ref{pthm:non-chiral-edge} automatically contains the classification of chiral gapless edges (see \cite[Theorem$^{\mathrm{ph}}$\,6.7]{kz4}) as special cases, in which $W=V_L\otimes_\Cb \overline{V}_R$ and $V_R=\Cb$.  
\end{rem}

\begin{thm} \label{pthm:bb-relation-non-chiral}
By Theorem\,\ref{thm:KZ2}, a non-chiral gapless edge $(W, {}^\CB\CX)$ automatically satisfies the boundary-bulk relation, i.e. 
$$
\FZ( {}^\CB\CX) \simeq \CC.
$$ 
\end{thm}

\begin{defn}
A non-chiral gapless edges $(W, {}^\CB\CX)$ is called {\it anomaly-free} if the bulk is trivial, i.e. $\FZ( {}^\CB\CX) \simeq \bh$; it is called {\it anomalous} if otherwise. It is called {\it trivial} if $(W, {}^\CB\CX)=(\Cb,{}^\bh\bh)$. In general, a non-chiral gapless edges $(W, {}^\CB\CX)$ can be factorized as a product: 
$$
(W_1,{}^{\CB_1}\CX_1) \boxtimes \cdots \boxtimes (W_k,{}^{\CB_k}\CX_k).
$$
If a non-chiral gapless edge can not be factorized as a product of two non-trivial edges, then it is called {\it primary}. For a given non-chiral gapless edge, we will call the product of all its primary anomalous factors as its {\it anomalous core}. 
\end{defn}

\begin{expl}
The anomalous cores of $(B,{}^{\Mod_B}\rep(\Zb_2))$ and $(B,{}^{\Mod_B}\vect_{\Zb_2})$ in (\ref{eq:BH-repZ2}) and (\ref{eq:BH-vecZ2}) are given by $\rep(\Zb_2)$ and $\vect_{\Zb_2}$, respectively. 
\end{expl}

\begin{rem}
It is clear that the the most interesting part of the classification of all non-chiral gapless edges of a given 2d topological order lies in the classification of the anomalous cores of all non-chiral gapless edges. 
\end{rem}

Recently, Ji and Wen showed in some concrete examples that the partition functions $[\one_\CX,\one_\CX]_\CB$ of gapless edges of 2d topological orders transform covariantly under the mapping class group $SL(2,\Zb)$ according to the $S$-,$T$-matrix of the bulk UMTC \cite{jw}. But this covariance does not hold in general. We give a precise statement of this covariance.  

\begin{pthm}
Let $(W, {}^\CB\CX)$ be a non-chiral gapless edge of a 2d topological order $(\CC,c)$. The partition functions $A=[\one_\CX,\one_\CX]_\CB$ of gapless edges of 2d topological orders transform covariantly under the mapping class group $SL(2,\Zb)$ according to the $S$-,$T$-matrices of the UMTC $\CB_A^0$ of local $A$-modules in $\CB$. 
\end{pthm}
\pf
This follows automatically from the Huang's proof of modular tensor category from rational VOA \cite{huang-mtc}.
\epf

\begin{expl}
Consider a conformal embedding $V\subsetneq A$ of unitary rational VOA's, e.g.
\begin{align*}
&su(m)_n \times su(n)_m \subset su(mn)_1, \quad 
sp(2m)_n \times sp(2n)_m  \subset so(4mn)_1,  \\
&so(m)_n \times so(n)_m \subset so(mn)_1, \quad 
so(m)_4 \times su(2)_m \subset sp(2m)_1, \cdots ,
\end{align*} 
or any embedding of unitary rational full field algebras $V\subsetneq A$. Then $A$ can be viewed as a condensable algebra in $\CB=\Mod_V$ \cite{KO,hkl} and we have $\Mod_A= \CB_A^0$. Therefore, the two topological orders $(\CB,c)$ and $(\Mod_A,c)$ can be connected by a gapped wall given by $\CB_A$. By topological wick rotations, we obtain a chiral gapless edge of $(\CB,c)$ defined by $(B, {}^{\Mod_A}(\CB_A))$, in which $M_{\one,\one}=A$. In this case, the modular transformations of the partition function of $M_{\one,\one}=A$ coincide with the $S$-,$T$-matrices of $\Mod_A$ instead of those of the bulk UMTC $\CB$. 
\end{expl}

\subsection{Purely edge phase transitions} \label{sec:edge-tpt}

\begin{figure} 
$$ 
\raisebox{-30pt}{
  \begin{picture}(140,75)
   \put(0,15){\scalebox{0.7}{\includegraphics{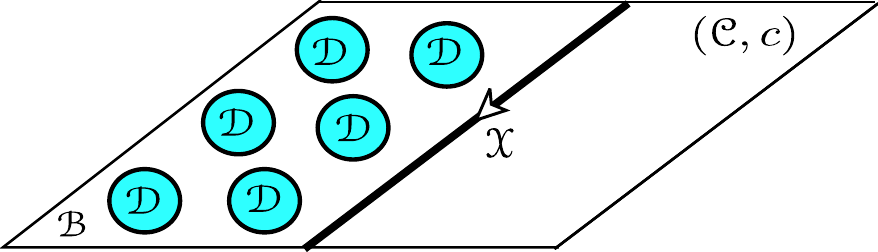}}}
   \put(0,15){
     \setlength{\unitlength}{.75pt}\put(0,0){
     }\setlength{\unitlength}{1pt}}
  \end{picture}} 
\quad \xrightarrow{\mbox{\footnotesize topological Wick rotation}} \quad 
\raisebox{-30pt}{
  \begin{picture}(100,75)
   \put(0,10){\scalebox{0.7}{\includegraphics{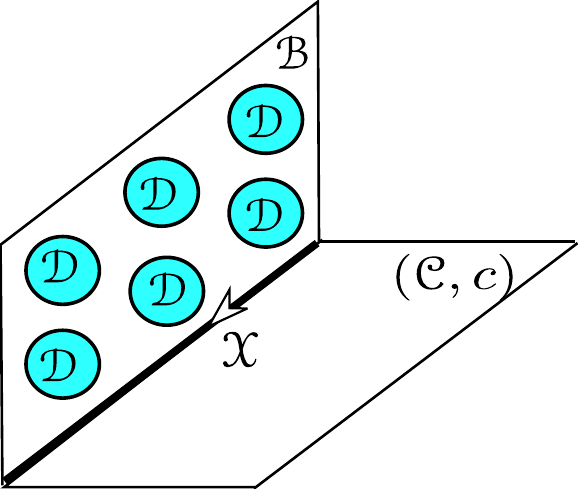}}}
   \put(0,10){
     \setlength{\unitlength}{.75pt}\put(0,0){
     }\setlength{\unitlength}{1pt}}
  \end{picture}}
$$
\caption{These two pictures depict a physical description of purely edge phase transition via a topological Wick rotation. 
}
\label{fig:phase-transition}
\end{figure}

We have mentioned that different pairs $(W,{}^\CB\CX)$ and $(W',{}^{\CB'}\CX')$ should represent different non-chiral gapless edges. It means that if we deform one edge by adding perturbations to get the other one, we need go through at least one phase transition points. We do not have a physical proof of this claim. Actually, as far as we known, there is no universal or model-independent definition of a phase transition between two gapless phases. As we have already pointed out for chiral gapless edges in \cite{kz4}, our mathematical theory of gapless edges actually provides such definitions. These definitions can automatically be generalized to include non-chiral gapless edges. More precisely, we propose that a 1+1D purely edge phase transition between two gapless chiral (resp. non-chiral) edges can be defined either
\bnu
\item as a process of changing or breaking chiral (resp. non-chiral) symmetries; or
\item as the topological Wick rotation of a 2d topological phase transition, which is defined by a process of closing the gap, as illustrated in Figure\,\ref{fig:phase-transition}. 
\enu

\begin{rem}
We do not know how to generalize the first definition to higher dimensional gapless phases
because the replacement for chiral or non-chiral symmetries is not so clear in higher dimensions.\footnote{We expect that an $n$-dimensional local quantum symmetries should be an analogue of an $E_n$-algebra.} But the second definition can be automatically generalized to higher dimensions. This provides a surprising and exciting implications to the study of higher dimensional gapless phases. 
\end{rem}

Actually, the story about purely edge phase transitions between chiral gapless edges become complete only when we include all non-chiral gapless edges because non-chiral gapless modes should appear at the critical points even if the initial and final edges are either chiral or gapped. For example, we can consider purely edge phase transitions between two gapped edges of a non-chiral topological order $(\CC,0)$. At the critical point, the gap is closed and necessarily non-chiral. Therefore,  
\begin{quote}
the critical point of a {\bf purely edge topological phase transition} is nothing but a gappable non-chiral gapless edge, and should be mathematically described by a non-chiral symmetry and an enriched fusion category, whose Drinfeld center coincides with the UMTC of the bulk. 
\end{quote}
For example, in \cite{cjkyz}, it was shown in great details that the non-chiral gapless edges given in Example\,\ref{expl:ising2+toric} and Example\,\ref{expl:canonical-edge-toric} precisely describe the critical points of purely edge topological phase transitions between the two well-known gapped edges of the 2d $\mathbb{Z}_2$ topological order \cite{bk}. 

We believe that chiral gapless/gapped edges and certain non-gappable non-chiral gapless edges are stable in the sense that they are RG fixed points. Other non-chiral gapless edges are unstable. For example, gappable non-chiral gapless edges are all unstable because they can be gapped. As a consequence, we should expect that the following result. 
\begin{quote}
The critical point of a {\bf purely edge phase transition} between two stable edges of a 2d topological order defines an unstable non-chiral gapless edge, and should be mathematically described by a non-chiral symmetry and an enriched fusion category, whose Drinfeld center coincides with the UMTC of the bulk. 
\end{quote}
For a given 2d topological order, it is an important problem to work out the complete phase diagram of all edges. A cell of the highest dimension in the phase diagram should represent a stable edge, and a cell of codimension 1 should represent an unstable non-chiral gapless edge. If higher codimensional cells exist, then it means that there are different levels of unstableness. This will be really interesting. Note that the physical fact of the 2d bulk being an invariant (as the gravitational anomaly) of the entire phase diagram is confirmed by the mathematical Theorem\,\ref{pthm:bb-relation-non-chiral}. We hope to study the phase diagram in the future.

\subsection{0+1D gapless walls} 
\label{sec:non-chiral-edge-chiral-wall}
\begin{figure} 
$$
 \raisebox{-50pt}{
  \begin{picture}(150,140)
   \put(-70,7){\scalebox{0.7}{\includegraphics{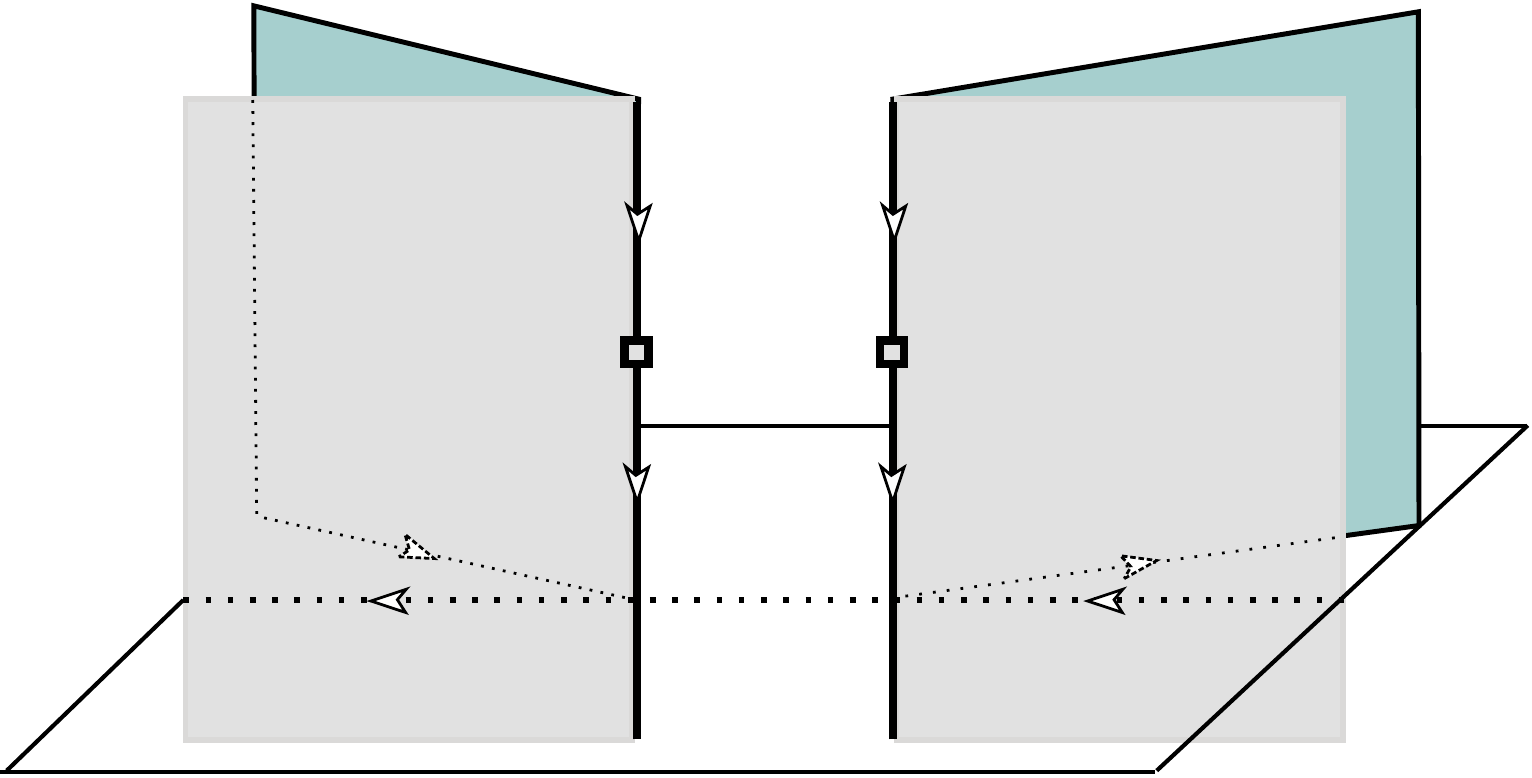}}}
   \put(-2,7){
     \setlength{\unitlength}{.75pt}\put(0,0){
     \put(10,36)   {\footnotesize$\CX$}
     \put(15,70)   {\footnotesize $\CY$}
     \put(205,34) {\footnotesize$\CX'$}
     \put(210,67)   {\footnotesize $\CY'$}
     
     \put(87,148) {\footnotesize $\CP$}
     \put(156,148) {\footnotesize $\CP'$}
     \put(110,80)  {\footnotesize$(\CC,c)$}
     \put(90,54)   {\footnotesize$\CC=(\Cb,{}^\bh\CC)$}
     
    \put(190,167) {\footnotesize $(V_{\CA'},{}^{\CA'}\CX')$}
    \put(200, 207) {\footnotesize $(V_{\CB'},{}^{\CB'}\CY')$}     
     \put(-10,167)  {\footnotesize $(V_\CA,{}^{\CA}\CX)$}
     \put(0, 210) {\footnotesize $(V_\CB,{}^{\CB}\CY)$}
     
     \put(70,192)   {\footnotesize $(V,X,{}^\CP\CM)$}
     \put(125,192) {\footnotesize $(V',X',{}^{\CP'}\CM')$}    
    
     \put(-70,10) {\footnotesize $(\CC,c)$}
     \put(46,35)  {\footnotesize $m\in \CM$}
     \put(156,35) {\footnotesize $m'\in \CM'$}
     
     \put(40,112)  {\footnotesize $[m,n]_\CP$}
     \put(160,112) {\footnotesize $[m',n']_{\CP'}$}


     }\setlength{\unitlength}{1pt}}
  \end{picture}}
$$
\caption{This picture depicts two chiral 0+1D walls connecting a gapped wall with two non-chiral gapless walls, the left of which is obtained by folding the chiral gapless edge $(V_\CB, {}^\CB\CY)$ in Figure\,\ref{fig:0d-defects} backwards, and the right one is obtained similarly. 
}
\label{fig:0d-defects-chiral}
\end{figure}

In this subsection, we study 0+1D gapless walls between two non-chiral gapless edges. We first illustrate 
three special types of 0+1D gapless walls in Figure\,\ref{fig:0d-defects-chiral}. 

\bnu

\item 0+1D walls between a non-chiral gapless edge/wall and a gapped edge/wall: for example, in Figure\,\ref{fig:0d-defects-chiral}, 
$(V,X,{}^\CP\CM)$ defines a 0+1D wall between the 1+1D non-chiral gapless wall $(V_\CA\otimes_\Cb \overline{V}_\CB, {}^{\CA\boxtimes\overline{\CB}}(\CX\boxtimes\CY^\rev))$ and the 1+1D gapped wall $\CC$; and $(V',X',{}^{\CP'}\CM')$ defines a 0+1D wall between the 1+1D gapped wall $\CC$ and the 1+1D non-chiral gapless wall $(V_{\CA'}\otimes_\Cb \overline{V}_{\CB'}, {}^{\CA'\boxtimes\overline{\CB'}}(\CX'\boxtimes(\CY')^\rev))$. 

\item If we spatially fuse two 0+1D gapless walls in Figure\,\ref{fig:0d-defects-chiral}, we obtain a 0+1D gapless wall
\be \label{eq:1D-wall-non-chiral-1}
(V,X,{}^\CP\CM) \boxtimes_{(\Cb,{}^\bh\CC)} (V',X',{}^{\CP'}\CM') := (V\otimes_\Cb V', X\boxtimes X', {}^{\CP\boxtimes\CP'}(\CM\boxtimes_\CC\CM'))
\ee
between the following two 1d non-chiral gapless walls: 
\be \label{eq:1D-wall-non-chiral-2}
(V_\CA\otimes_\Cb \overline{V}_\CB, {}^{\CA\boxtimes\overline{\CB}}(\CX\boxtimes\CY^\rev))\quad \mbox{and} \quad (V_{\CA'}\otimes_\Cb \overline{V}_{\CB'}, {}^{\CA'\boxtimes\overline{\CB'}}(\CX'\boxtimes(\CY')^\rev)).
\ee

\item (\ref{eq:1D-wall-non-chiral-1}) can also be viewed as a 0+1D gapless wall between the following two 1d non-chiral gapless walls: 
\be \label{eq:1D-wall-non-chiral-3}
(V_\CA\otimes_\Cb \overline{V}_{\CA'}, {}^{\CA\boxtimes\overline{\CA'}}(\CX\boxtimes(\CX')^\rev))\quad \mbox{and} \quad (V_{\CB}\otimes_\Cb \overline{V}_{\CB'}, {}^{\CB\boxtimes\overline{\CB'}}(\CY\boxtimes(\CY')^\rev));
\ee

\enu

Type-3 is relatively easier to understand because the chiral parts and the anti-chiral parts are completely separated. Therefore, mathematical description follows from that of a 0+1D chiral gapless wall between two 1+1D chiral gapless edges. 

Type-2 is new. In general, there are more 0+1D gapless walls between the two 1+1D non-chiral gapless walls in (\ref{eq:1D-wall-non-chiral-2}) than just (\ref{eq:1D-wall-non-chiral-1}). They can be classified by reducing the problem to an old one. Indeed, by flipping the orientations of the anti-chiral parts of two non-chiral edges in (\ref{eq:two-non-chiral-edges}) and, at the same time, changing $(\overline{V}_\CB,{}^{\overline{\CB}}\CY^\rev)$ and $(\overline{V}_{\CB'},{}^{\overline{\CB'}}(\CY')^\rev)$ to $(V_\CB,{}^{\CB}\CY)$ and $(V_{\CB'},{}^{\CB'}\CY')$, respectively, we see that a 0+1D gapless wall between two 1+1D non-chiral gapless edges
\be \label{eq:two-non-chiral-edges}
(V_\CA\otimes_\Cb \overline{V}_\CB, {}^{\CA\boxtimes\overline{\CB}}(\CX\boxtimes\CY^\rev))
\quad \mbox{and} \quad
(V_{\CA'}\otimes_\Cb \overline{V}_{\CB'}, {}^{\CA'\boxtimes\overline{\CB'}}(\CX'\boxtimes(\CY')^\rev))
\ee
is precisely a 0+1D chiral gapless wall between the following two chiral gapless walls
\be \label{eq:two-chiral-edges}
(V_\CA\otimes_\Cb V_{\CB'}, {}^{\CA\boxtimes \CB'} (\CX\boxtimes \CY')) 
\quad \mbox{and} \quad
(V_\CB\otimes_\Cb V_{\CA'}, {}^{\CB\boxtimes \CA'} (\CY\boxtimes \CX')).  
\ee

Type-1 walls actually covers all 0+1D walls if we apply the folding trick in in temporal direction. In this context, 
there is no need to distinguish ``chiral'' and ``non-chiral'' for gapless 0+1D walls because we can always view the boundary CFT's on a 0+1D world line as chiral under the folding trick.

\begin{figure}
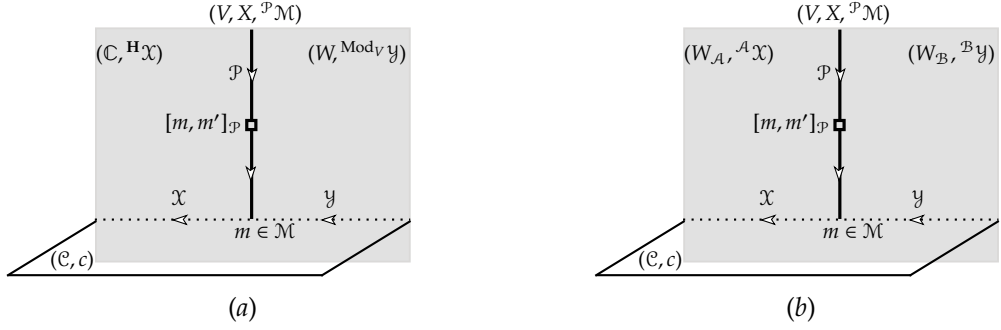
 
$$
 \raisebox{-50pt}{
  \begin{picture}(180,100)
   \put(-10,0){\scalebox{0.5}{\includegraphics{pic-two-0d-wall-op-2-eps-converted-to.pdf}}}
   \put(-10,0){
     \setlength{\unitlength}{.75pt}\put(0,0){
     \put(22,6) {\footnotesize $(\CC,c)$}
     \put(83,37)  {\footnotesize $\CX$}
     \put(45,110)  {\footnotesize $(\Cb,{}^{\bh}\CX)$}
     \put(160,37)  {\footnotesize $\CY$}
     \put(152,110){\footnotesize $(W,{}^{\Mod_V}\CY)$}

     \put(115,20) {\footnotesize $m\in\CM$}
     \put(112,100)  {\footnotesize $\CP$}
     \put(102,130)  {\footnotesize $(V,X,{}^{\CP}\CM)$}
     \put(80,75)  {\footnotesize $[m,m']_{\CP}$}
     }\setlength{\unitlength}{1pt}}
  \end{picture}}
  \quad\quad\quad\quad
 \raisebox{-50pt}{
  \begin{picture}(150,100)
   \put(-10,0){\scalebox{0.5}{\includegraphics{pic-two-0d-wall-op-2-eps-converted-to.pdf}}}
   \put(-10,0){
     \setlength{\unitlength}{.75pt}\put(0,0){
     \put(22,6) {\footnotesize $(\CC,c)$}
     \put(83,37)  {\footnotesize $\CX$}
     \put(45,110)  {\footnotesize $(W_\CA,{}^{\CA}\CX)$}
     \put(160,37)  {\footnotesize $\CY$}
     \put(158,110){\footnotesize $(W_\CB,{}^{\CB}\CY)$}

     \put(115,20) {\footnotesize $m\in\CM$}
     \put(112,100)  {\footnotesize $\CP$}
     \put(102,130)  {\footnotesize $(V,X,{}^{\CP}\CM)$}
     \put(80,75)  {\footnotesize $[m,m']_{\CP}$}
     }\setlength{\unitlength}{1pt}}  \end{picture}}
$$
$$
(a) \quad\quad\quad\quad\quad\quad\quad\quad\quad\quad\quad\quad\quad\quad\quad\quad\quad\quad\quad\quad
(b)
$$
\caption{These pictures depicts two 0+1D gapless walls.}
\label{fig:two-0d-walls}
\end{figure}

\medskip
Using both the folding tricks in temporal and spatial directions, we can reduce the problem of classifying all 0+1D walls (without any gappable parts (see Remark\,\ref{rem:gappable-excluded})) to the classification of all 0+1D gapless walls between a gapped edge $\CX=(\Cb,^\bh\CX)$ and a 1d non-chiral gapless edge $(W, {}^{\Mod_W}\CY)$ as depicted in Figure\,\ref{fig:two-0d-walls} (a), where $W$ is a unitary rational full field algebra over $V_1\otimes_\Cb \overline{V}_2$. In this case, we still have a 1+1D chiral symmetry $V$ and a 0+1D chiral symmetry $X$. More precisely, $V$ is a unitary rational sub-VOA of $V_1$ and $V_2$.\footnote{If such $V$ does not exist, then no 0+1D gapless wall exists between these two 1d edges.} The full field algebra $W$ can be viewed as a condensable algebra in $\FZ(\Mod_V)=\Mod_V\boxtimes \overline{\Mod_V}$. Note that $\Mod_V$ is a closed right fusion $\FZ(\Mod_V)$-module. We denote the category of right $W$-modules in $\Mod_V$ by $\RMod_W(\Mod_V)$. If we denote the image of $W$ in $\Mod_V$ under the forgetful functor $\forget: \FZ(\Mod_V) \to \Mod_V$ by $\forget(W)$, which is an algebra in $\Mod_V$, we have $\RMod_W(\Mod_V)=(\Mod_V)_{\forget(W)}$, where $(\Mod_V)_{\forget(W)}$ is the category of right $\forget(W)$-modules in $\Mod_V$. 
It is clear that all 0D defects on the 0+1D world line of the wall are objects in $\RMod_W(\Mod_V)$. Therefore, the 0+1D chiral symmetry $X$ must be a symmetric separable $\dagger$-Frobenius algebra in $\RMod_W(\Mod_V)$. The relation between $V$ and $X$ can be summarized by the commutative diagram in (\ref{diag:relation-W-Y}). By \cite[Theorem\,3.20]{dmno}, $(\Mod_V)_{\forget(W)}$ is a closed right multi-fusion $\Mod_W$-module, so is $((\Mod_V)_{\forget(W)})_{X|X}$. 
\begin{pthm} \label{pthm:chiral-wall-non-chiral-edges}
For a unitary rational non-trivial full field algebra $W$ over $V_1\otimes_\Cb \overline{V}_2$, 0+1D gapless walls (without any gappable parts (see Remark\,\ref{rem:gappable-excluded})) between a gapped edge $\CX=(\Cb,^\bh\CX)$ and a 1d non-chiral gapless edge $(W,{}^{\Mod_W}\CY)$ of a 2d topological order $(\CC,c)$, as depicted in Figure\,\ref{fig:two-0d-walls} (a), are mathematically described and classified by triples $(V,X,{}^\CP\CM)$, where 
\bnu
\item $V$ is the 1+1D chiral symmetry, i.e. a unitary rational VOA; $X$ is the 0+1D chiral symmetry, i.e. a symmetric separable $\dagger$-Frobenius algebra in $(\Mod_V)_{\forget(W)}$. They are equipped with algebra homomorphisms in $\Mod_V$ rendering the following diagram commutative
\be \label{diag:relation-W-Y}
\raisebox{3em}{\xymatrix@R=1em{
& V \ar@{^{(}->}[ld] \ar@{^{(}->}[d] \ar@{^{(}->}[rd] & \\
V_1 \ar[r]^{\iota_L} \ar@{^{(}->}[dr] & X & V_2 \ar[l]_{\iota_R} \ar@{^{(}->}[dl] \\
& \forget(W) \ar[u]_{\iota_Y} & 
}} 
\ee
where $\iota_Y$ is an algebra homomorphism between two algebras in $(\Mod_V)_{\forget(W)}$ and defines the unit of the algebra $X$. 

\item ${}^\CP\CM$ is an enriched category defined by the canonical construction from the pair $(\CP,\CM)$, where $\CP$ is a closed right multi-fusion $\Mod_W$-module defined by 
$\CP=((\Mod_V)_{\forget(W)})_{X|X}$, and the category $\CM$ of topological excitations is a finite unitary category uniquely determined by the unitary monoidal equivalence:  $(\CX^\rev\boxtimes \CP\boxtimes_{\Mod_W} \CY)\boxtimes_{\FZ(\CC)} \CC \simeq \fun(\CM,\CM)$. Note that $\CM$ has a canonical left $\CP$ structure defined by $\CP \to (\CX^\rev\boxtimes \CP\boxtimes_{\Mod_W} \CY)\boxtimes_{\FZ(\CC)} \CC \simeq \fun(\CM,\CM)$. In particular, the space of chiral fields living on the world line between two wall excitations $m,m'\in\CM$ is given by $M_{m,m'}=[m,m']_\CP$ for $m,m'\in\CM$. 
\enu
Moreover, all these 0+1D gapless walls are spatially equivalent and define the same 0d wall. When $W=\Cb$, we must have $V=\Cb$, and this 0+1D wall is gapped. 

\end{pthm}

\begin{rem}
Similar to Remark\,\ref{rem:include-gappable}, if we ignore $V$ and $X$, the pure categorical description ${}^\CP\CM$ automatically covers 0+1D gappable factors or parts. 
\end{rem}

\begin{rem}
If we want to emphasize a particular spatial slide of the 0+1D wall, we can specify a wall excitation $m\in\CM$ in the spatial slide, thus obtain a quadruple $(V,X,{}^\CP\CM,m)$. 
\end{rem}


\begin{expl}
Recall Example\,\ref{expl:ising2+toric}, \ref{expl:canonical-edge-toric} and \ref{expl:edge-toric-III}. Let $W$ and $B$ be the full field algebras defined by (\ref{eq:def-A}) and (\ref{eq:def-B}), respectively. We have the following two non-chiral gapless edges of the $\Zb_2$ 2d topological order: 
\be \label{eq:two-non-chiral-edge-toric}
(V_\ising \otimes_\Cb \overline{V}_\ising, {}^{\FZ(\ising)} \FZ(\ising)_W), \quad (W, {}^{\Mod_W}\toric)
\ee
and two gapped edges $\rep(\Zb_2)$ and $\vect_{\Zb_2}$. 
\bnu
\item For a proper $\CM$, the triple $(V_\ising,V_\ising,{}^\ising\CM)$ defines a 0+1D gapless wall between $\rep(\Zb_2)$ (or $\vect_{\Zb_2}$) and $(V_\ising \otimes_\Cb \overline{V}_\ising, {}^{\FZ(\ising)} \FZ(\ising)_W)$; 

\item For a proper $\CM$, the triple $(V_\ising,\forget(W), {}^{(\ising)_{\forget(W)}}\CM)$ defines a 0+1D gapless wall between $\rep(\Zb_2)$ (or $\vect_{\Zb_2}$) and $(W, {}^{\Mod_W}\toric)$.  
\enu
\end{expl}

It is not so convenient to see 0+1D gapless walls between two non-chiral gapless edges in (\ref{eq:two-non-chiral-edge-toric}) because we need to apply the folding trick first in order to reduce the problem to the situation in Theorem${}^{\mathrm{ph}}$\,\ref{pthm:chiral-wall-non-chiral-edges}. For readers' convenience, we work out a special case of Theorem${}^{\mathrm{ph}}$\,\ref{pthm:chiral-wall-non-chiral-edges} depicted in Figure\,\ref{fig:two-0d-walls} (b) and summarize it as the following physical theorem.

\begin{pthm} \label{pthm:wall-chiral-edges}
For two unitary rational full field algebras $W_\CA$ and $W_\CB$ over $V_L\otimes_\Cb \overline{V}_R$, 0+1D non-chiral gapless walls (without gappable parts) between two non-chiral gapless edges $(W_\CA,{}^{\CA}\CX)$ and $(W_\CB,{}^\CB\CY)$ of a 2d topological order $(\CC,c)$ (see Figure\,\ref{fig:two-0d-walls} (b)), preserving the 1+1D non-chiral symmetry $V_L\otimes_\Cb \overline{V}_R$, are mathematically described and classified by triples $(W,X,{}^\CP\CM)$, 
where 
\bnu
\item $W$ is the 1+1D non-chiral symmetry defined by a unitary rational full field algebra $W$ over $V_L\otimes_\Cb \overline{V}_R$; $X$ is the 0+1D non-chiral symmetry, i.e. a symmetric separable $\dagger$-Frobenius algebra in $(\Mod_{W})_{W_\CA|W_\CB}$. They are equipped with algebra homomorphisms in $\Mod_W$ rendering the following diagram commutative
\be \label{diag:relation-W-Y-2}
\raisebox{3em}{\xymatrix@R=1em{
& W \ar@{^{(}->}[ld] \ar@{^{(}->}[d] \ar@{^{(}->}[rd] & \\
W_\CA \ar[r]^{\iota_L} \ar@{^{(}->}[dr] & X & W_\CB \ar[l]_{\iota_R} \ar@{^{(}->}[dl] \\
& W_\CA \otimes_W W_\CB \, ,\ar[u]_{\iota_Y} & 
}} 
\ee
where $\iota_Y$ is an algebra homomorphism between two algebras in $(\Mod_{W})_{W_\CA|W_\CB}$. 

\item ${}^\CP\CM$ is an enriched category defined by the canonical construction from the pair $(\CP,\CM)$, where
$\CP=((\Mod_{W})_{W_\CA|W_\CB})_{X|X}$, and the underlying category $\CM$ is uniquely determined by a unitary monoidal equivalence $(\CX^\rev\boxtimes_\CA \CP\boxtimes_\CB \CY)\boxtimes_{\FZ(\CC)} \CC \simeq \fun(\CM,\CM)$, and has a canonical left $\CP$ structure defined by
$\CP \to (\CX^\rev\boxtimes_\CA \CP\boxtimes_\CB \CY)\boxtimes_{\FZ(\CC)} \CC \simeq \fun(\CM,\CM)$.
\enu
\end{pthm}

Note that $(W_\CA,W_\CA,{}^\CA\CX)$ defines the trivial 0+1D wall between $(W_\CA,{}^{\CA}\CX)$ and $(W_\CA,{}^{\CA}\CX)$. 

\begin{expl} \label{expl:non-chiral-0d-walls}
We give a few concrete examples. Recall Example\,\ref{expl:ising2+toric}, \ref{expl:canonical-edge-toric} and \ref{expl:edge-toric-III}. Let $W$ and $B$ be the full field algebras defined by (\ref{eq:def-A}) and (\ref{eq:def-B}), respectively. We have the following four non-chiral gapless edges of the $\Zb_2$ 2d topological order: 
$$
(V_\ising \otimes_\Cb \overline{V}_\ising, {}^{\FZ(\ising)} \FZ(\ising)_W), \quad (W, {}^{\Mod_W}\toric), \quad 
(B,{}^{\Mod_B}\rep(\Zb_2)), \quad (B, {}^{\Mod_B}\vect_{\Zb_2}). 
$$
Then we have
\bnu
\item $(V_\ising \otimes_\Cb \overline{V}_\ising, W, {}^{\FZ(\ising)_W}(\FZ(\ising)_W))$
is a wall between $(V_\ising \otimes_\Cb \overline{V}_\ising, {}^{\FZ(\ising)} \FZ(\ising)_W)$ and $(W, {}^{\Mod_W}\toric)$;

\item $(W,B,{}^{\rep(\Zb_2)}\rep(\Zb_2))$ is a wall between $(W, {}^{\Mod_W}\toric)$ and $(B,{}^{\Mod_B}\rep(\Zb_2))$; 

\item $(W,B,{}^{\vect_{\Zb_2}}\vect_{\Zb_2})$ is a wall between $(W, {}^{\Mod_W}\toric)$ and $(B,{}^{\Mod_B}\vect_{\Zb_2})$; 

\item $(W,B,{}^{\rep(\Zb_2)}\vect)$ is a wall between $(W, {}^{\Mod_W}\toric)$ and $(B,{}^{\Mod_B}\vect_{\Zb_2})$;

\item $(V_\ising \otimes_\Cb \overline{V}_\ising, B, {}^{\FZ(\ising)_B}\CM)$
is a wall between $(V_\ising \otimes_\Cb \overline{V}_\ising, {}^{\FZ(\ising)} \FZ(\ising)_W)$ and $(B,{}^{\Mod_B}\rep(\Zb_2))$ for a proper $\CM$.  
\enu
\end{expl}

\subsection{Spatial fusion of 0+1D walls and anomalies} \label{sec:fusion-0d-wall-non-chiral}

Spatial fusion of two 0+1D gapless walls between two non-chiral gapless edges/walls are similar to that of walls between chiral gapless edges/walls (see Section\,\ref{sec:fusion-anomaly}).

\medskip
We first consider the spatial fusion of two 0+1D gapless walls covered in Theorem${}^{\mathrm{ph}}$\,\ref{pthm:wall-chiral-edges}. By first breaking the 1+1D non-chiral symmetries of two walls to a smaller but the same one, we reduce the problem to a special case, in which the 1+1D non-chiral symmetries is preserved during the spatial fusion. 

\begin{figure} 
$$
 \raisebox{-50pt}{
  \begin{picture}(180,100)
   \put(-10,0){\scalebox{0.5}{\includegraphics{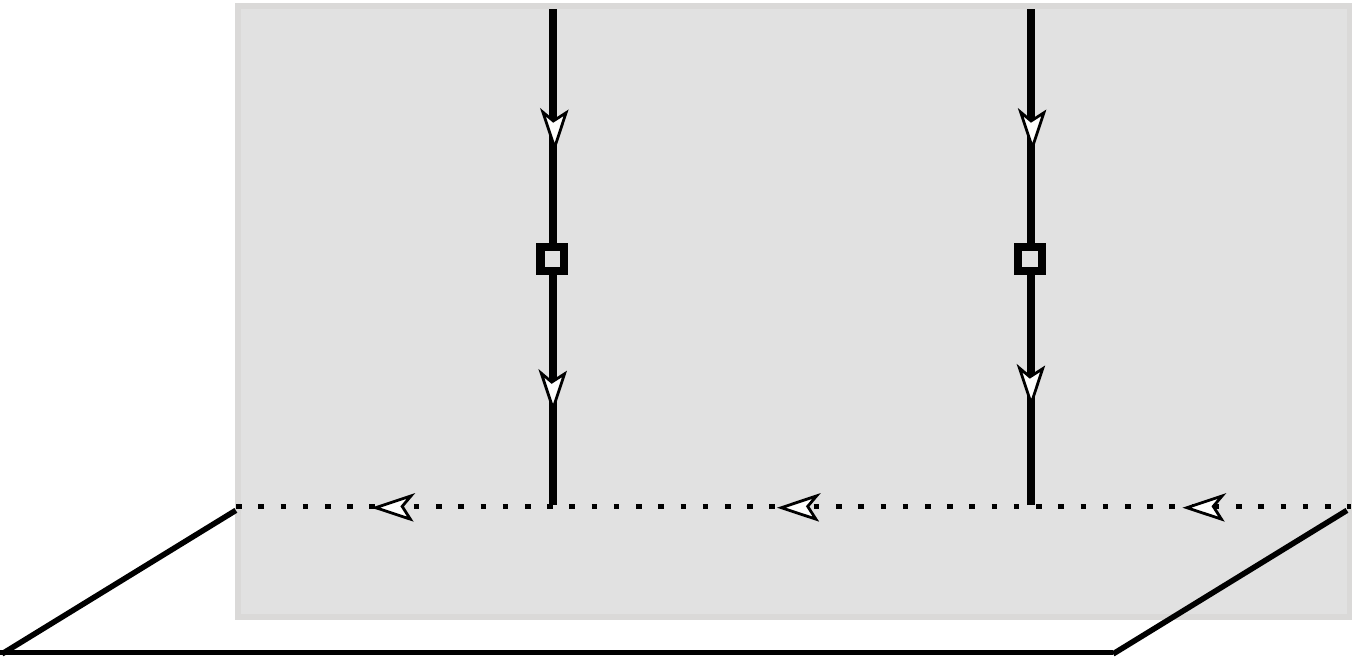}}}
   \put(-10,0){
     \setlength{\unitlength}{.75pt}\put(0,0){
     \put(22,6) {\footnotesize $(\CC,c)$}
     \put(73,37)  {\footnotesize $\CX_1$}
     \put(15,100)  {\footnotesize $(W_1,{}^{\CB_1}\CX_1)$}
     \put(147,37)  {\footnotesize $\CX_2$}
     \put(130,100){\footnotesize $(W_2,{}^{\CB_2}\CX_2)$}
     \put(228,37)  {\footnotesize $\CX_3$}
     \put(240,100) {\footnotesize $(W_3,{}^{\CB_3}\CX_3)$}
     \put(95,20) {\footnotesize $m\in\CM$}
     \put(95,100)  {\footnotesize $\CP$}
     \put(203,100)  {\footnotesize $\CQ$}
     \put(190,20) {\footnotesize $n\in \CN$}
     \put(82,130)  {\footnotesize $(W_{12},X,{}^{\CP}\CM)$}
     \put(175,130)  {\footnotesize $(W_{23},Y,{}^{\CQ}\CN)$}
     \put(63,75)  {\footnotesize $[m,m']_{\CP}$}
     \put(205,75) {\footnotesize $[n,n']_{\CQ}$}
     }\setlength{\unitlength}{1pt}}
  \end{picture}}
  \quad\quad \longrightarrow \quad\quad
 \raisebox{-50pt}{
  \begin{picture}(150,100)
   \put(-10,0){\scalebox{0.5}{\includegraphics{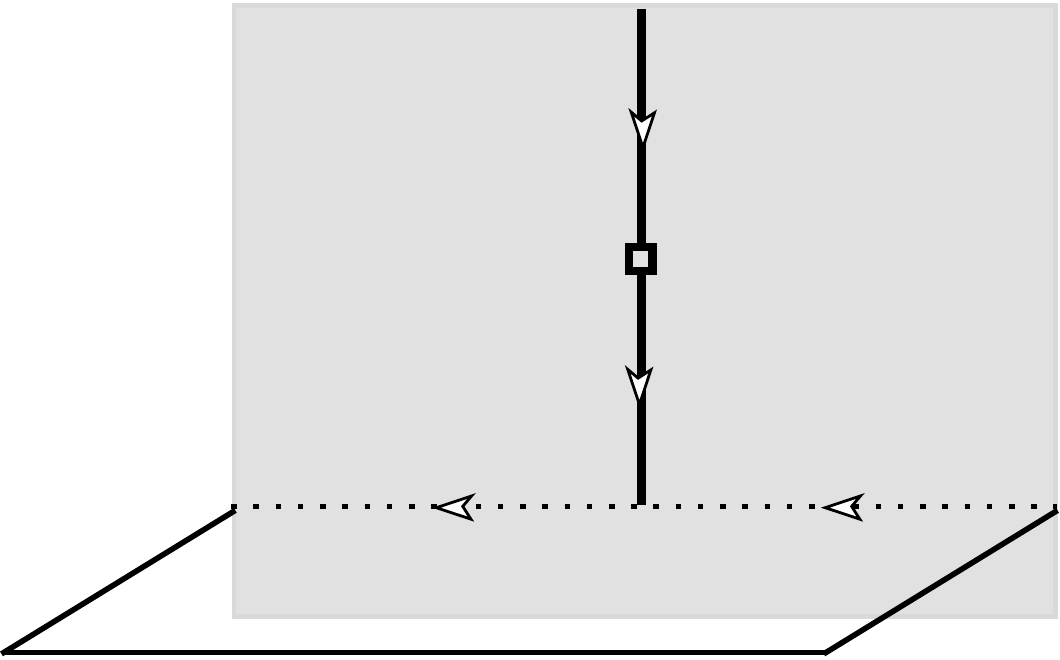}}}
   \put(-10,0){
     \setlength{\unitlength}{.75pt}\put(0,0){
      \put(22,6) {\footnotesize $(\CC,c)$}
     \put(60,37)  {\footnotesize $\CX_1$}
     \put(25,103)  {\footnotesize $(V_1,{}^{\CB_1}\CX_1)$}
     \put(178,37)  {\footnotesize $\CX_3$}
     \put(175,103) {\footnotesize $(V_3,{}^{\CB_3}\CX_3)$}
     \put(105,20) {\footnotesize $m\boxtimes_{\CX_2} n$}
     \put(80,130)  {\footnotesize $(W_{123}, X\boxtimes_{\CB_2}Y, {}^{\CP\boxtimes_{\CB_2}\CQ}(\CM\boxtimes_{\CX_2}\CN))$}
     \put(23,75)  {\footnotesize $[m\boxtimes_{\CX_2} n,m'\boxtimes_{\CX_2} n']$}
     }\setlength{\unitlength}{1pt}}  \end{picture}}
$$
\caption{This picture illustrates the fusion of two 0d gapless walls $(V_{12},X,{}^\CP\CM)$ and $(V_{23},Y,{}^\CQ\CN)$. This fusion is defined by (\ref{eq:fusion-0d-walls}).
}
\label{fig:0d-walls-non-chiral}
\end{figure}

More precisely, we consider a 0+1D non-chiral (resp. chiral) gapless wall $(W_{12},X,{}^\CP\CM)$ between $(W_1,{}^{\CB_1}\CX_1)$ and $(W_2,{}^{\CB_2}\CX_2)$ and a 0+1D non-chiral (resp. chiral) gapless wall $(W_{23},Y,{}^\CP\CM)$ between $(W_2,{}^{\CB_2}\CX_2)$ and $(W_3,{}^{\CB_3}\CX_3)$ as depicted in Figure\,\ref{fig:0d-walls-non-chiral}. Now we assume that all $W_1,W_{12}$, $W_2$,$W_{23}$,$W_3$ are unitary rational full field algebras over $V_L\otimes_\Cb \overline{V}_R$. 
Without loss of generality, we assume $W=W_{12}=W_{23}$. 
In this case, we have the following fusing formula: 
\be \label{eq:fusion-0d-wall-non-chiral}
\left( W,X,{}^\CP\CM \right) \boxtimes_{(W_2,{}^{\CB_2}\CX_2)} \left( W,Y,{}^{\CQ}\CN \right) = \left( W, X\boxtimes_{\CB_2} Y, {}^{\CP\boxtimes_{\CB_2}\CQ} (\CM\boxtimes_{\CX_2}\CN) \right).
\ee
This formula automatically includes (\ref{eq:fusion-0d-walls}) in the chiral cases as special cases. We will give some interesting example of spatial fusions of 0+1D walls in Section\,\ref{sec:gappable-edges-morita}.

\medskip
We want to point out again that the canonical morphism
$$
f: [m,m']_\CP \boxtimes_{\FZ(\CX_2)} [n,n']_\CQ \to [m\boxtimes_{\CX_2} n,m'\boxtimes_{\CX_2} n']
$$
is not an isomorphism in general when the 2d bulk $(\CC,c)$ is non-trivial. This failure of being an isomorphism is called spatial fusion anomaly, which reflects the fact that the edge is an anomalous 1+1D phase. On the other hand, when $(\CC,c)=(\bh,0)$, we should expect that $f$ is an isomorphism because the 1+1D edge is anomaly-free now. This result is proved in \cite{kyz}.

\begin{rem}
The vanishing of the spatial fusion anomaly when $(\CC,c)=(\bh,0)$ implies the functoriality of the full center, a special case of which was proved in \cite{dkr}. Moreover, one can show that this full center functor is fully faithful \cite{kyz}. This results generalize many earlier results in boundary-bulk RCFT's \cite{ffrs, kr1,dkr}, and 
provides a complete mathematical description of boundary-bulk duality in RCFT's. 
\end{rem}

Let us consider another spatial fusion of 0+1D gapless walls. Let $V$ be a unitary rational VOA. Let $W$ be a unitary rational full field algebra over $V\otimes_\Cb \overline{V}$, i.e. a simple symmetric separable $\dagger$-Frobenius algebra in $\FZ(\Mod_V)=\Mod_V\boxtimes\overline{\Mod_V}$. Let 
$$
\CX_1 = (\Cb,{}^\bh\CX_1), \quad (V\otimes_\Cb \overline{V},{}^{\FZ(\Mod_V)}\CX_2), \quad (W,{}^{\Mod_W}\CX_3)
$$
be a 1d gapped edge and two non-chiral 1d gapless edges, respectively, of the same 2d bulk. 
\bnu
\item For a proper $\CM$, the triple $(V,V,{}^{\Mod_V}\CM)$ defines a 0+1D gapless wall between $\CX_1$ and $(V\otimes_\Cb \overline{V},{}^{\FZ(\Mod_V)}\CX_2)$; 
\item For a proper $\CN$, the triple $(V\otimes_\Cb \overline{V}, W, {}^{\FZ(\Mod_V)_W}\CN)$ defines a 0+1D gapless wall between
$(V\otimes_\Cb \overline{V},{}^{\FZ(\Mod_V)}\CX_2)$ and $(W,{}^{\Mod_W}\CX_3)$. 
\enu
The spatial fusion of these two 0+1D walls produces a 0+1D gapless wall between $\CX_1$ and $(W,{}^{\Mod_W}\CX_3)$. It is defined by 
$$
(V,V,{}^{\Mod_V}\CM) \boxtimes_{(V\otimes_\Cb \overline{V},{}^{\FZ(\Mod_V)}\CX_2)} (V\otimes_\Cb \overline{V}, W, {}^{\FZ(\Mod_V)_W}\CN) = (V, \forget(W), {}^{\RMod_W(\Mod_V)}\CM\boxtimes_{\CX_2}\CN), 
$$
where we have used the fact that $V\boxtimes_{\FZ(\Mod_V)} W\simeq \forget(W) \in \RMod_W(\Mod_V)$ and 
$$
\Mod_V \boxtimes_{\FZ(\Mod_V)} \, \FZ(\Mod_V)_W \simeq \RMod_W(\Mod_V)
$$
as closed right multi-fusion $\Mod_W$-modules \cite[Theorem\,3.20]{dmno}. Note that such obtained 0+1D wall is precisely one of those given in Theorem${}^{\mathrm{ph}}$\,\ref{pthm:chiral-wall-non-chiral-edges}.

\void{
\begin{figure} 
$$
 \raisebox{-50pt}{
  \begin{picture}(180,100)
   \put(-10,0){\scalebox{0.5}{\includegraphics{pic-two-0d-wall-CFT.eps}}}
   \put(-43,-5){
     \setlength{\unitlength}{.75pt}\put(0,0){
     \put(73,37)  {\footnotesize $\CX_1$}
     \put(15,100)  {\footnotesize $(W_1,{}^{\FZ(\CX_1)}\CX_1)$}
     \put(147,37)  {\footnotesize $\CX_2$}
     \put(123,100){\footnotesize $(W_2,{}^{\FZ(\CX_2)}\CX_2)$}
     \put(228,37)  {\footnotesize $\CX_3$}
     \put(240,100) {\footnotesize $(W_3,{}^{\FZ(\CX_3)}\CX_3)$}
     \put(95,20) {\footnotesize $m\in\CM$}
     \put(95,100)  {\footnotesize $\CP$}
     \put(203,100)  {\footnotesize $\CQ$}
     \put(190,20) {\footnotesize $n\in \CN$}
     \put(92,130)  {\footnotesize $(W_{12},{}^{\CP}\CM)$}
     \put(185,130)  {\footnotesize $(W_{23},{}^{\CQ}\CN)$}
     \put(63,75)  {\footnotesize $[m,m']_{\CP}$}
     \put(205,75) {\footnotesize $[n,n']_{\CQ}$}
     }\setlength{\unitlength}{1pt}}
  \end{picture}}
$$
\caption{This picture depicts three non-chiral gapless edges $(V_i,{}^{\FZ(\CX_i)}\CX_i)$, $i=1,2,3$ of the trivial 2d topological order, and two 0d gapless walls $(V_{12},{}^\CP\CM)$ and $(V_{23},{}^\CQ\CN)$. 
}
\label{fig:0d-walls-bulk-CFT}
\end{figure}
}

\section{Computing physical processes} \label{sec:compute}
A gapless edge is gappable if it shares the same bulk with a gapped edge. Mathematically, by Theorem\,\ref{thm:morita}, a non-chiral gapless edge $(W, {}^\CB\CX)$ is gappable if and only if the enriched multi-fusion category ${}^\CB\CX$ is spatially Morita equivalent to a UFC. We will illustrate this phenomenon by examples. 

\subsection{Shrinking and gapping a gapless hole} \label{sec:gappable-edges-morita}

\begin{figure} 
$$
 \raisebox{-50pt}{
  \begin{picture}(150,140)
   \put(-95,7){\scalebox{0.7}{\includegraphics{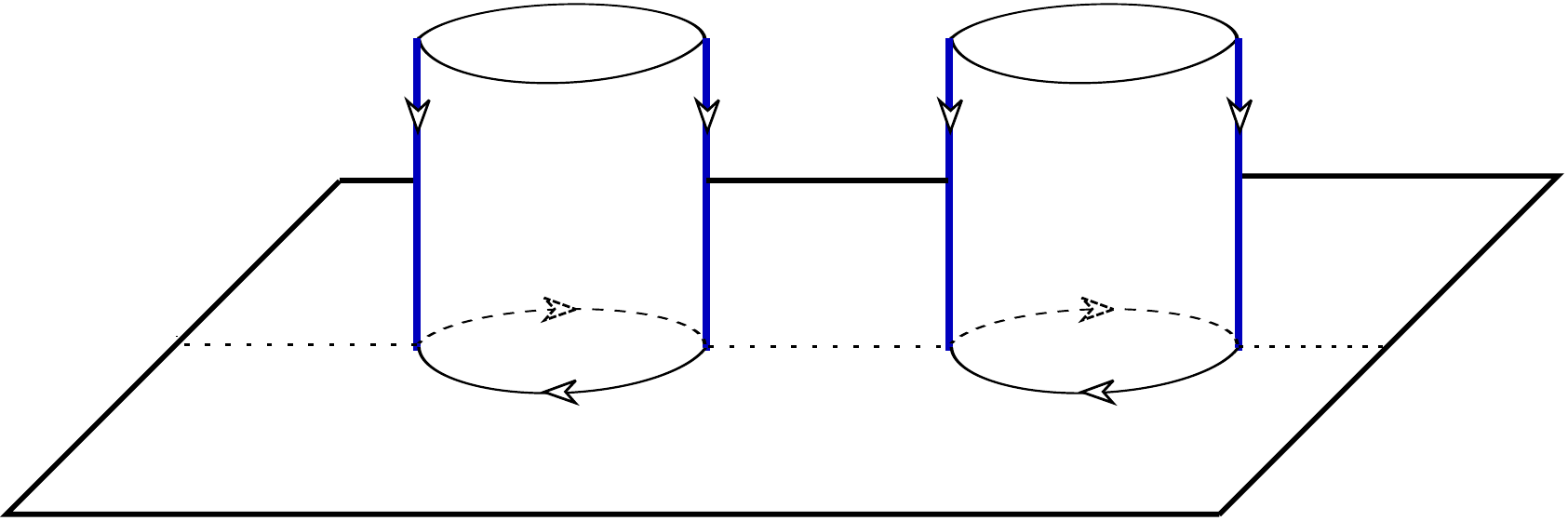}}}
   \put(-2,7){
     \setlength{\unitlength}{.75pt}\put(0,0){
     \put(33,24)   {\footnotesize$\CX$}
     \put(33,67)   {\footnotesize $\CY$}
     \put(27,47)   {\footnotesize $(\CD,c')$}
     \put(190,24) {\footnotesize$\CX'$}
     \put(190,67)   {\footnotesize $\CY'$}
     \put(185,47)   {\footnotesize $(\CD,c')$}
     
     \put(87,115) {\footnotesize $\CP_2$}
     \put(130,115) {\footnotesize $\CP_3$}
     \put(-20,115) {\footnotesize $\CP_1$}
     \put(242,115) {\footnotesize $\CP_4$}
     
     \put(288,88)  {\footnotesize$(\CC,c)$}
     \put(114,52)   {\footnotesize$\CC$}
     
    \put(170,115) {\footnotesize $(V_{\CA'},{}^{\CA'}\CX')$}
    \put(170, 155) {\footnotesize $(V_{\CB'},{}^{\CB'}\CY')$}     
     \put(20,115)  {\footnotesize $(V_\CA,{}^{\CA}\CX)$}
     \put(20, 155) {\footnotesize $(V_\CB,{}^{\CB}\CY)$}
     
     \put(83,80)   {\footnotesize $(V_2,X_2{}^{\CP_2}\CM_2)$}
     \put(-65,145)   {\footnotesize $(V_1,X_1,{}^{\CP_1}\CM_1)$}   
     \put(236,145)   {\footnotesize $(V_4,X_4,{}^{\CP_4}\CM_4)$}
     \put(90,145) {\footnotesize $(V_3,X_3,{}^{\CP_3}\CM_3)$}    
    
     \put(-70,10) {\footnotesize $(\CC,c)$}
     \put(82,40)  {\footnotesize $\CM_2$}
     \put(237,40)  {\footnotesize $\CM_4$}
     \put(137,40) {\footnotesize $\CM_3$}
     \put(-13,40) {\footnotesize $\CM_1$}
     


     }\setlength{\unitlength}{1pt}}
  \end{picture}}
$$
\caption{This picture depicts a 2d topological order $(\CC,c)$ with two holes filled with $(\CD,c')$ and two gapless walls on the boundaries of two filled holes. Two cylinders in the picture depict the 1+1D world sheet of two gapless walls. On each cylinder, there are two chiral gapless walls separated by two 0+1D gapless walls. 
}
\label{fig:two-cylinders}
\end{figure}

Consider the physical configuration depicted in Figure\,\ref{fig:two-cylinders}. It depicts a 2d topological order $(\CC,c)$ with two holes filled with the same 2d topological order $(\CD,c')$. On the boundary of the left hole in Figure\,\ref{fig:two-cylinders}, there are two chiral gapless edges $(V_{\CA},{}^{\CA}\CX)$ and $(V_\CB,{}^{\CB}\CY)$, separated by two 0d gapless walls $(V_i,X_i,{}^{\CP_i}\CM_i)$ for $i=1,2$. The boundary of the right filled hole is similar. 

One can also view Figure\,\ref{fig:two-cylinders} as a configuration for five 1d walls between two 2d topological orders $(\CC,c)$ and $(\CC,c)$. These five walls include three trivial gapped walls $\CC=(\Cb,{}^\bh\CC)$ and two non-chiral gapless walls defined by 
\be \label{eq:non-chiral-wall-1}
(V_\CA,{}^\CA\CX) \boxtimes_{(\CD,c')}  (\overline{V}_\CB,{}^{\overline{\CB}}\CY^\rev)
=(V_\CA\otimes_\Cb \overline{V}_\CB,\, {}^{\CA\boxtimes \overline{\CB}}(\CX \boxtimes_{\overline{\CD}} \CY^\rev)) 
\ee
and 
\be \label{eq:non-chiral-wall-2}
(V_{\CA'},{}^{\CA'}\CX') \boxtimes_{(\CD,c')}  (\overline{V}_{\CB'},{}^{\overline{\CB'}}(\CY')^\rev)
=(V_{\CA'}\otimes_\Cb \overline{V}_{\CB'}, \, {}^{\CA'\boxtimes \overline{\CB'}}(\CX' \boxtimes_{\overline{\CD}} (\CY')^\rev)), 
\ee
respectively. They are separated by four 0+1D gapless walls, which are defined by $(V_i,X_i,{}^{\CP_i}\CM_i)$ for $i=1,2,3,4$.

In this subsection, we study how to shrink and gap out the left gapless hole in Figure\,\ref{fig:two-cylinders}. We will use spatial equivalence seriously. For this reason, there is no need keep track of chiral symmetries $V_i, X_i$ for $i=1,2,3,4$. We simply abbreviate $(V_i,X_i,{}^{\CP_i}\CM_i)$ to ${}^{\CP_i}\CM_i$ for $i=1,2,3,4$. 

\medskip
Physically, we know that if we shrink the left hole to a point. The spectrum of the edge modes become gapped in this limit because tunneling effects or backscattering process between two sides of the hole becoming local operators as the size of the hole getting small. Remember that our mathematical description of the gapless edge only works in the thermodynamics limit and in the long wave length limit. Both limits break down if we shrink the hole to a point.

Mathematically, we can fuse the first and the second 0+1D gapless walls along the non-chiral gapless wall defined in Eq.\,(\ref{eq:non-chiral-wall-1}). This mathematical fusion is, however, completely independent of the size of the hole. 
As a consequence, such obtained 0+1D wall remains gapless after the naive mathematical fusion. The mathematical structure that characterizes the gapping-out process is precisely the spatial equivalence. More precisely, we have 
\begin{align}
{}^{\CP_1}\CM_1 \boxtimes_{{}^{\CA\boxtimes \overline{\CB}}(\CX \boxtimes_{\overline{\CD}} \CY^\rev)} {}^{\CP_2}\CM_2 
&\simeq {}^{\CP_1\boxtimes_{\CA\boxtimes\overline{\CB}}\CP_2}(\CM_1\boxtimes_{\CX\boxtimes_{\overline{\CD}}\CY^\rev}\CM_2) \nn
&\simeq {}^{\fun_\bh(\CR,\CR)}(\CM_1\boxtimes_{\CX\boxtimes_{\overline{\CD}}\CY^\rev}\CM_2) \nn
& \ssimeq {}^\bh(\CR^{\mathrm{op}}\boxtimes_{\fun_\bh(\CR,\CR)}(\CM_1\boxtimes_{\CX\boxtimes_{\overline{\CD}}\CY^\rev}\CM_2)) \nn
& \simeq (\Cb, {}^\bh\CC),  \label{eq:fusing-12-walls}
\end{align}
where all the four steps are explained below. 
\bnu
\item The first ``$\simeq$'' is obvious. 

\item In the second ``$\simeq$'', since $\CP_1$ and $\CP_2$ are both closed $\CA\boxtimes\overline{\CB}$-modules, there is a unique finite unitary category $\CR$ such that $\CP_1\boxtimes_{\CA\boxtimes\overline{\CB}}\CP_2 \simeq \fun_\bh(\CR,\CR)$ as UMFC's. 

\begin{figure} 
$$
 \raisebox{-50pt}{
  \begin{picture}(270,80)
   \put(45,7){\scalebox{0.6}{\includegraphics{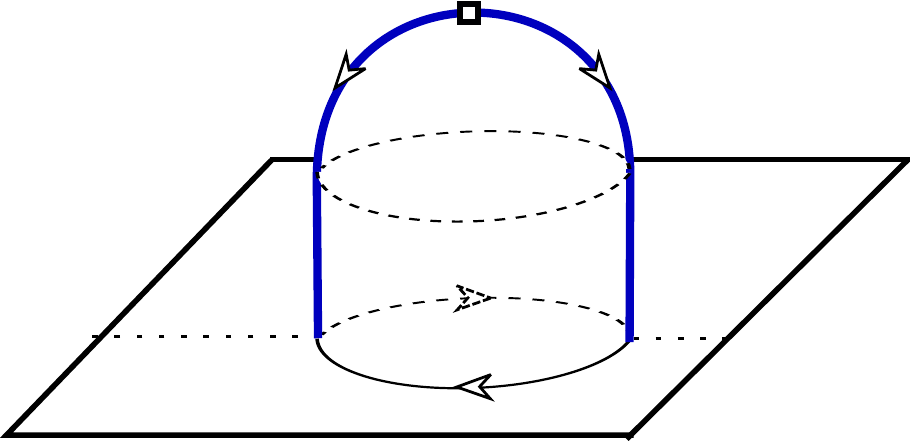}}}
   \put(48,7){
     \setlength{\unitlength}{.75pt}\put(90,0){

     
     
     \put(0,17)  {\footnotesize $\CX$}
     \put(-5,56) {\footnotesize $(\CA,c)$}
     \put(10, 106) {\footnotesize $\CR^{\mathrm{op}}$}
     
     \put(25,38)   {\footnotesize $\CY$}
     \put(15,76) {\footnotesize $(\CB,c)$}
     \put(-30,85)   {\footnotesize $\CP_1$}   
     \put(50,85) {\footnotesize $\CP_2$}
     \put(-35,15) {\footnotesize $\CM_1$}
     \put(43,10){\footnotesize $\CM_2$}
    
     \put(-70,10) {\footnotesize $(\CC,c)$}
 


     }\setlength{\unitlength}{1pt}}
  \end{picture}}
$$
\caption{This picture depicts how to gapped out a 0+1D non-chiral gapless wall mathematically by a spatial equivalence, 
i.e. by inserting an invertible 0D wall $\CR^\op$ to the world line. 
}
\label{fig:cylinders-cap}
\end{figure}

\item In the spatial equivalence``$\ssimeq$'', since $\CR$ is an invertible $\fun_\bh(\CR,\CR)$-$\bh$-bimodule and $\CR^{\mathrm{op}}$ is an invertible $\bh$-$\fun_\bh(\CR,\CR)$-bimodule, then the pair $(\CR^{\mathrm{op}},\id)$, where 
$$
\id: \CR^\op\boxtimes_{\fun_\bh(\CR,\CR)}(\CM_1\boxtimes_{\CX\boxtimes_{\overline{\CD}}\CY^\rev}\CM_2)) \to \CR^\op \boxtimes_{\fun_\bh(\CR,\CR)}(\CM_1\boxtimes_{\CX\boxtimes_{\overline{\CD}}\CY^\rev}\CM_2))
$$
is the identity functor, defines a spatial equivalence 
$$
{}^{\fun_\bh(\CR,\CR)}(\CM_1\boxtimes_{\CX\boxtimes_{\overline{\CD}}\CY^\rev}\CM_2)\simeq 
{}^\bh(\CR^{\mathrm{op}}\boxtimes_{\fun_\bh(\CR,\CR)}(\CM_1\boxtimes_{\CX\boxtimes_{\overline{\CD}}\CY^\rev}\CM_2)
$$ 
as ${}^\bh\CC$-${}^\bh\CC$-bimodules. The physical meaning of this spatial equivalence is illustrated in Figure\,\ref{fig:cylinders-cap}. 

\item In the last ``$\simeq$'', we have used the fact that 
$$
\CR^{\mathrm{op}} \boxtimes_{\fun_\bh(\CR,\CR)}(\CM_1\boxtimes_{\CX\boxtimes_{\overline{\CD}}\CY^\rev}\CM_2) \simeq \CC
$$ 
as unitary categories. We prove this fact below. To compute $\CR^{\mathrm{op}} \boxtimes_{\fun_\bh(\CR,\CR)}(\CM_1\boxtimes_{\CX\boxtimes_{\overline{\CD}}\CY^\rev}\CM_2)$ amounts to push the whole ``cap'' down to the horizontal plane. This produces a 0d defect in $(\CC,c)$ (recall Example\,\ref{expl:ai}). Since this process preserves the anomaly-free condition by Theorem\,\ref{thm:ai}, the resulting anomaly-free 0d defect has to be given by $(\CC,x)$ for some object $x\in \CC$. In particular, we obtain an equivalence of unitary categories: 
\begin{align*}
\CR^{\mathrm{op}} \boxtimes_{\fun_\bh(\CR,\CR)}(\CM_1\boxtimes_{\CX\boxtimes_{\overline{\CD}}\CY^\rev}\CM_2) &\xrightarrow{\simeq} \CC \\
r \boxtimes_{\fun_\bh(\CR,\CR)} (m_1 \boxtimes_{\CX\boxtimes_{\overline{\CD}}\CY^\rev} m_2) &\mapsto x 
\end{align*} 
where $x$ depends on the choices $r\in\CR$, $m_1\in\CM_1$ and $m_2\in\CM_1$. 

Actually, to describe a particular gapping-out process, $m_1$ and $m_2$ need to be fixed as an initial data. $\CR$ is uniquely fixed by the anomaly-free condition. Therefore, the gapping-out process is completely determined by a choice of an object $r\in\CR$. The particle $x$ obtained after the gapping-out process is uniquely determined by $r$. 
\enu

In a summary, we have shown that a fusion of two 0+1D gapless walls produces a 0+1D gappable gapless wall, and the gapping-out process is determined by an instanton, i.e. a pair $(\CR,r)$, where $\CR$ is uniquely fixed and $r$ is an object in $\CR$.

\subsection{Fusing two gapless filled holes} \label{sec:fusion-holes}

In this subsection, we study how to fuse the second and the third 0+1D gapless walls along the trivial gapped wall $\CC$ in Figure\,\ref{fig:two-cylinders}. We claim that this fusion produces a 0+1D gapless wall 
\be \label{eq:fusing-23-walls}
{}^{\CP_2}\CM_2 \boxtimes_{{}^\bh\CC} {}^{\CP_3}\CM_3 \ssimeq {}^\CS\CS_0 \boxtimes_{\overline{\CD}} {}^\CT\CT_0 = {}^{\CS\boxtimes \CT}(\CS_0\boxtimes_{\overline{\CD}}\CT_0),
\ee
as illustrated in the following picture.
\be \label{fig:saddle-result}
 \raisebox{-50pt}{
  \begin{picture}(270,105)
   \put(15,7){\scalebox{0.6}{\includegraphics{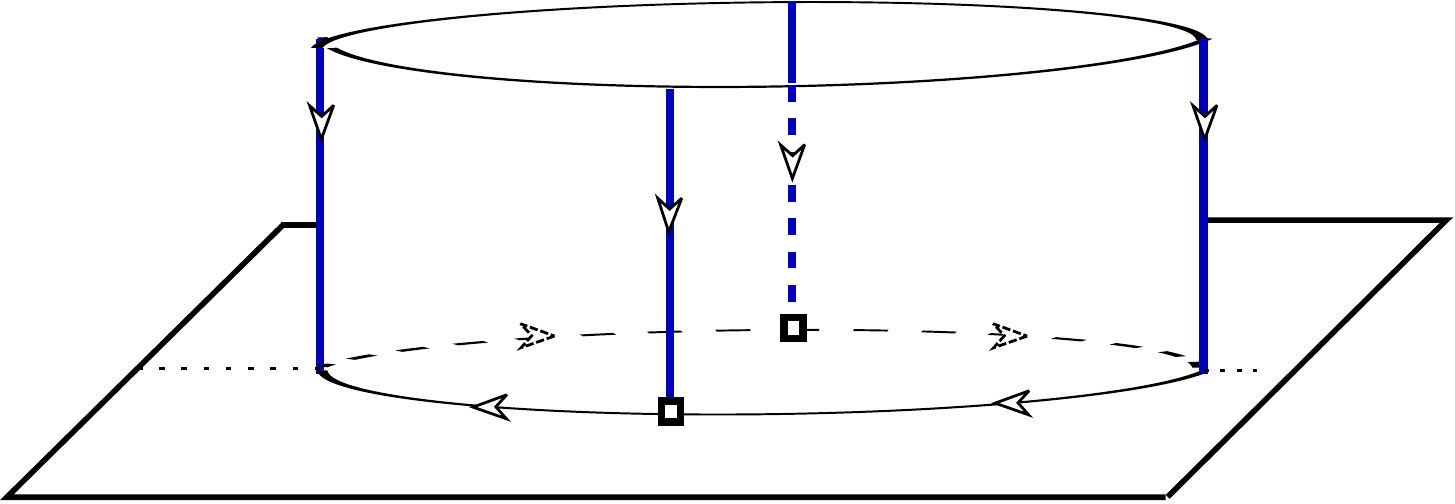}}}
   \put(18,12){
     \setlength{\unitlength}{.75pt}\put(90,0){
     \put(65,115) {\footnotesize $(U_2,Y_2,{}^\CT\CT_0)$}
     \put(30,95) {\footnotesize $(U_1,Y_1,{}^\CS\CS_0)$}
     
     \put(203,48)  {\footnotesize$(\CC,c)$}
     \put(86,20)   {\footnotesize$\CT_0$}
     \put(58,2)   {\footnotesize$\CS_0$}
     
    \put(120,75) {\footnotesize $(V_{\CA'},{}^{\CA'}\CX')$}
    \put(130, 113) {\footnotesize $(V_{\CB'},{}^{\CB'}\CY')$}     
     \put(0,75)  {\footnotesize $(V_\CA,{}^{\CA}\CX)$}
     \put(-10, 112) {\footnotesize $(V_\CB,{}^{\CB}\CY)$}
     
     \put(50,56)   {\footnotesize $\CS$}
     \put(-90,80)   {\footnotesize $(V_1, X_1,{}^{\CP_1}\CM_1)$}   
     \put(191,80)   {\footnotesize $(V_4,X_4,{}^{\CP_4}\CM_4)$}
     \put(95,70) {\footnotesize $\CT$}    
    
     \put(-75,0) {\footnotesize $(\CC,c)$}
     \put(30,20) {\tiny $(\CD,c')$}
 


     }\setlength{\unitlength}{1pt}}
  \end{picture}}
\ee
We will prove the formula (\ref{eq:fusing-23-walls}) by the following equivalences:
\begin{align}  \label{eq:saddle-0}
{}^{\CP_2}\CM_2 \boxtimes_{{}^\bh\CC} {}^{\CP_3}\CM_3 &\simeq {}^{\CP_2\boxtimes \CP_3}(\CM_2\boxtimes_\CC \CM_3)) \nn
&\ssimeq {}^{\CS\boxtimes \CT}(\CO\boxtimes_{\CP_2\boxtimes \CP_3} (\CM_2\boxtimes_\CC \CM_3) )) \nn
&\simeq {}^{\CS\boxtimes \CT} (\CS_0\boxtimes_{\overline{\CD}} \CT_0)), 
\end{align}
which will be explained below. 
\bnu

\begin{figure}[htbp] 
$$
 \raisebox{-50pt}{
  \begin{picture}(250,95)
   \put(10,7){\scalebox{0.6}{\includegraphics{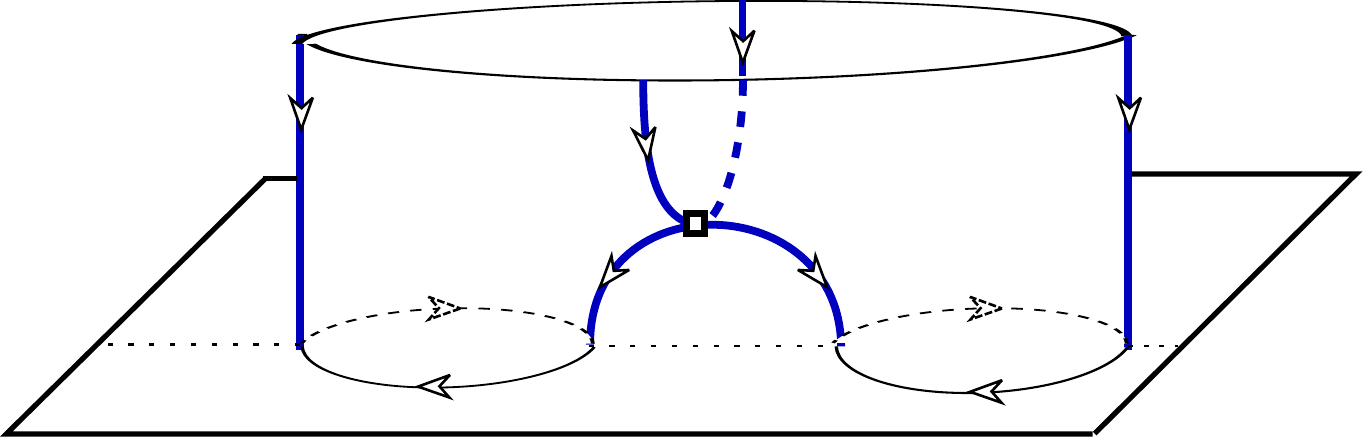}}}
   \put(5,5){
     \setlength{\unitlength}{.75pt}\put(90,0){
     \put(94,95) {\footnotesize $\CT$}
     \put(55,70) {\footnotesize $\CS$}
     
     \put(195,53)  {\footnotesize$(\CC,c)$}
     \put(78,20)   {\footnotesize$\CC$}
     \put(74,40)   {\footnotesize$\CO$}
     
    \put(150,80) {\footnotesize $\CA'$}
    
    \put(-14,8)   {\footnotesize $\CX$}
    \put(17,39)   {\footnotesize $\CY$}
         
     \put(5,80)  {\footnotesize $\CA$}
     \put(17, 93) {\footnotesize $\CB$}
     \put(135, 94) {\footnotesize $\CB'$}
     
     \put(41,40)   {\footnotesize $\CP_2$}
     \put(52,13)   {\footnotesize $\CM_2$}
     \put(98,13)   {\footnotesize $\CM_3$}
     \put(164,6)   {\footnotesize $\CX'$}
     \put(138,39)   {\footnotesize $\CY'$}
     \put(110,40) {\footnotesize $\CP_3$}    
    
     \put(-65,8) {\footnotesize $(\CC,c)$}
     \put(5,21) {\tiny $(\CD,c')$}
     \put(140,21) {\tiny $(\CD,c')$}
 


     }\setlength{\unitlength}{1pt}}
  \end{picture}}
\quad\quad\quad
\raisebox{-50pt}{
  \begin{picture}(150,95)
   \put(5,10){\scalebox{0.6}{\includegraphics{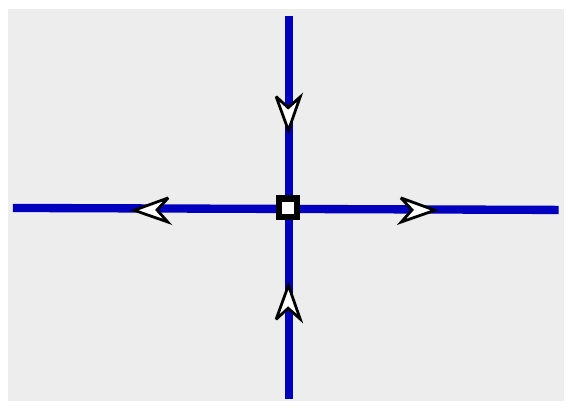}}}
   \put(8,10){
     \setlength{\unitlength}{.75pt}\put(0,0){
     \put(27,34)   {\footnotesize$\CP_2$}
     \put(52,67)   {\footnotesize $\CT$}
     \put(90,34) {\footnotesize$\CP_3$}
     \put(52,20)   {\footnotesize $\CS$}
     \put(66,37)  {\footnotesize $\CO$}
     \put(3,10) {\footnotesize $\CA$}
     \put(3,80) {\footnotesize $\CB$}
     \put(105,10) {\footnotesize $\CA'$}
     \put(105,80) {\footnotesize $\CB'$}
 }\setlength{\unitlength}{1pt}}
  \end{picture}}  
$$
$$
(a) \quad\quad\quad\quad\quad\quad\quad\quad\quad\quad\quad\quad\quad\quad\quad\quad\quad\quad\quad\quad (b)
$$
\caption{This picture depicts a fusion of two holes (with gapless edges) inside a 2d topological order $(\CC,c)$. 
}
\label{fig:saddle}
\end{figure}

\item ``$\ssimeq$'': First, consider Figure\,\ref{fig:saddle} (b). Since $\CA,\overline{\CB}, \CA', \overline{\CB'}$ are all Witt equivalent to $\CC$, there is a gapped wall, defined by a UFC $\CS$, between $\CA$ and $\CA'$, and a gapped wall, defined by a UFC $\CT$, between $\CB$ and $\CB'$. Then all these data $\CA,\CB,\CA',\CB', \CS,\CT$ determines a unique (up to equivalences) finite unitary category $\CO$, which defines an anomaly-free 0d defect in Figure\,\ref{fig:saddle} (b). We can rearrange the neighborhood of this 0D defect $\CO$ looks like the saddle point depicted in Figure\,\ref{fig:saddle} (a). In particular, $\CO$ is a $(\CS\boxtimes \CT)$-$(\CP_2\boxtimes\CP_3)$-bimodule. Therefore, we obtain 
$$
{}^{\CP_2\boxtimes \CP_3}(\CM_2\boxtimes_\CC \CM_3) \ssimeq {}^{\CS\boxtimes \CT}(\CO\boxtimes_{\CP_2\boxtimes \CP_3} (\CM_2\boxtimes_\CC \CM_3)), 
$$
which implies the ``$\ssimeq$'' in (\ref{eq:saddle-0}). 

\item The last ``$\simeq$'': To compute $\CO\boxtimes_{\CP_2\boxtimes \CP_3} (\CM_2\boxtimes_\CC \CM_3)$ amounts to squeeze the part below the saddle point (see (\ref{eq:six-squares})) to a point-like anomaly-free defect, which is uniquely determined by its environment. By the uniqueness, it is enough to show that $\CS_0\boxtimes \CT_0$ is a solution to the anomaly-free condition. Since $\CS$ is a gapped wall between $(\CA,c)$ and $(\CA',c)$, $\CX$ is a gapped wall between $(\CA,c)$ and $(\CC\boxtimes\overline{\CD},c-c')$ and $\CX'$ is gapped wall between $(\CA',c)$ and $(\CC\boxtimes\overline{\CD},c-c')$. These is a unique 0D defect $S_0$, which is uniquely determined by $\CA,\CX,\CC,\CD,\CA',\CX',\CS$, as shown in the picture in (\ref{fig:saddle-result}). Similarly, we obtain another 0D defect $\CT_0$, which is uniquely determined by $\CB,\CY,\CC,\CD,\CB',\CY',\CT$. As a consequence, $\CS_0\boxtimes_{\overline{\CD}} \CT_0$ is a solution, i.e. 
\be \label{eq:s0-t0}
\CO\boxtimes_{\CP_2\boxtimes \CP_3} (\CM_2\boxtimes_\CC \CM_3) \simeq  \CS_0\boxtimes_{\overline{\CD}} \CT_0.
\ee

This argument via anomaly-free condition might looks mysterious. We would like to provide a more physical proof. We first illustrated the part below the saddle point labeled by $\CO$ by the following picture.
\be \label{eq:six-squares}
\raisebox{-37pt}{
  \begin{picture}(150,95)
   \put(5,10){\scalebox{0.7}{\includegraphics{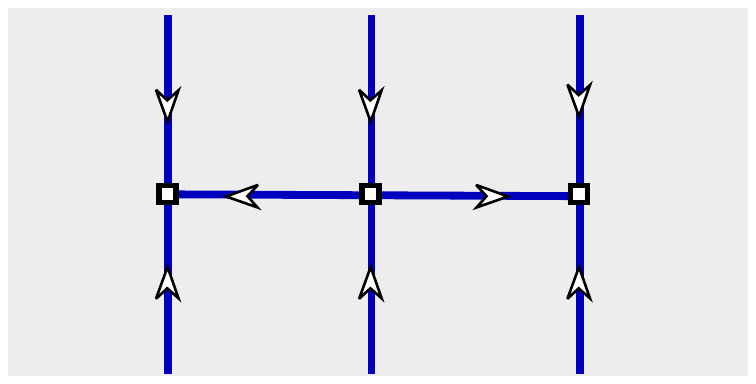}}}
   \put(8,10){
     \setlength{\unitlength}{.75pt}\put(0,0){
     \put(0,88)  {\footnotesize$\CD$}
     \put(22,48) {\footnotesize$\CM_2$}
     \put(29,23) {\footnotesize$\CX$}
      \put(30,73) {\footnotesize$\CY$}
     \put(187,88) {\footnotesize$\CD$}
     \put(158,48) {\footnotesize$\CM_3$}
     \put(158,23){\footnotesize$\CX'$}
     \put(158,73){\footnotesize$\CY'$}
     \put(57,57)   {\footnotesize $\CP_2\boxtimes\CC$}
     \put(66,73)   {\footnotesize $\CT\boxtimes\CC$}
     \put(112,57) {\footnotesize$\CP_3\boxtimes\CC$}
     \put(66,23)   {\footnotesize $\CS\boxtimes\CC$}
     \put(100,39)  {\footnotesize $\CO\boxtimes\CC$}
     \put(57,5) {\footnotesize $\CA\boxtimes\CC$}
     \put(57,92) {\footnotesize $\CB\boxtimes\CC$}
     \put(112,5) {\footnotesize $\CA'\boxtimes\CC$}
     \put(112,92) {\footnotesize $\CB'\boxtimes\CC$}
 }\setlength{\unitlength}{1pt}}
  \end{picture}}  
\ee
If we squeeze it horizontally, we obtain a 1d gapped wall between $(\CD,c')$ and $(\CD,c')$ as illustrated in (\ref{fig:saddle-1d-wall}), in which we have pushed 0d gapped wall $\CM_2\boxtimes_{(\CP_2\boxtimes\CC)^\rev}(\CO\boxtimes\CC) \boxtimes_{\CP_3\boxtimes \CC} \CM_3$ on this 1d gapped wall to one of its two end points. 
\begin{figure}[htbp] 
$$
 \raisebox{-50pt}{
  \begin{picture}(270,105)
   \put(15,7){\scalebox{0.6}{\includegraphics{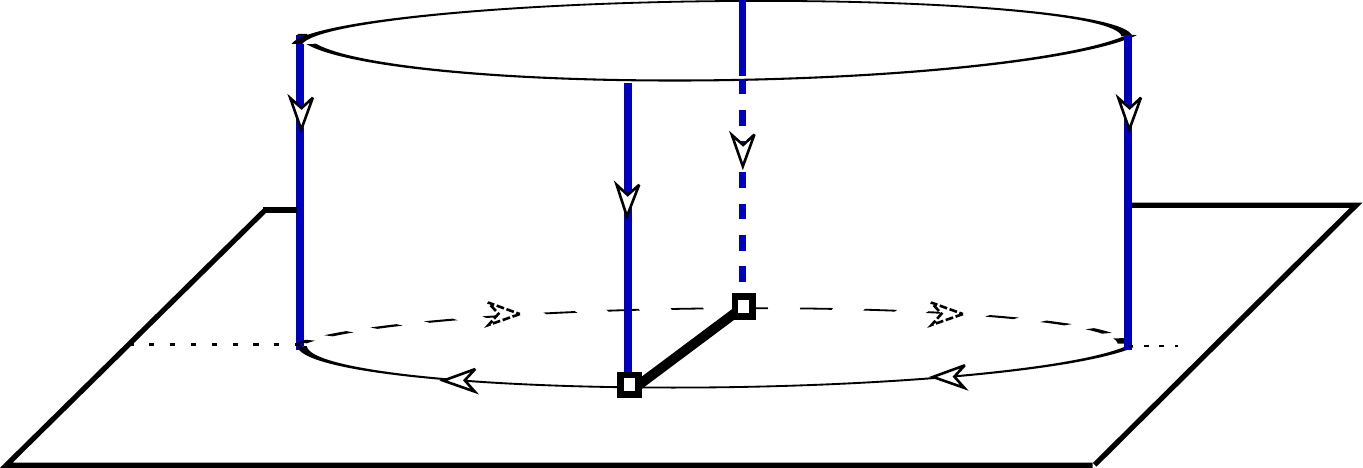}}}
   \put(10,8){
     \setlength{\unitlength}{.75pt}\put(90,0){
     \put(65,112) {\footnotesize $(U_2,Y_2,{}^\CT\CT_0)$}
     \put(28,92) {\footnotesize $(U_1,Y_1,{}^\CS\CS_0)$}
     
     \put(194,48)  {\footnotesize$(\CC,c)$}
     
    \put(115,65) {\footnotesize $(V_{\CA'},{}^{\CA'}\CX')$}
    \put(130, 110) {\footnotesize $(V_{\CB'},{}^{\CB'}\CY')$}     
     \put(0,65)  {\footnotesize $(V_\CA,{}^{\CA}\CX)$}
     \put(-10, 110) {\footnotesize $(V_\CB,{}^{\CB}\CY)$}
     
     \put(50,56)   {\footnotesize $\CS$}
     \put(-90,80)   {\footnotesize $(V_1, X_1,{}^{\CP_1}\CM_1)$}   
     \put(185,80)   {\footnotesize $(V_4,X_4,{}^{\CP_4}\CM_4)$}
     \put(95,70) {\footnotesize $\CT$}    
    
     \put(-72,3) {\footnotesize $(\CC,c)$}
     
     \put(83,23) {\tiny $\fun_\CD(\CK,\CK)$}
     \put(27,24) {\tiny $(\CD,c')$}
     \put(140,26) {\tiny $(\CD,c')$}
 


     }\setlength{\unitlength}{1pt}}
  \end{picture}}
$$
\caption{This picture depicts a intermediate step in the process of fusing of two filled holes (filled with $(\CD,c')$ and a 1d gapless wall) inside a 2d topological order $(\CC,c)$. 
}
\label{fig:saddle-1d-wall}
\end{figure}
Then this 1d gapped wall can be described by the UFC $\fun_\CD(\CK,\CK)$ for a right $\CD$-module $\CK$. By $\fun_\CD(\CK,\CK)\simeq \CK^\op\boxtimes_\CD \CK$, we know that we can cut this 1d wall according to the first two picture in (\ref{fig:split-wall}).
\be \label{fig:split-wall}
\raisebox{-20pt}{
  \begin{picture}(80,50)
   \put(0,5){\scalebox{0.6}{\includegraphics{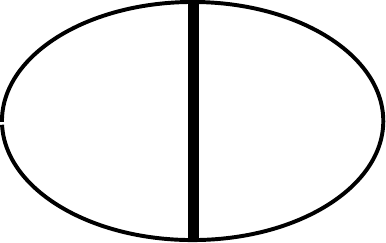}}}
   \put(0,5){
     \setlength{\unitlength}{.75pt}\put(0,0){
          
     \put(50,30)  {\tiny $\fun_\CD(\CK,\CK)$}
     \put(55,15)  {\tiny $(\CD,c')$}
     \put(15,15)  {\tiny $(\CD,c')$}
     
     }\setlength{\unitlength}{1pt}}
  \end{picture}}
\quad \rightsquigarrow \quad
 \raisebox{-20pt}{
  \begin{picture}(80,50)
   \put(0,5){\scalebox{0.6}{\includegraphics{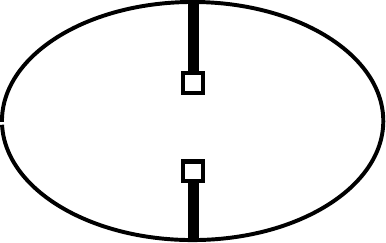}}}
   \put(0,5){
     \setlength{\unitlength}{.75pt}\put(0,0){
   
     \put(50,35)  {\scriptsize $\CK^\op$}
     \put(50,12)  {\scriptsize $\CK$}
     
     \put(10,25)  {\scriptsize $(\CD,c')$}
     
     }\setlength{\unitlength}{1pt}}
  \end{picture}}
\quad \rightsquigarrow \quad
 \raisebox{-20pt}{
  \begin{picture}(80,50)
   \put(0,5){\scalebox{0.6}{\includegraphics{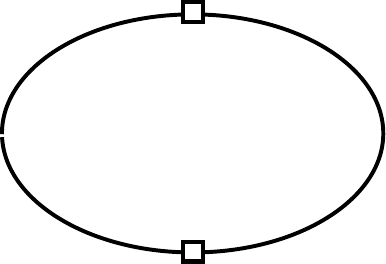}}}
   \put(0,5){
     \setlength{\unitlength}{.75pt}\put(0,0){
     
     \put(41,45) {\scriptsize $\CT_0$}
     \put(41,10) {\scriptsize $\CS_0$}
     
     \put(10,28)  {\scriptsize $(\CD,c')$}
     
     }\setlength{\unitlength}{1pt}}
  \end{picture}}
\ee
This cutting produces two end points labeled by $\CK$ and $\CK^\op$, respectively. Then absorbing these two ends to two sides of the hole, we obtain a single filled hole as depicted in the third picture. 

\enu

\medskip
We consider a special case $\CA'=\CA$, $\CB'=\CB$, $\CX'=\CX$, $\CY'=\CY$, $\CP_1=\CP_2^\rev=\CP_3$, $\CM_1=\CM_2^\op=\CM_3$ and $V_i=V_j$ and $\CM_2=\CM_4$. In this case, the formula (\ref{eq:fusing-12-walls}) and (\ref{eq:fusing-23-walls}) become
\begin{align}
({}^{\CP_2^\rev}\CM_2^{\mathrm{op}}) \boxtimes_{{}^{\CA\boxtimes \overline{\CB}}(\CX \boxtimes_{\overline{\CD}} \CY^\rev)} ({}^{\CP_2}\CM_2) \simeq {}^{\fun_\bh(\CR,\CR)}(\CM_1\boxtimes_{\CX\boxtimes_{\overline{\CD}}\CY^\rev}\CM_1^\op) &\ssimeq (\Cb,{}^\bh\CC), \label{eq:morita-1} \\
({}^{\CP_2}\CM_2) \boxtimes_{{}^\bh\CC} ({}^{\CP_2^\rev}\CM_2^{\mathrm{op}}) &\ssimeq 
{}^{\CA\boxtimes\overline{\CB}}(\CX\boxtimes_{\overline{\CD}}\CY^\rev). 
\label{eq:morita-2}
\end{align}
These two spatial equivalences simply say that ${}^{\CP_2}\CM_2$ is a spatially invertible ${}^{\CA\boxtimes\overline{\CB}}(\CX\boxtimes_{\overline{\CD}}\CY^\rev)$-$\CC$-bimodule and defines a spatial Morita equivalence between ${}^{\CA\boxtimes\overline{\CB}}(\CX\boxtimes_{\overline{\CD}}\CY^\rev)$ and $\CC$. Therefore, the gappability of the non-chiral 1d gapless wall $(V_\CA\otimes_\Cb \overline{V}_{\CB}, {}^{\CA\boxtimes\overline{\CB}}(\CX\boxtimes_{\overline{\CD}}\CY^\rev))$ is precisely captured by the fact that ${}^{\CA\boxtimes\overline{\CB}}(\CX\boxtimes_{\overline{\CD}}\CY^\rev)$ and $\CC$ are spatially Morita equivalent.

\begin{rem} \label{rem:equivalence-two-scenarios}
Figure\,\ref{fig:saddle} (a) show the spatial equivalence between two 0+1D gapless walls. The condition that all UMTC's $\CA,\CA',\CB,\CB'$ are all Witt equivalent automatically holds in the situation depicted in Figure\,\ref{fig:two-cylinders}, and plays a crucial role in the proof. Indeed, the Witt equivalence of $\CA,\CA',\CB,\CB'$ implies that we can form Figure\,\ref{fig:saddle} (b), which further implies that Figure\,\ref{fig:saddle} (a) is physically realizable.
\end{rem}

\begin{rem}
We can certainly fill two holes by two different 2d topological orders $(\CD,c')$ and $(\CE,c'')$. The same arguments again lead us to Figure\,\ref{fig:saddle-1d-wall} or the first picture in (\ref{fig:split-wall}). But the spliting the 1d gapped wall is not possible any more because there should always be some non-trivial 1d wall separating $(\CD,c')$ and $(\CE,c'')$. 
\end{rem}

\begin{rem}
When $\CA=\CB=\CA'=\CB'=\CP_1=\CP_3=\CP_2^\rev=\CP_4^\rev$ and $\CX=\CX'=\CY=\CY'=\CM_1=\CM_2^\op=\CM_3=\CM_4^\op$.
The spatial fusion of two filled holes endows the filled hole with an algebraic structure. This leads to a generalized anyon condensation theory, which will be developed elsewhere. 
\end{rem}

\subsection{Dimensional reduction to boundary-bulk CFT's} \label{sec:cft}
Recall that we have used in \cite[Section\,3.3]{kz4} a dimensional reduction process to prove the appearance of boundary CFT's on a chiral gapless edge based on a ``No-Go Theorem'' as depicted in \cite[Figure\,5]{kz4}. In this subsection, using the precisely mathematical description of chiral gapless edges and their 0+1D gapless walls, we are able to compute this dimensional reduction process precisey. In particular, we will work out explicitly which boundary-bulk CFT is produced by this dimensional reduction process. It turns out that all boundary-bulk RCFT's can be obtained in this way (first announced in \cite{kz3}).

\begin{figure} 
$$
 \raisebox{-50pt}{
  \begin{picture}(150,100)
   \put(-10,5){\scalebox{0.5}{\includegraphics{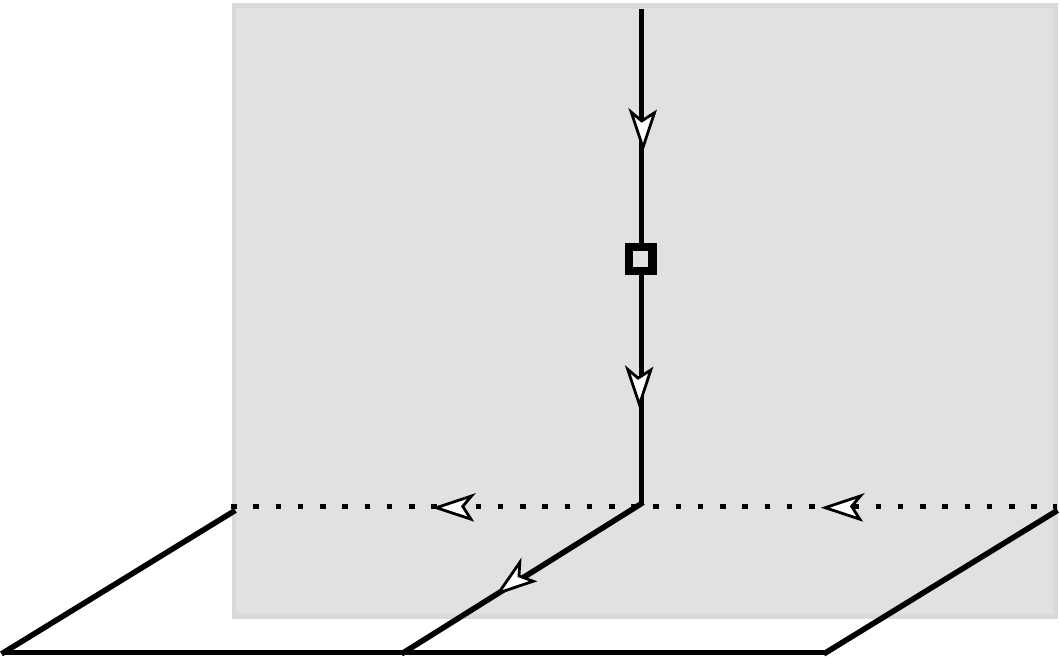}}}
   \put(-10,5){
     \setlength{\unitlength}{.75pt}\put(0,0){
      \put(22,6) {\footnotesize $(\CC,c)$}
      \put(142,6) {\footnotesize $(\CC,c)$}
     \put(85,34)  {\footnotesize $\CC$}
     \put(55,110)  {\footnotesize $(V,{}^{\CC}\CC)$}
     \put(100,103)  {\footnotesize $\CC_{B|B}$}
     \put(160,34)  {\footnotesize $\CC$}
          \put(155,110) {\footnotesize $(V,{}^{\CC}\CC)$}
     \put(121,20) {\footnotesize $x\in\CC_{B|A}$}
     \put(100,6) {\footnotesize $\CC_{A|A}$}
     \put(110,130)  {\footnotesize $(V,B,{}^{\CC_{B|B}}\CC_{B|A}$)}
     \put(75,75)  {\footnotesize $[x,x']_{\CC_{B|B}}$}  
     }\setlength{\unitlength}{1pt}}  \end{picture}}
 \quad\quad\quad\quad
  \raisebox{-50pt}{
  \begin{picture}(150,100)
   \put(-10,5){\scalebox{0.5}{\includegraphics{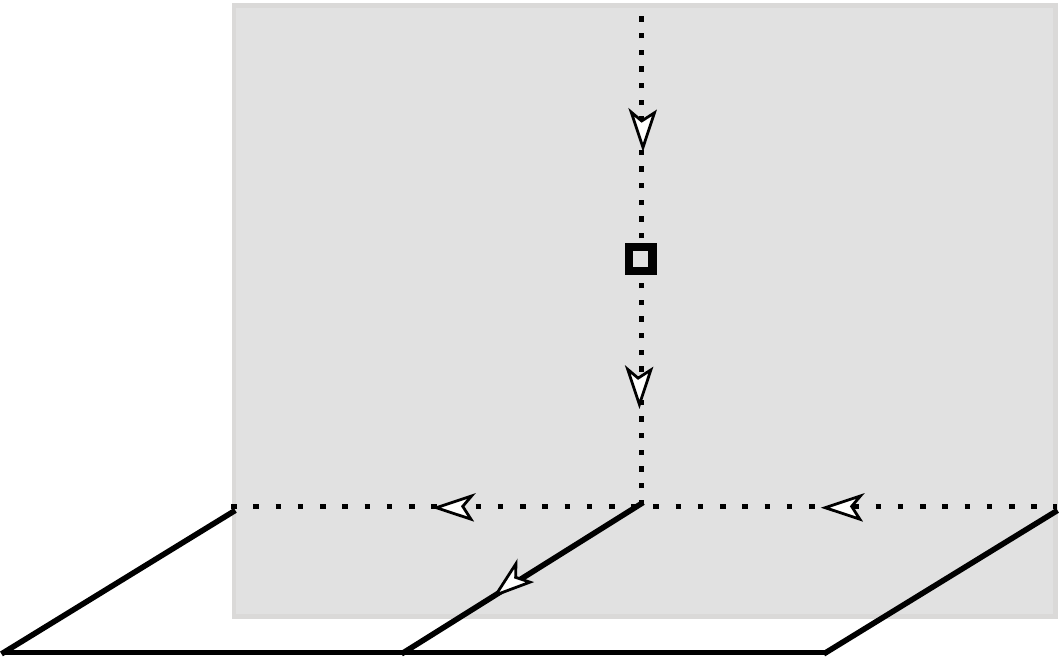}}}
   \put(-10,5){
     \setlength{\unitlength}{.75pt}\put(0,0){
      \put(22,6) {\footnotesize $(\CC,c)$}
      \put(142,6) {\footnotesize $(\CC,c)$}
     \put(85,34)  {\footnotesize $\CC$}
     \put(55,113)  {\footnotesize $(V,{}^{\CC}\CC)$}
     \put(112,100)  {\footnotesize $\CC$}
     \put(160,34)  {\footnotesize $\CC$}
          \put(155,113) {\footnotesize $(V,{}^{\CC}\CC)$}
     \put(121,20) {\footnotesize $x\in \CC_A$}
     \put(100,6) {\footnotesize $\CC_{A|A}$}
     \put(110,130)  {\footnotesize $(V,V,{}^{\CC}\CC_A$)}
     \put(85,75)  {\footnotesize $[x,x']_\CC$}  
     }\setlength{\unitlength}{1pt}}  \end{picture}}    
$$
$$
(a) \quad\quad\quad\quad\quad\quad\quad\quad\quad\quad\quad\quad\quad\quad\quad\quad\quad\quad (b)
$$
\caption{Picture (a) depicts a gapped wall between 2d topological order $(\CC,c)$ and $(\CC,c)$ and its gapless boundary $(V,{}^{\CC_{B|B}}\CC_{B|A})$; Picture (b) depicts a special case when $B=\one_\CC$. 
}
\label{fig:0d-defects-example}
\end{figure}

\medskip
Let us consider the situation depicted in Figure\,\ref{fig:0d-defects-example} (a). Let $\CC=\Mod_V$ be a UMTC. Let $A$ and $B$ be two $\dagger$-SSSFA's in $\CC$. Then the category $\CC_{A|A}$ of $A$-$A$-bimodules in $\CC$ and $\CC_{B|B}$ are UFC's and define two 1d gapped walls between two 2d topological orders $(\CC,c)$ and $(\CC,c)$. The category $\CC_{B|A}$ defines a 0d gapped wall between $\CC_{B|B}$ and $\CC_{A|A}$. By a topological wick rotation, we obtain the canonical gapless edge $(V,{}^\CC\CC)$ of $(\CC,c)$ and a 0+1D gapless relative boundary of the 1d gapped wall $\CC_{A|A}$, defined by $(V,B,{}^{\CC_{B|B}}\CC_{B|A})$, which is also a 0+1D gapless wall between two canonical 1d gapless edges $(V,{}^\CC\CC)$ and $(V,{}^\CC\CC)$. Figure\,\ref{fig:0d-defects-example} (b) depicts a special case when $B=\one_\CC$. When $x=x'=A\in \CC_A$, we have $[x,x']_\CC=A$.

\begin{figure} 
$$
 \raisebox{-70pt}{
  \begin{picture}(60,150)
   \put(-30,10){\scalebox{0.5}{\includegraphics{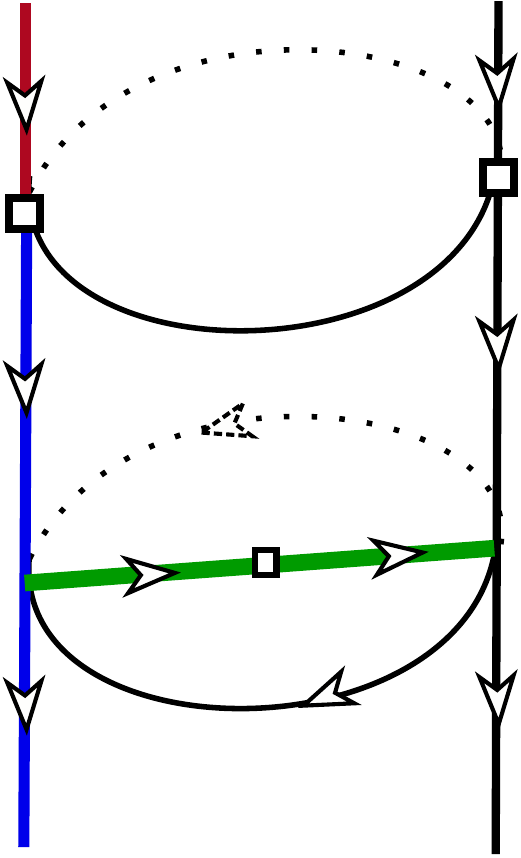}}}
   \put(-30,10){
     \setlength{\unitlength}{.75pt}\put(0,0){
     \put(-40,122)  {\footnotesize $[x,y]_{\CC_{B|B}}$}
     \put(-35,50)  {\footnotesize $\CC_{B|A}\ni x$}
     \put(-37,20)  {\footnotesize $[x,x]_{\CC_{B|B}}$}
     \put(-40,150) {\footnotesize $[y,y]_{\CC_{B|B}}$}
     \put(102,20)  {\footnotesize $[x',x']_{\CC_{B'|B'}}$}
     \put(102,150)  {\footnotesize $[y',y']_{\CC_{B'|B'}}$}
     \put(102,130)  {\footnotesize $[x',y']_{\CC_{B'|B'}}$}
     \put(100, 59)  {\footnotesize $x'\in \CC_{A'|B'}$}
     \put(88, -12) {\footnotesize $(V,(B')^\ast,{}^{\CC_{B'|B'}^\rev}\CC_{A'|B'})$}
     \put(-30, -12) {\footnotesize $(V,B,{}^{\CC_{B|B}}\CC_{B|A})$}
     \put(60, 73)  {\footnotesize $(\CC,c)$}
     \put(28,35)  {\footnotesize$(\CC,c)$}
     \put(15, 64) {\tiny $\CC_{A|A}$}
     \put(40,66)  {\tiny $\CC_{A|A'}$}
     \put(63,49) {\tiny $\CC_{A'|A'}$}
     \put(35,162) {\footnotesize $(V,{}^\CC\CC)$}
     \put(35,15) {\footnotesize $(\overline{V}, {}^{\overline{\CC}}\CC^\rev)$}
     }\setlength{\unitlength}{1pt}}
  \end{picture}}
\quad \xrightarrow{\mbox{\footnotesize dimensional reduction}} \quad\quad
 \raisebox{-70pt}{
  \begin{picture}(90,150)
   \put(0,10){\scalebox{0.5}{\includegraphics{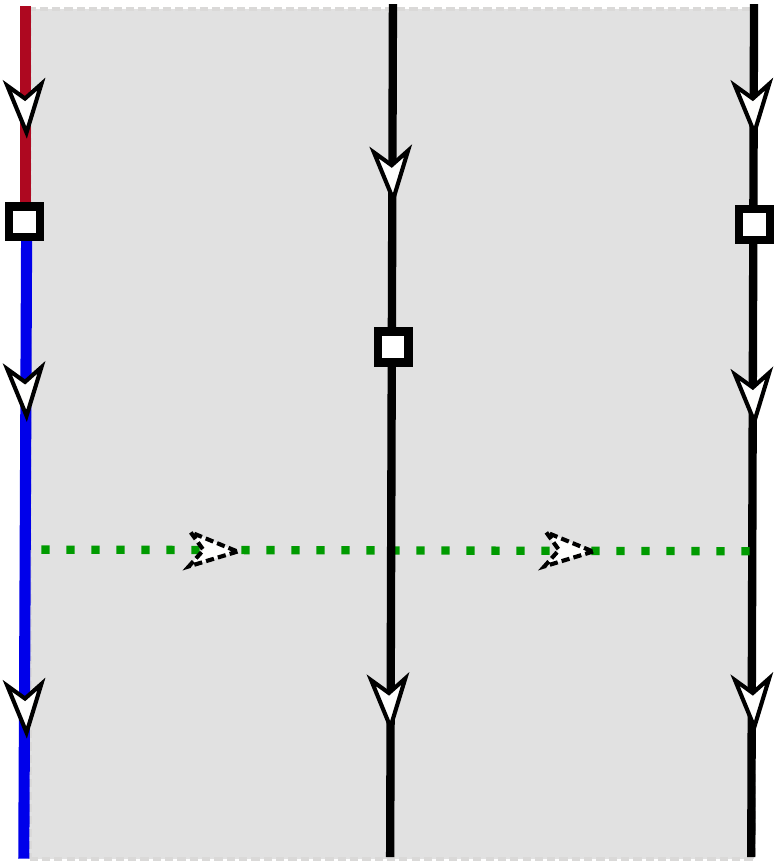}}}
   \put(0,10){
     \setlength{\unitlength}{.75pt}\put(0,0){
     \put(-40,122)  {\footnotesize $[x,y]_{\CC_{B|B}}$}
     \put(-35,58)  {\footnotesize $\CC_{B|A}\ni x$}
     \put(-35,20)  {\footnotesize $[x,x]_{\CC_{B|B}}$}
     \put(-38,145)  {\footnotesize $[y,y]_{\CC_{B|B}}$}
     \put(152,145)  {\footnotesize $[y',y']_{\CC_{B'|B'}}$}
     \put(145, -13) {\footnotesize $(V,(B')^\ast,{}^{\CC_{B'|B'}^\rev}\CC_{A'|B'})$}
     \put(-40,-13) {\footnotesize $(V,B,{}^{\CC_{B|B}}\CC_{B|A})$}
     \put(35,-11)  {\tiny $(V\otimes_\Cb \overline{V},V\otimes_\Cb \overline{V}, {}^{\FZ(\CC)}\CC_{A|A'})$}
      \put(-25, 183) {\footnotesize $(V\otimes_\Cb \overline{V},{}^{\FZ(\CC)}\CC_{A|A})$}
       \put(95, 183) {\footnotesize $(V\otimes_\Cb \overline{V},{}^{\FZ(\CC)}\CC_{A'|A'})$}
      \put(30, 170) {\footnotesize $\searrow$}
      \put(110, 170) {\footnotesize $\swarrow$}
     \put(152,122)  {\footnotesize $[x',y']_{\CC_{B'|B'}}$}
     \put(152,20)  {\footnotesize $[x',x']_{\CC_{B'|B'}}$}
     \put(28,98)    {\footnotesize $[a,a']_{\FZ(\CC)}$}
     \put(32,28)    {\footnotesize $[a,a]_{\FZ(\CC)}$}
     \put(150, 59)  {\footnotesize $x'\in \CC_{A'|B'}$}
     \put(30,69)  {\footnotesize $\CC_{A|A}$}
     \put(95,69)  {\footnotesize $\CC_{A'|A'}$}
     \put(20,130) {\footnotesize $[A,A]_{\FZ(\CC)}$}
     \put(85,130) {\footnotesize $[A',A']_{\FZ(\CC)}$}
     \put(90,43)  {\footnotesize $a\in\CC_{A|A'}$}
     \put(78,50)  {\footnotesize $\nwarrow$}
     }\setlength{\unitlength}{1pt}}
  \end{picture}}  
$$
$$
(a) \quad\quad\quad\quad \quad \quad\quad\quad\quad \quad\quad \quad\quad\quad\quad\quad
\quad\quad\quad
\quad\quad\quad\quad  (b)
$$
\caption{These pictures depict physical configurations before and after the process of dimensional reduction. 
}
\label{fig:bcft-2}
\end{figure}

\medskip
Consider the physical configuration depicted in Figure\,\ref{fig:bcft-2} (a). Two 2d topological orders $(\CC,c)$ and $(\CC,c)$ are separated by two 1d gapped walls, which are defined by two UFC's $\CC_{A|A}$ and $\CC_{A'|A'}$ for two $\dagger$-SSSFA's  $A$ and $A'$ in $\CC$. These two 1d gapped walls are separated by a 0d gapped wall defined by a finite unitary category $\CC_{A|A'}$, and they also have 0+1D gapless relative boundaries defined by $(V,B,{}^{\CC_{B|B}}\CC_{B|A})$ and $(V,B',{}^{\CC_{B'|B'}}\CC_{A|B'})$, respectively, where $B$ and $B'$ are $\dagger$-SSSFA's in $\CC$. Moreover, each of these two 0+1D gapless relative boundaries can also be viewed as a 0+1D gapless wall between two canonical chiral gapless edge $(V,{}^\CC\CC)$. $x,y$ are objects in $\CC_{B|A}$ and $x',y'$ are objects in $\CC_{A'|B'}$. Note that we have flipped the orientation of one of the canonical gapless edges and changed the label from $(V,V,{}^\CC\CC)$ to $(\overline{V},\overline{V},{}^{\overline{\CC}}\CC^\rev)$ without altering the physics. 

By the same dimensional reduction process as in \cite[Figure\,5]{kz4}, i.e. fusing of two gapless edges in Figure \,\ref{fig:bcft-2} (a), we obtain the physical configuration in Figure\,\ref{fig:bcft-2} (b). The 1+1D world sheet in Figure\,\ref{fig:bcft-2} (b) contains five parts:
\begin{align*}
(V,B,{}^{\CC_{B|B}}\CC_{B|A}), \quad &(V\otimes_\Cb \overline{V},{}^{\FZ(\CC)}\CC_{A|A}), \quad (V\otimes_\Cb \overline{V},V\otimes_\Cb \overline{V}, {}^{\FZ(\CC)}\CC_{A|A'}), \\
&(V\otimes_\Cb \overline{V},{}^{\FZ(\CC)}\CC_{A'|A'}), \quad (V,(B')^\ast, {}^{\CC_{B'|B'}^\rev}\CC_{A'|B'}),
\end{align*}
where two gapless 0+1D boundaries $(V,B,{}^{\CC_{B|B}}\CC_{B|A})$ and $(V,(B')^\ast,{}^{\CC_{B'|B'}^\rev}\CC_{A'|B'})$ remains the same during the dimensional reduction process, and the remaining three are obtained from the following fusion formula:
\begin{align} \label{eq:C-M-C}
(V, {}^\CC\CC) \boxtimes_{(\CC,c)} (\Cb,{}^\bh\CM) \boxtimes_\CC (\overline{V}, {}^{\overline{\CC}}\CC^\rev)
&= (V\otimes_\Cb \overline{V}, {}^{\CC\boxtimes \overline{\CC}}\CM) 
\end{align}
for $\CM=\CC_{A|A}, \CC_{A|A'}, \CC_{A'|A'}$. We would like to show that the physical configuration depicted in Figure\,\ref{fig:bcft-2} (b) is physically consistent according to the mathematical theory of boundary-bulk RCFT's.

\medskip
The space of non-chiral fields that can live on 1+1D world sheet on the left (resp. right) side is given by internal hom 
$$
[\one_{\CC_{A|A}}, \one_{\CC_{A|A}}]_{\FZ(\CC)}=[A,A]_{\FZ(\CC)} \quad\quad \mbox{(resp. $[A',A']_{\FZ(\CC)}$)},
$$ 
which is nothing but the full center of $A$, i.e. $Z(A) = [\one_{\CC_{A|A}},\one_{\CC_{A|A}}]_{\FZ(\CC)}$ \cite{davydov}.
By results in \cite{frs,kr2,davydov}, this internal hom $[A,A]_{\FZ(\CC)}$ (resp. $[A',A']_{\FZ(\CC)}$) is precisely a modular invariant bulk CFT with a boundary CFT given by $A$ \cite{ffa-mod-inv,kr2}. By \cite[Theorem\,3.4]{kr2}, a modular invariant bulk CFT is equivalent to a Lagrangian algebra in $\FZ(\CC)$. By \cite[Theorem\,1.1]{kr1}, there is a one-to-one correspondence between the set of Morita classes of SSSFA's in $\CC$ that of Lagrangian algebras in $\FZ(\CC)$ defined by $A \mapsto [\one_{\CC_{A|A}},\one_{\CC_{A|A}}]_{\FZ(\CC)}$. 

Internal homs $[x,x]_{\CC_{B|B}}, [x,y]_{\CC_{B|B}}, [y,y]_{\CC_{B|B}}$ for $x\in \CC_{B|A}$ define boundary CFT's and 0D walls. According to mathematical theory of RCFT \cite{frs,kr2}, these boundary CFT's must share a unique bulk given by their full center. Therefore, to show that physical configuration defined in Figure\,\ref{fig:bcft-2} (b) defines consistent boundary-bulk CFT's, it is enough to show that their full center is precisely given by $[A,A]_{\FZ(\CC)}$.

By \cite[Theorem\,3.3.1]{ostrik}, we have $(\CC_{B|B})_{[x,x]} \simeq \CC_{B|A} \simeq (\CC_{B|B})_{[y,y]}$. It implies that $[x,x]_{\CC_{B|B}}$ and $[y,y]_{\CC_{B|B}}$ are Morita equivalent. By \cite{kr1,davydov}, they must share the same full center in $\FZ(\CC_{B|B}) \simeq \FZ(\CC)$. By \cite[Theorem\,7.12.11]{egno}, we have the following monoidal equivalences of UFC's 
$$
(\CC_{B|B})_{[x,x]|[x,x]} \simeq \fun_{\CC_{B|B}}(\CC_{B|A}, \CC_{B|A}) \simeq \CC_{A|A}.
$$ 
In particular, the tensor unit $\one_{(\CC_{B|B})_{[x,x]|[x,x]}}=[x,x]$ is mapped to the tensor unit $\one_{\CC_{A|A}}=A$. By the definition of full center, we obtain 
$$
Z([x,x]) = [\one_{\CC_{A|A}}, \one_{\CC_{A|A}}]_{\FZ(\CC)} = Z(A)
$$ 
for $x\in \CC_{B|A}$. Similarly, the boundary CFT's $[x',x']_{\CC_{B'|B'}}$ for $x'\in \CC_{A'|B'}$ share the same bulk that is given by $Z([x',x']_{\CC_{B'|B'}})\simeq Z(A')=[\one_{\CC_{A'|A'}}, \one_{\CC_{A'|A'}}]_{\FZ(\CC)}$. Therefore, Figure\,\ref{fig:bcft-2} (b) gives consistent boundary-bulk CFT's. 

\begin{rem}
It is very interesting to work out a few special cases of above discussion.  
\bnu
\item When $A=B=\one_\CC$, $\CC_{A|A}=\CC_{B|A}=\CC_{B|B}=\CC$ and $(V, {}^{\CC_{B|B}}\CC_{B|A})=(V,{}^\CC\CC)$.  In this case, $[x,x]_\CC=x\otimes x^\ast$ and $Z([x,x]_\CC)\simeq Z(\one_\CC)=[\one_\CC,\one_\CC]_{\FZ(\CC)}=\oplus_{i\in\mathrm{Irr}(\CC)} i \boxtimes i^\ast$ is the famous charge conjugate modular invariant CFT. 

\item When $A=\one$, we still have $[x,x]_{\CC_{B|B}} = x\otimes x^\ast$ for $x\in \CC_{B|\one_\CC}$. On the one hand, by the definition of full center, we have  $Z([x,x]_{\CC_{B|B}}):=[\one_{(\CC_{B|B})_{[x,x]|[x,x]}}, \one_{(\CC_{B|B})_{[x,x]|[x,x]}}]\simeq [\one_\CC,\one_\CC]_{\FZ(\CC)}$. On the other hand, the $\dagger$-SSSFA $x\otimes x^\ast$ in $\CC_{B|B}$ viewed as boundary CFT's (via the forgetful functor $\forget: \CC_{B|B} \to \CC$) are nothing but those appeared in Cardy case. They share the same bulk (or equivalently, the full center) with the trivial boundary CFT $V=A$.

\item When $B=\one$ and $A$ is not Morita equivalent to $\one_\CC$, for $x\in\CC_A$, $[x,x]_\CC\simeq (x\otimes_A x^\ast)^\ast$ is not Morita equivalent to $\one_\CC$. Instead, $[x,x]_\CC$ is Morita equivalent to $[A,A]_\CC=A$ because $[A,-]_\CC: \CC_A \to \CC$ is the forgetful functor. In this case, the bulk of $[x,x]_\CC$ is a modular invariant bulk CFT different from $Z(\one_\CC)$. By taking $A$ from all Morita classes, we recover all possible modular invariant bulk RCFT's satisfying the $V$-invariant boundary condition. 
\enu
\end{rem}

The observables on the 0+1D world line in the middle of Figure\,\ref{fig:bcft-2} (b) form a triple $(V\otimes_\Cb\overline{V},V\otimes_\Cb\overline{V},{}^{\FZ(\CC)}\CC_{A|A'})$, which defines a 0+1D wall between the bulk CFT $Z(A)$ and $Z(A')$. By the folding trick, non-chiral fields $[a,a]_{\FZ(\CC)}$ for $a\in \CC_{A|A'}$ should be viewed as a boundary CFT of a double layered bulk CFT $Z(A)\boxtimes Z(A') \in \overline{\FZ(\CC)} \boxtimes \FZ(\CC)$. One can prove this by proving that the full center of $[a,a]_{\FZ(\CC)}$ in $\FZ(\FZ(\CC))=\overline{\FZ(\CC)} \boxtimes \FZ(\CC)$ is precisely given by $Z(A)\boxtimes Z(A')$ as shown below.
\bnu

\item It is clear that $\CC_{A|A}^\rev\boxtimes \CC_{A'|A'}$ are Morita equivalent to $\FZ(\CC)$, and the Morita equivalence is defined by the invertible $(\CC_{A|A}\boxtimes \CC_{A'|A'}^\rev)$-$\FZ(\CC)$-bimodule $\CC_{A|A'}$. 

\item Then we obtain a monoidal equivalence $\FZ(\CC)_{[a,a]|[a,a]} \simeq (\CC_{A|A}\boxtimes \CC_{A'|A'}^\rev)$. Therefore, we obtain
\begin{align}
Z([a,a]_{\FZ(\CC)}) &:=[\one_{(\FZ(\CC))_{[a,a]|[a,a]}}, \one_{(\FZ(\CC))_{[a,a]|[a,a]}}]_{\overline{\FZ(\CC)} \boxtimes \FZ(\CC)} \nn
&\simeq [\one_{\CC_{A|A}}\boxtimes \one_{\CC_{A'|A'}}, \one_{\CC_{A|A}}\boxtimes \one_{\CC_{A'|A'}}]_{\overline{\FZ(\CC)} \boxtimes \FZ(\CC)} = Z(A) \boxtimes Z(A'). \nonumber
\end{align}
\enu
Therefore, we have shown that the physical configuration in Figure\,\ref{fig:bcft-2} (b) gives physically consistent wall-boundary-bulk RCFT's.


\begin{rem}
We have seen that the bulk CFT $[A,A]_{\FZ(\CC)}$ is independent of the choice of $B$, and $[x,y]_{\CC_{B|B}}$ on the 0+1D gapless wall $(V,{}^{\CC_{B|B}}\CC_{B|A})$ are all consistent with the same bulk CFT $[A,A]_{\FZ(\CC)}$. Moreover, when $[x,y]_{\CC_{B|B}}$ is viewed as a  wall between boundary CFT's, it is physically indistinguishable with an object in $\CC$. In other words, one can identify $[x,y]_{\CC_{B|B}}$ with an object in $\CC$ via the forgetful functor $\forget: \CC_{B|B} \to \CC$.
\end{rem}

\begin{rem}
We have proved the claims in \cite[Section\,3.3,Figure\,5]{kz4}. Moreover, above computation of dimensional reduction also provides a rigorous proof and a non-trivial generalization of the physical results and claims in \cite{cz,levin}. 
\end{rem}

\begin{figure}
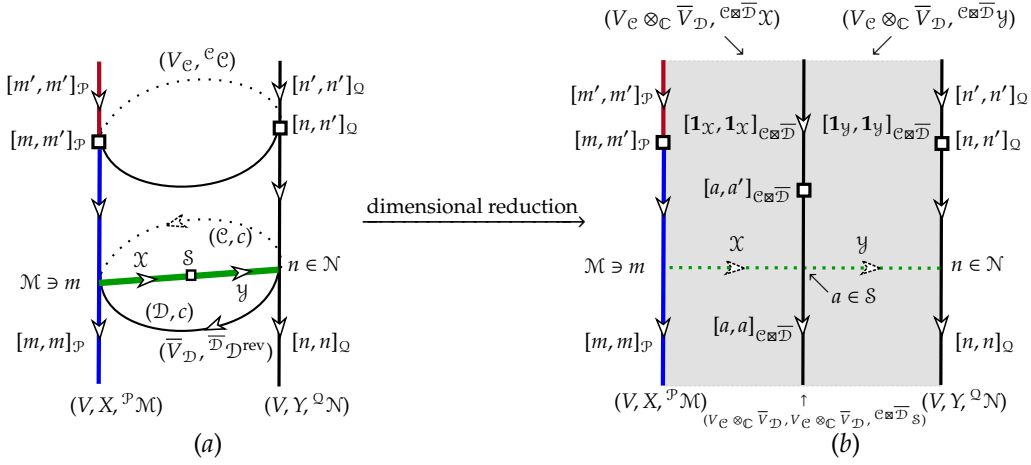

$$
 \raisebox{-70pt}{
  \begin{picture}(60,130)
   \put(-30,10){\scalebox{0.5}{\includegraphics{pic-fh-bcft-1-eps-converted-to.pdf}}}
   \put(-30,10){
     \setlength{\unitlength}{.75pt}\put(0,0){
     \put(-40,122)  {\footnotesize $[m,m']_\CP$}
     \put(-35,50)  {\footnotesize $\CM\ni m$}
     \put(-37,20)  {\footnotesize $[m,m]_\CP$}
     \put(-42,150) {\footnotesize $[m',m']_\CP$}
     \put(102,20)  {\footnotesize $[n,n]_\CQ$}
     \put(102,150)  {\footnotesize $[n',n']_\CQ$}
     \put(102,130)  {\footnotesize $[n,n']_\CQ$}
     \put(100, 59)  {\footnotesize $n\in \CN$}
     \put(88, -12) {\footnotesize $(V,Y,{}^\CQ\CN)$}
     \put(-10, -12) {\footnotesize $(V,X,{}^\CP\CM)$}
     \put(60, 73)  {\footnotesize $(\CC,c)$}
     \put(28,35)  {\footnotesize$(\CD,c)$}
     \put(22, 61) {\footnotesize $\CX$}
     \put(47,62)  {\footnotesize $\CS$}
     \put(74,45) {\footnotesize $\CY$}
     \put(35,162) {\footnotesize $(V_\CC,{}^\CC\CC)$}
     \put(35,15) {\footnotesize $(\overline{V}_\CD, {}^{\overline{\CD}}\CD^\rev)$}
     }\setlength{\unitlength}{1pt}}
  \end{picture}}
\quad \xrightarrow{\mbox{\footnotesize dimensional reduction}} \quad\quad
 \raisebox{-70pt}{
  \begin{picture}(90,130)
   \put(0,10){\scalebox{0.5}{\includegraphics{pic-fh-bcft-2-eps-converted-to.pdf}}}
   \put(0,10){
     \setlength{\unitlength}{.75pt}\put(0,0){
     \put(-40,122)  {\footnotesize $[m,m']_\CP$}
     \put(-35,58)  {\footnotesize $\CM\ni m$}
     \put(-35,20)  {\footnotesize $[m,m]_\CP$}
     \put(-42,145)  {\footnotesize $[m',m']_\CP$}
     \put(152,145)  {\footnotesize $[n',n']_\CQ$}
     \put(135, -10) {\footnotesize $(V,Y,{}^\CQ\CN)$}
     \put(-20,-10) {\footnotesize $(V,X,{}^\CP\CM)$}
     \put(25,-18)  {\tiny $(V_\CC\otimes_\Cb \overline{V}_\CD, V_\CC\otimes_\Cb \overline{V}_\CD, {}^{\CC\boxtimes \overline{\CD}}\CS)$}
     \put(74,-8) {\tiny $\uparrow$}
      \put(-25, 183) {\footnotesize $(V_\CC\otimes_\Cb \overline{V}_\CD,{}^{\CC\boxtimes \overline{\CD}}\CX)$}
      \put(35, 170) {\footnotesize $\searrow$}
      \put(110, 170) {\footnotesize $\swarrow$}
      \put(95, 183) {\footnotesize $(V_\CC\otimes_\Cb \overline{V}_\CD,{}^{\CC\boxtimes \overline{\CD}}\CY)$} 
     \put(152,122)  {\footnotesize $[n,n']_\CQ$}
     \put(152,20)  {\footnotesize $[n,n]_\CQ$}
     \put(25,98)    {\footnotesize $[a,a']_{\CC\boxtimes \overline{\CD}}$}
     \put(30,28)    {\footnotesize $[a,a]_{\CC\boxtimes \overline{\CD}}$}
     \put(150, 59)  {\footnotesize $n\in \CN$}
     \put(38,69)  {\footnotesize $\CX$}
     \put(102,69)  {\footnotesize $\CY$}
     \put(15,130) {\footnotesize $[\one_\CX,\one_\CX]_{\CC\boxtimes \overline{\CD}}$}
     \put(85,130) {\footnotesize $[\one_\CY,\one_\CY]_{\CC\boxtimes \overline{\CD}}$}
     \put(90,43)  {\footnotesize $a\in\CS$}
     \put(78,50)  {\footnotesize $\nwarrow$}
     }\setlength{\unitlength}{1pt}}
  \end{picture}}  
$$
$$
(a) \quad\quad\quad\quad \quad \quad\quad\quad\quad \quad\quad \quad\quad\quad\quad\quad
\quad\quad\quad
\quad\quad\quad\quad  (b)
$$
\caption{These pictures depict physical configurations before and after the process of dimensional reduction.}
\label{fig:bcft-3}
\end{figure}

\medskip
More generally, bulk phases on the two sides of the gapped wall $\CM$ in Figure\,\ref{fig:bcft-2} (a) can be different, say $(\CC,c)$ and $(\CD,c)$ as illustrated in Figure\,\ref{fig:bcft-3} (a). The $V_\CC$ and $V_\CD$ are unitary rational VOA's such that $\CC=\Mod_{V_\CC}$ and $\CD=\Mod_{V_\CD}$ are UMTC's. The UFC's $\CX$ and $\CY$ describe two gapped wall between two topological orders $(\CC,c)$ and $(\CD,c)$. The finite unitary category $\CS$ describes a gapped 0d wall between $\CX$ and $\CY$. Two 0d gapless walls between two canonical chiral gapless edges are necessarily preserve a chiral symmetry $V$ (i.e. a sub-VOA of both $V_\CC$ and $V_\CD$), which will be assumed to be unitary and rational. Then these two 0d walls are given by $(V,X,{}^\CP\CM)$ and $(V,Y,{}^\CQ\CN)$, where 
\bnu
\item For $\CE=:(\Mod_V)_{V_\CD|V_\CC}$ and $\CF:=(\Mod_V)_{V_\CC|V_\CD}$, the UMFC $\CP$ (resp. $\CQ$) is given by $\CE_{X|X}$ (resp. $\CF_{Y|Y}$) for a symmetric special $\dagger$-Frobenius algebra $X \in \CE$ (resp. $Y\in\CF$);
\item $\CP$ (resp. $\CQ$) is a UMFC Morita equivalent to $\CX$ (resp. $\CY$) with the Morita equivalence defined by the invertible bimodule $\CM$ (resp. $\CN$). 
\enu
By similar argument, one can show that Figure\,\ref{fig:bcft-3} (b) give consistent physical configurations. More precisely, we have 
$$
Z([m,m]) \simeq [\one_\CX,\one_\CX]_{\CC\boxtimes\overline{\CD}}, \quad\quad
Z([n,n]) \simeq [\one_\CY,\one_\CY]_{\CC\boxtimes\overline{\CD}}, \quad\quad
Z([a,a]) \simeq [\one_\CX,\one_\CX]_{\CC\boxtimes\overline{\CD}} \boxtimes [\one_\CY,\one_\CY]_{\CC\boxtimes\overline{\CD}}
$$
for $m\in\CM$, $n\in\CN$, $a\in\CS$. 

In summary, we have shown that dimensional reduction processes of 2d topological orders naturally recovers all boundary-bulk RCFT's. Perhaps, a more interesting point of view is that the physics of 2d topological orders provide a physical reconstruction of the entire mathematical theory of wall-boundary-bulk RCFT's \cite{frs,fjfrs,ffrs,kr1,kr2,dkr}. These processes also explain why there are one-to-one correspondences among the following four sets: (1) the set of modular invariant bulk RCFT's \cite{ffa-mod-inv,kr2}, (2) the set of Lagrangian algebras in $\FZ(\CC)$ \cite{dmno}, (3) the set of indecomposable module categories of $\CC$ \cite{ostrik}, (4) the set of monoidal equivalence classes of UFC's that are Morita equivalent to $\CC$ \cite{ostrik,dmno}.

\begin{rem}
As a by-product, we have proved Gapped-gapless Correspondence between the set of all 2+1D anomaly-free non-chiral topological orders and that of all 1+1D anomaly-free boundary-bulk-wall RCFT's (up to the missing chiral and non-chiral symmetries).  
\end{rem}

\section{Conclusions and outlooks} \label{sec:outlooks}
In this work and \cite{kz4}, we have developed the mathematical theory of gapped/gapless edges of 2d topological orders and higher codimensional gapped/gapless defects based on enriched (multi-)fusion categories and their representations. In this section, we discuss a few lessons we have learned from these two works. 

\medskip
The first lesson is that the mathematical description of a potentially anomalous gapped/gapless phase $X$ depending on its codimension with respect to an anomaly-free topological order, in which $X$ is realized as a defect with a non-trivial codimension \cite{kong-wen-zheng-1}. For example, an anomaly-free 1+1D modular invariant RCFT has a precise mathematical description (see for example \cite[Theorem\ 4.17]{geometry}), which is a 0-codimensional description. If we want to regard it as a boundary of the trivial 2d topological order, then we need add all possible defects that are allowed by the local quantum symmetries (i.e. a non-chiral symmetry in this case). These defects form an enriched fusion category, which provides a 1-codimensional description of the anomaly-free 1+1D modular invariant RCFT. Moreover, the center of this enriched fusion category is precisely the 0-codimensional description of the trivial 2d topological order. We believe that this is a special case of a general principle for topological orders and its gapped/gapless boundaries in all dimensions. 
\begin{quote}
{\bf Boundary-bulk relation}: The center of the 1-codimensional categorical description, which contains all possible topological defects that can be obtained from elementary ones via condensations (called condensation descendants \cite{klwzz}), of a gapped/gapless boundary of an anomaly-free $n$d topological order $\CX$ coincides with the 0-codimensional categorical description of $\CX$. 
\end{quote}
See more discussion in \cite[Section\,3.3]{klwzz} and mathematical motivations in \cite{gjf}. 

\medskip
The second lesson is that the study of gapped phases is that of gapless phases in disguise. Indeed, a general gapless phase can be obtained by stacking a gapless phase with a gapped phase. Therefore, the mathematical structure of a gapped phase, such as the higher category of topological excitations \cite{kong-wen-zheng-1,jf20,klwzz}, is also an indispensable ingredient of that of a generic gapless phase. This structure might be changed if we introduce interactions between two stacked layers of gapped and gapless phases, but its higher categorical nature remains intact. On the other hand, instead of using stacking, we can also describe this structure intrinsically. For a potentially anomalous gapless phase, it is possible to have gapped excitations, which are topological sectors of the complete Hilbert space. These topological sectors should form a higher categorical structure similar to those topological excitations in a topological order. We will call this higher categorical structure the ``topological skeleton'' of the gapless phase. 
\begin{itemize}
\item For example, in the triple $(V,X,{}^\CP\CM)$ described in Theorem$^{\mathrm{ph}}$\ \ref{pthm:wall-chiral-edges}, the enriched category ${}^\CP\CM$ is the topological skeleton. For a complete mathematical description of the gapless phase, one need add local quantum symmetries to the topological skeleton, such as $V$ and $X$ in the triple $(V,X,{}^\CP\CM)$. 
\end{itemize}

We believe that this example has revealed the general features of gapless phases in all dimensions. More precisely, let us consider an anomalous gapped/gapless $n$D phase $\CX_n$ realized as a defect in an $(n+k)$D anomaly-free gapped/gapless phase $\CY_{n+k}$. More precisely, $\CX_n$ should be viewed as an $n$D domain wall between two $n+$1D defects $\CS_{n+1}$ and $\CT_{n+1}$ in $\CY_{n+k}$, so on and so forth. There is an $n$D local quantum symmetry $V_n$ defined on $\CX_n$, where $V_n$ is a (topological or conformal or geometrical) analogue of an $E_n$-algebra and will be called an $n$-disk algebra. There should be an $n+$1D local quantum symmetry $V_{n+1}$ defined in the $n+$1D neighborhood of $\CX_n$ determined by $\CS_{n+1}$ and $\CT_{n+1}$, where $V_{n+1}$ is an ($n+$1)-disk algebra and a common subalgebra of two ($n+$1)-disk algebras that define the $n+$1D local quantum symmetries in $\CS_{n+1}$ and $\CT_{n+1}$, respectively. Moreover, $V_n$ is an $n$-disk algebra over $V_{n+1}$ in the sense of Lurie \cite{lurie}. So on and so forth. 
As a consequence, we expect a $k$-codimensional mathematical description of $\CX_n$ to be given by $(V_{n+k}, \cdots, V_n, \CXs)$, where $\CXs$ is the topological skeleton. 

In this context, the Gapped-Gapless Correspondence proposed in \cite[Section\ 7]{kz4} can be restated as follows: the topological skeletons of gapless phases can all be obtained by topological Wick rotations from gapped phases. Also note that local quantum symmetries $V_i$ are not independent of and must be compatible with the topological skeletons, which provide a severe constraint to possible local quantum symmetries. In this sense, the topological skeleton can be viewed as a categorical symmetry of a gapless phase, and provides a powerful tool and a systematic way to study all gapless phases.


\begin{thebibliography}{99}

\bibitem[AKZ]{ai}
Y.H.~Ai, L.~Kong, H.~Zheng, 
{\it Topological orders and factorization homology}, Adv. Theor. Math. Phys. Vol. 21, Number 8, (2017) 1845-1894 [arXiv:1607.08422]


\bibitem[AF]{af}
D.~Ayala, J.~Francis,
{\it A factorization homology primer}, [arXiv:1903.10961]


\bibitem[AFT]{aft}
D.~Ayala, J.~Francis, H.~L.~Tanaka,
{\it Factorization homology of stratified spaces}, Selecta Math. (N.S.) 23, no. 1, 293--362 (2017)



\bibitem[BK]{bk}
S.B.~Bravyi, A.Y.~Kitaev, 
{\it Quantum codes on a lattice with boundary}, [arXiv:quant-ph/9811052]


\bibitem[BS]{bs}
F.A. Bais, J.K. Slingerland, 
{\it Condensate induced transitions between topologically ordered phases}, Phys. Rev. B 79, 045316 (2009). 


\bibitem[CCBCN]{ccbcn}
J.~Cano, M.~Cheng, M.~Barkeshli, D.J.~Clarke, C.~Nayak
{\it Chirality-Protected Majorana Zero Modes at the Gapless boundary of Abelian Quantum Hall States}, 
Phys. Rev. B 92, 195152 (2015) arXiv:1505.07825 [cond-mat.str-el]


\bibitem[CZ]{cz}
A.~Cappelli, G.~Zemba, 
{\it Modular invariant partition functions in the quantum Hall effect}, Nucl. Whys. B 490, 595-632 (1997)


\bibitem[CJKYZ]{cjkyz}
W.-Q.~Chen, C.-M.~Jian, L.~Kong, Y.-Z.~You, H.~Zheng, 
{\it A topological phase transition on the edge of the 2d $\Zb_2$ topological order}, [arXiv:1903.12334]


\bibitem[Da]{davydov}
A. Davydov, 
{\it Centre of an algebra}, Adv. Math. 225 (2010) 319-348. 


\bibitem[DKR]{dkr}
A.~Davydov, L.~Kong, I.~Runkel,
{\it Functoriality of the center of an algebra}, 
Adv. Math. 285 (2015) 811-876. 

\bibitem[DMNO]{dmno}
A.~Davydov, M.~M\"{u}ger, D.~Nikshych, V.~Ostrik,
{\it The Witt group of nondegenerate braided fusion categories},
 J. Reine Angew. Math. 677 (2013), 135-177 [arXiv:1009.2117]



\bibitem[EGNO]{egno}
P.~Etingof, S.~Gelaki, D.~Nikshych, V.~Ostrik, 
{\it Tensor Categories}, American Mathematical Society, Providence, RI, 2015.



\bibitem[ENO2]{eno2008} P. Etingof, D. Nikshych, V. Ostrik,
{\it Weakly group-theoretical and solvable fusion categories},
Adv. Math. 226 (2010), no. 1, 176--205. [arXiv:0809.3031]


\bibitem[ENO3]{eno2009}
P.I. Etingof, D. Nikshych, V. Ostrik, 
{\it Fusion categories and homotopy theory}, 
Quantum Topol. 1 (2010) 209-273.



\bibitem[FjFRS]{fjfrs}
J. Fjelstad, J. Fuchs, I. Runkel, C. Schweigert, Uniqueness of open/closed rational CFT with given algebra of open states, Adv. Theor. Math. Phys. 12 (2008) 1283-1375. 



\bibitem[FFRS]{ffrs}
J.~Fr\"{o}hlich, J.~Fuchs, I.~Runkel, C.~Schweigert,
{\it Duality and defects in rational conformal field theory}, Nucl. Phys. B 763, 354-430 (2007)


\bibitem[FRS]{frs}
J. Fuchs, I. Runkel, C. Schweigert, 
{\it TFT construction of RCFT correlators. I: partition functions}, Nuclear Phys. B 646 (2002) 353-497. 


\bibitem[FSV]{fsv}
J.~Fuchs, C.~Schweigert, A.~Valentino,
{\it Bicategories for boundary conditions and for surface defects in 3-d TFT}, 
Commun. Math. Phys., Volume 321, Issue 2, 543-575

\bibitem[GJF]{gjf}
D. Gaiotto, T. Johnson-Freyd, 
{\it Condensations in higher categories}, [arXiv:1905.09566].

\bibitem[Hu]{huang-mtc}
Y.-Z.~Huang,
{\it Vertex operator algebras and the Verlinde conjecture}, Comm. Contemp.
Math. 10 (2008), 103-154.



\bibitem[HKL]{hkl}
Y.-Z.~Huang, A.~Kirillov, Jr. J.~Lepowsky, 
{\it Braided tensor categories and extensions of vertex operator algebras}, 
Comm. Math. Phys. 337 (2015), no. 3, 1143-1159 [arXiv:1406.3420]

\bibitem[HK1]{osvoa}
Y.-Z.~Huang, L~Kong, 
{\it Open-string vertex algebras, tensor categories and operads}, 
Commun. Math. Phys. {\bf 250} (2004) 433-471 [math.QA/0308248]

\bibitem[HK2]{ffa}
Y.-Z.~Huang, L~Kong,
{\it Full field algebras}, Comm. Math. Phys. 272 (2007)
345-396, [arXiv:math/0511328]

\bibitem[HK3]{ffa-mod-inv}
Y.-Z. Huang and L. Kong, 
{\it Modular invariance for conformal full field algebras}, 
Trans. Amer. Math. Soc. 362 (2010), no. 6, 3027-3067.


\bibitem[JW]{jw}
W.~Ji, X.-G.~Wen,
{\it Non-invertible anomalies and mapping-class-group transformation of anomalous partition functions}, Phys. Rev. Research 1, 033054 (2019) [arXiv:1905.13279 ]

\bibitem[JF]{jf20}
T.~Johnson-Freyd, 
{\it On the classification of topological orders}, (2020), [arXiv:2003.06663].

\bibitem[KO]{KO}
A. Kirillov Jr., V. Ostrik, 
{\it On q-analog of McKay correspondence and ADE classification of $\mathrm{sl}_2$ conformal field
theories}, Adv. Math. 171 (2) (2002) 183-227.



\bibitem[Ki1]{kitaev}
A.Y.~Kitaev,
{\it Anyons in an exactly solved model and beyond},
Ann. Phys. 321 (2006) 2-111. [arXiv:cond-mat/0506438]



\bibitem[KK]{kk}
A.~Kitaev, L.~Kong,
{\it Models for gapped edges and domain walls}, Commun. Math. Phys.
313 (2012) 351-373, [arXiv:1104.5047].


\bibitem[Ko1]{kong-ffa}
L.~Kong
{\it Full field algebras, operads and tensor categories}, Adv. Math. 213 (2007)
271-340, [arXiv:math/0603065]

\bibitem[Ko2]{geometry}
L.~Kong, 
{\it Conformal field theory and a new geometry}, Mathematical Foundations of Quantum Field and Perturbative String Theory, Hisham Sati, Urs Schreiber (eds.), Proceedings of Symposia in Pure Mathematics, AMS, Vol. 83 (2011) 199-244 [arXiv:1107.3649]


\bibitem[Ko3]{anyon}
L.~Kong,
{\it Anyon condensation and tensor categories}, Nucl. Phys. B 886 (2014) 436-482 

\bibitem[KLWZZ]{klwzz}
L.~Kong, T.~Lan, X.-G.~Wen, Z.-H.~Zhang, H.~Zheng,
{\it Classification of topological phases with finite internal symmetries in all dimensions}, [arXiv:2003.08898]

\bibitem[KR1]{kr1}
L.~Kong, I.~Runkel, 
{\it Morita classes of algebras in modular tensor categories}, Adv. Math. 219,
1548-1576 (2008), 


\bibitem[KR2]{kr2}
L.~Kong, I.~Runkel, 
{\it Cardy Algebras and Sewing Constraints, I}, 
Commun. Math. Phys. 292, 871-912 (2009)


\bibitem[KWZ1]{kong-wen-zheng-1}
L.~Kong, X.-G.~Wen, H.~Zheng,
{\it Boundary-bulk relation for topological orders as the functor mapping higher categories to their centers}, [arXiv:1502.01690]

\bibitem[KWZ2]{kong-wen-zheng-2}
L.~Kong, X.-G.~Wen, H.~Zheng,
{\it Boundary-bulk relation in topological orders}, Nucl. Phys. B 922 (2017), 62-76 [arXiv:1702.00673]


\bibitem[KYZ1]{kyz}
L.~Kong, W.~Yuan, H.~Zheng, 
{\it Pointed Drinfeld center functor}, [arXiv:1912.13168]


\bibitem[KYZ2]{kyz2}
L.~Kong, W.~Yuan, H.~Zheng,
{\it Enriched monoidal categories}, in preparation. 

\bibitem[KZ1]{kz1}
L.~Kong, H.~Zheng,
{\it The center functor is fully faithful}, Adv. Math. 339 (2018) 749-779 [arXiv:1507.00503]

\bibitem[KZ2]{kz2}
L.~Kong, H.~Zheng,
{\it Drinfeld center of enriched monoidal categories}, Adv. Math. 323 (2018) 411-426 [arXiv:1704.01447]

\bibitem[KZ3]{kz3}
L.~Kong, H.~Zheng,
{\it Gapless edges of 2d topological orders and enriched monoidal categories}, Nucl. Phys. B 927 (2018) 140-165 [arXiv:1705.01087]


\bibitem[KZ4]{kz4}
L.~Kong, H.~Zheng,
{\it A mathematical theory of gapless edges of 2d topological orders. Part I}, J. High Energ. Phys. 2020, 150 (2020) [arXiv:1905.04924] 

\bibitem[Le]{levin}
M. Levin, 
{\it Protected edge modes without symmetry}, Phys. Rev. X, 3 021009 (2013)

\bibitem[Lu]{lurie} J. Lurie, 
	Higher algebras, 
	a book available at: http://www.math.harvard.edu/\textasciitilde lurie/papers/ higheralgebra.pdf.


\bibitem[O]{ostrik}
 V. Ostrik, 
 {\it Module categories, weak Hopf algebras and modular invariants}, Transform. Groups 8 (2003) 177-206
	



\bibitem[Zhe]{zheng}
H.~Zheng,
{\it Extended TQFT's arising from enriched multi-fusion categories}, [arXiv:1704.05956]



\end{thebibliography}
\end{document}